\documentclass[aps,prl,reprint,groupedaddress]{revtex4-2}

\usepackage{amsmath}
\usepackage{graphicx}
\usepackage[bookmarks=false]{hyperref}
\usepackage{mathtools}
\usepackage{stmaryrd}
\usepackage{wasysym}
\usepackage{bm}

\hypersetup{colorlinks,citecolor=blue,linkcolor=blue,urlcolor=blue}

\newcommand{\tr}[1]{\ensuremath{\textrm{tr}\left(#1\right)}}
\renewcommand{\vec}[1]{\mathbf{#1}}

\DeclarePairedDelimiter\norm{\lVert}{\rVert}

\begin{document}

\title{Numerical evidence for many-body localization in two and three dimensions}

\author{Eli Chertkov, Benjamin Villalonga, and Bryan K. Clark}
\affiliation{Institute for Condensed Matter Theory and IQUIST and Department of Physics, University of Illinois at Urbana-Champaign, Urbana, Illinois 61801, USA}

\begin{abstract}
Disorder and interactions can lead to the breakdown of statistical mechanics in certain quantum systems, a phenomenon known as many-body localization (MBL).
Much of the phenomenology of MBL emerges from the existence of $\ell$-bits, a set of conserved quantities that are quasilocal and binary (i.e., possess only $\pm 1$ eigenvalues).
While MBL and $\ell$-bits are known to exist in one-dimensional systems, their existence in dimensions greater than one is a key open question.
To tackle this question, we develop an algorithm that can find approximate binary $\ell$-bits in arbitrary dimensions by adaptively generating a basis of operators in which to represent the $\ell$-bit.
We use the algorithm to study four models: the one-, two-, and three-dimensional disordered Heisenberg models and the two-dimensional disordered hard-core Bose-Hubbard model.
For all four of the models studied, our algorithm finds high-quality $\ell$-bits at large disorder strength and rapid qualitative changes in the distributions of $\ell$-bits in particular ranges of disorder strengths, suggesting the existence of MBL transitions.
These transitions in the one-dimensional Heisenberg model and two-dimensional Bose-Hubbard model coincide well with past estimates of the critical disorder strengths in these models which further validates the evidence of MBL phenomenology in the other two and three-dimensional models we examine. 
In addition to finding MBL behavior in higher dimensions, our algorithm can be used to probe MBL in various geometries and dimensionality.
\end{abstract}

\maketitle

\emph{Introduction.}--- It is natural to expect quantum systems to obey statistical mechanics. However, there is increasing evidence that there exist disordered strongly interacting quantum systems that do not obey the laws of statistical mechanics and never reach thermal equilibrium -- a phenomenon known as many-body localization (MBL) \cite{Anderson1958,fleishman1980interactions,gornyi2005interacting,basko2006metal,Nandkishore2015,Abanin2017,Abanin2019}. A key feature of MBL systems is they exhibit robust emergent integrability, i.e., they possess many quasilocal \footnote{In the MBL literature, a ``quasilocal'' operator refers to an operator that has compact support over a finite region and exponentially decaying tails beyond that region. In other contexts, such as when discussing Anderson localization, such operators would be called local or localized instead.} conserved quantities (known as $\ell$-bits) \cite{Serbyn2013,Huse2014,Imbrie2017}. The existence of these robust conserved quantities is strongly related to other well-known properties of MBL, such as area-law entanglement of excited states and logarithmic growth of entanglement entropy under time-evolution \cite{Nandkishore2015,Abanin2017,Abanin2019}. Numerical methods have been key to studying MBL \cite{Luitz2015,Villalonga2018,Luitz2015,Serbyn2016,bauer2013area,kjall2014many,Yu2016,Yu2017,luitz2016long,Khemani2016,lim2016many}, but have mostly been limited to small finite-size systems and one spatial dimension.

A key open question that remains is the role of dimensionality in MBL \cite{Abanin2019}. In one-dimension, there is significant numerical and analytic evidence for MBL phenomena (although even this is still controversial \cite{panda2020can}). In higher dimensions, the situation is less clear. Cold-atom experiments show some signatures of slow thermalization in two and three dimensions \cite{Choi2016,Bordia2017,Kondov2015}. Some have argued that MBL phases are unstable to rare ergodic regions that trigger thermalizing avalanches \cite{DeRoeck2017a,DeRoeck2017b}. Others have suggested that an MBL phase might survive but only in nonstandard thermodynamic limits \cite{Chandran2016,Agarwal2017,Gopalakrishnan2019}.
In this work we take a pragmatic approach and numerically search for $\ell$-bits in higher dimensions, which we take as a practical signature of MBL.  Being able to predict properties of MBL in higher dimensions is also key to making the connection to two and three dimensional cold-atom experiments.    While some numerical approaches exist in two-dimensions \cite{BarLev2016,Inglis2016,Thomson2018,Kennes2018,Wahl2019,Geissler2019,DeTomasi2019,Theveniaut2019,Kshetrimayum2019,Pietracaprina2020,Doggen2020}, simulating MBL in higher dimensions is still largely intractable and it is important to develop new numerical techniques, particularly in three-dimensions, where to our knowledge no numerical studies have been done.

In this work, we present a new algorithm for finding approximate $\ell$-bits (or $\ell$-bit-like operators \cite{Chandran2016}) in interacting disordered systems of arbitrary dimensions. In MBL systems, an exact $\ell$-bit is an operator that (1) is quasilocal, (2) commutes with the Hamiltonian, and (3) has a binary spectrum, i.e., a spectrum of half $+1$ and half $-1$ eigenvalues. Our algorithm constructs an \emph{approximate} $\ell$-bit by finding an operator that satisfies these three properties as closely as possible. Property (1) is approximated by representing the approximate $\ell$-bit as a linear combination of finitely many local Pauli strings, while properties (2) and (3) are approximated by minimizing an objective function using gradient descent. 
Some previously developed numerical methods for finding $\ell$-bits in MBL systems have attempted to enforce these properties exactly \cite{Pekker2017b,Kulshreshta2018,Goihl2018,Yu2019,Varma2019,Peng2019}. Other methods have attempted to numerically construct operators that approximately satisfy properties (1) and (2) and either exactly enforce the binary property (3) \cite{Thomson2018,Kelly2020} or do not enforce that property at all \cite{Kim2015,Chandran2015, OBrien2016, Inglis2016, Lin2017, Mierzejewski2018, Pancotti2018}. Many of these methods have required numerically expensive calculations, e.g., exact diagonalization or large bond-dimension tensor networks, and, except for the methods of Refs.~\onlinecite{Inglis2016,Thomson2018,Wahl2019}, have been limited to the study of one-dimensional chains. Our algorithm can efficiently produce operators that are reasonable approximations of binary, quasilocal $\ell$-bits in arbitrary dimensions.

Using our algorithm, we study four model Hamiltonians: the disordered Heisenberg model in one, two, and three-dimensions, and the disordered hard-core Bose-Hubbard model in two-dimensions (also examined in Refs.~\onlinecite{Wahl2019,Geissler2019}). In all models studied, we find high quality $\ell$-bits at high disorder strengths suggesting MBL behavior and see statistical signatures of a potential transition from localized to delocalized integrals of motions. Our results provide new evidence for the existence of MBL phenomenology in two and three-dimensions.

\emph{Background.}--- In this work, we investigate two different types of Hamiltonians. First, we consider the disordered spin-$1/2$ Heisenberg model
\begin{align}
H = \sum_{\langle i j \rangle} \mathbf{S}_i \cdot \mathbf{S}_j + \sum_i h_i S^z_i \label{eq:Hheisenberg}
\end{align}
where the first summation is over nearest neighbor sites of a 1D, 2D, or 3D lattice, $h_i \in [-W, W]$ are random numbers drawn from a uniform distribution, and $W$ is the disorder strength. The 1D model has been extensively investigated numerically, mostly using exact diagonalization \cite{Oganesyan2007,Pal2010,Luitz2015,Serbyn2016,Yu2016} and tensor networks \cite{Znidaric2008,Bardarson2012,DeLuca2013,Pollman2016,Yu2017,Khemani2016,Pekker2017a,Wahl2017}. However, the model in higher dimensions has, up to this point, been largely unexplored \cite{Inglis2016,Kshetrimayum2019}.

Second, we consider the disordered Bose-Hubbard model
\begin{align}
H = -\sum_{\langle i j \rangle} \left(a_i^\dagger a_j  + \textrm{H.c.}\right) + \frac{U'}{2}\sum_{i}n_i (n_i - 1)  + \sum_i \delta_i n_i \label{eq:Hbosehubbard}
\end{align}
where the first summation is over nearest neighbor sites of a two-dimensional square lattice, $a_i^\dagger$ and $a_i$ are bosonic creation and annihilation operators, $n_i \equiv a^\dagger_i a_i$, and $\delta_i$ are random on-site potentials drawn from a Gaussian distribution with full-width half-maximum $\Delta$. This model approximately describes the interactions between bosonic $^{87}$Rb atoms in a two-dimensional disordered optical lattice experiment~\cite{Choi2016}, where a potential MBL-ergodic transition was observed at $\Delta_c^{exp} \approx 5.5(4)$ with $U'=24.4$. Refs.~\onlinecite{Wahl2019} numerically studied this model in the hard-core limit using tensor networks, where they found a transition at $\Delta_c^{tn} \approx 19$; we too work in this limit.

Generically, a Hamiltonian such as Eq.~(\ref{eq:Hheisenberg})~or~(\ref{eq:Hbosehubbard}) can be represented as
\begin{align}
H = \sum_i \tilde{h}_i \tau^z_i + \sum_{i,j} \tilde{J}_{ij} \tau^z_i \tau^z_j + \sum_{i,j,k} \tilde{J}_{ijk} \tau^z_i \tau^z_j\tau^z_k + \cdots
\end{align}
where $\tilde{h}_i,\tilde{J}_{ij},\ldots$ are coupling constants and $\tau^z_i = U^\dagger \sigma^z_i U$ where $U$ is a unitary that diagonalizes the Hamiltonian. The $\tau^z_i$ operators are integrals of motion ($[H, \tau^z_i ] = 0$) that mutually commute ($[\tau^z_i , \tau^z_j ] = 0$) and have a binary spectrum ($(\tau^z_i)^2 = I$ and $\tr{\tau^z_i}=0$). Note that these operators are not unique since there exist many unitaries that diagonalize $H$. In MBL systems, the $\tau^z_i$ operators can be made quasilocal, so that the support of the operators decays rapidly away from a single site on which they are localized, and are known as  $\ell$-bits. A $\tau^z_i$ operator can be written as
\begin{align}
\tau^z_i &= \sum_{a=1}^{|B|} c_a \mathcal{O}_a, \label{eq:tauz}
\end{align}
where $c_a$ is a real coefficient, $\mathcal{O}_a$ is a Pauli string (a product of Pauli matrices, such as $\sigma^x_1 \sigma^x_3 \sigma^z_5$), and $B=\{\mathcal{O}_a\}_{a=1}^{|B|}$ is a basis of Pauli strings of size $|B|$. The quasilocality of $\ell$-bits make it possible to accurately represent them using a small, finite basis $B$ of local Pauli strings.

\begin{figure}
\begin{center}
\includegraphics[width=\columnwidth]{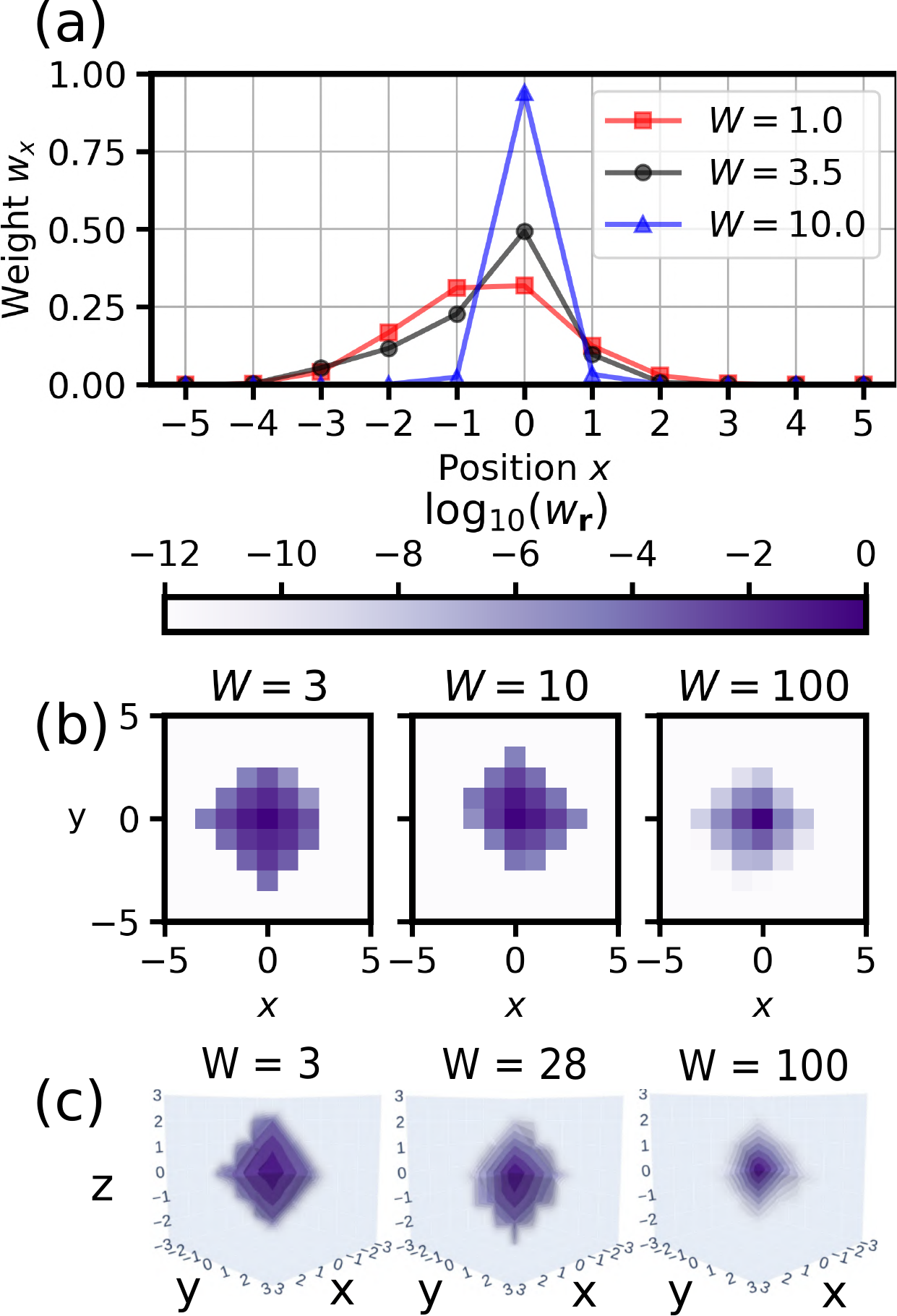}
\end{center}
\caption{Typical weights $w_{\vec{r}}$ of random $\tau^z_i$ for the (a) 1D, (b) 2D, and (c) 3D disordered Heisenberg models at different disorder strengths.} \label{fig:fig1}
\end{figure}

To quantify quasilocality, we can define the weight $w_{\vec{r}}$ of a $\tau^z_i$ operator \cite{Pancotti2018,Kulshreshta2018} as
\begin{align}
w_{\vec{r}} = \frac{\sum_{a \in B_{\vec{r}}} |c_a|^2}{\sum_{\vec{r}'} \sum_{b \in B_{\vec{r}'}} |c_b|^2} \label{eq:weight}
\end{align}
where $\vec{r}$ is the spatial coordinate of a site in the lattice and $B_{\vec{r}}$ is the set of (labels of) Pauli strings in the basis $B$ with (non-identity) support on lattice coordinate $\vec{r}$. The weight $w_{\vec{r}}$
decays rapidly in MBL phases, as shown in Fig.~\ref{fig:fig1}.

\emph{Method.}--- Our algorithm constructs quasilocal operators $\tau^z_i$ that approximately commute with the Hamiltonian and are approximately binary. In particular, the algorithm optimizes the $c_a$ parameters in Eq.~(\ref{eq:tauz}) to minimize the objective function
\begin{align}
Z[\{c_a\}] = \alpha \norm{[H, \tau^z_i ]}^2 + \beta \norm{\left( \tau^z_i \right)^2 - I}^2, \label{eq:obj}
\end{align}
where $\alpha,\beta > 0$, $\norm{O}^2 \equiv \tr{O^\dagger O}/\tr{I}$ is the Frobenius norm, and $I$ is the identity operator. As described in the supplement \footnote{See Supplemental Material for additional details on the methods used and for additional data obtained in this work. The supplement includes Refs.~\onlinecite{scipy2020,Canovi2011,Rademaker2017}.}, this minimization is done using gradient descent and Newton's method. Note that if the second term of Eq.~(\ref{eq:obj}) is zero, then the eigenvalues of $\tau^z_i$ have exactly equal sectors of $\pm 1$ eigenvalues because $\tau_i^z$ is traceless. Also note that while we do not constrain $\tau^z_i$ to be normalized ($\norm{\tau^z_i}^2 = \sum_a c_a^2=1$), it stays approximately normalized during the optimization because of the second term of Eq.~(\ref{eq:obj}). We set $\alpha=\beta=1$.

Rather than perform a single minimization of Eq.~(\ref{eq:obj}) in a fixed basis $B$, we iteratively and adaptively build the basis during the minimization (similar in spirit to selected configuration interaction, an adaptive basis technique in quantum chemistry \cite{Bender1969,Whitten1969,Holmes2016,Tubman2016}). The steps of the algorithm are:
\begin{enumerate}
    \item Initialize $B = \{\sigma_i^z \}$.
    \item Expand $B$ by adding new Pauli strings.
    \item Minimize Eq.~(\ref{eq:obj}) in basis $B$.
    \item Repeat steps 2--3 while $|B| \leq |B|_{max}$.
\end{enumerate}
In step 1, we initialize the basis with a single Pauli matrix at site $i$. In step 2, we expand the basis by including new Pauli strings that are important for minimizing the objective in Eq.~(\ref{eq:obj}). In particular, our heuristic expansion procedure is two-step: (a) first, we compute $[H,[H,\tau^z_i]] = \sum_{a} c_a' \mathcal{O}_a$ and add $M_1$ new Pauli strings $\mathcal{O}_a$ to $B$ with the largest amplitudes $|c_a'|$ \footnote{In order to save memory and time in our calculations, we modified step (a) so that only the largest 2000 terms of $[H, \tau^z_i]$ were kept before computing $[H, [H, \tau^z_i]]$.}; (b) then, we compute $(\tau^z_i)^2 - I = \sum_{a} c_a'' \mathcal{O}_a$ and add $M_2$ new Pauli strings to $B$ with the largest amplitudes $|c_a''|$. The logic behind step (a) is that, to cancel the remainder of $[H, \tau^z_i]$, we need to add Pauli strings that, when commuted through the Hamiltonian, coincide with the remainder. These are the terms in $[H, [H, \tau^z_i]]$. The logic is similar for step (b). In our calculations, we set $M_1=M_2=100$ and perform $11$ basis expansions, so that we expand by up to $200$ Pauli strings per iteration to a maximum basis size of $|B|_{max}=2201$. In step 3, we perform gradient descent with the $c_a$ parameters in Eq.~(\ref{eq:tauz}) initialized to the optimized values obtained in the previous basis size, but rescaled so they are normalized to one.

We execute our algorithm on 1D, 2D, and 3D periodic lattices of size $101$, $21 \times 21$ and $11 \times 11 \times 11$, respectively. It is important to note that, because of the basis sizes $|B|$ considered, the optimized $\tau^z_i$ never reach the lattice boundaries, indicating that our calculations do not exhibit any finite system-size effects or boundary effects, but do exhibit finite \emph{basis-size} effects.

Our code is available online \cite{bioms} and is based on the Qosy package \cite{qosy}.

\begin{figure}
\begin{center}
\includegraphics[width=0.45\textwidth]{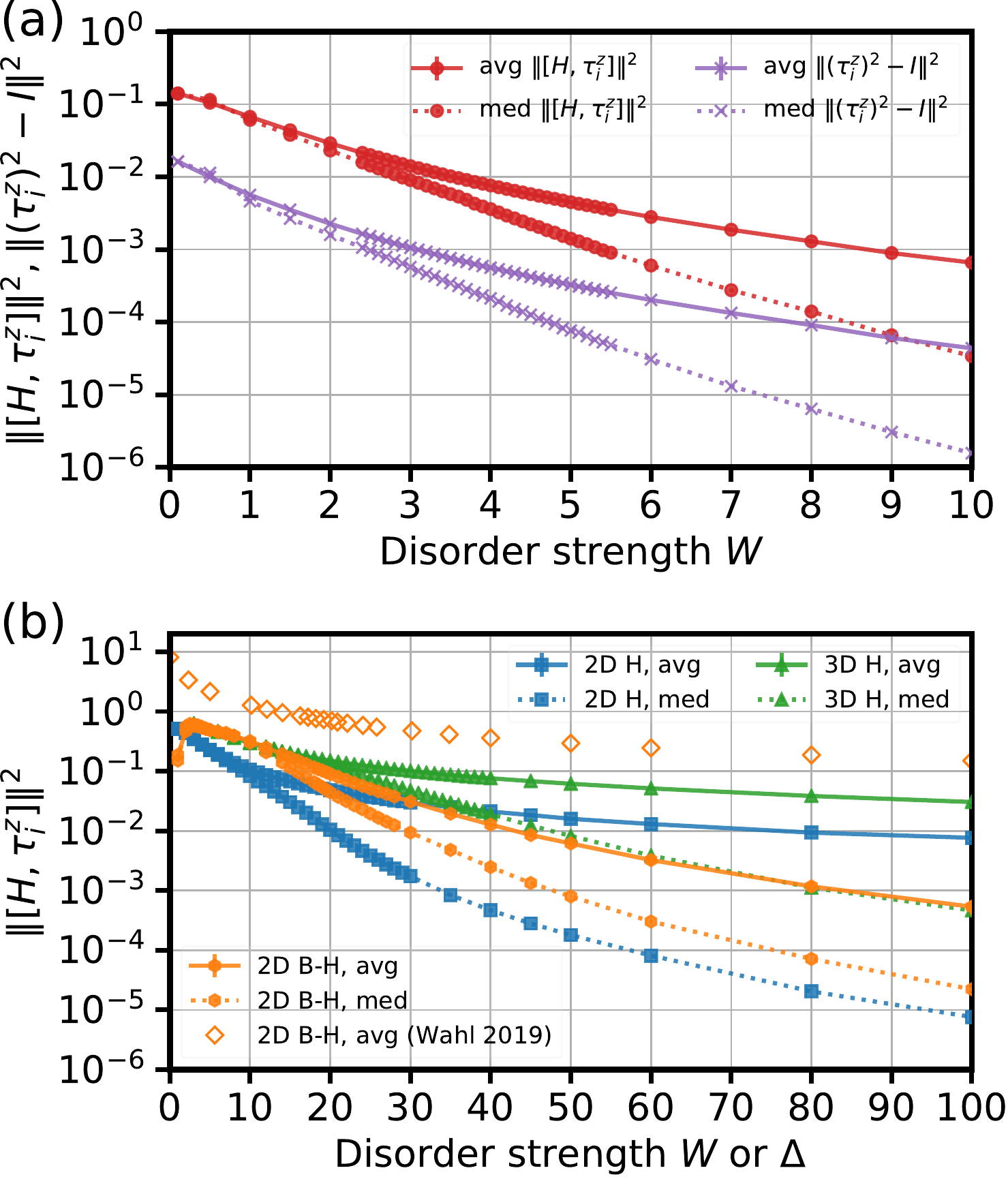}
\end{center}
\caption{The average and median commutator norms $\norm{[H, \tau^z_i]}^2$ and binarities $\norm{(\tau^z_i)^2 - I}^2$ (only for (a)) of our optimized $\tau^z_i$ operators for the disordered (a) 1D Heisenberg model and (b) 2D and 3D Heisenberg models and 2D hard-core Bose-Hubbard model. The average commutator norms obtained by Ref.~\onlinecite{Wahl2019} (Wahl 2019) using shallow 2D tensor networks for the 2D Bose-Hubbard model are also shown. Note that the method of Ref.~\onlinecite{Wahl2019} finds all $\tau^z_i$ in a $10 \times 10$ lattice, while our method finds only a single $\tau^z_i$.} \label{fig:fig2}
\end{figure}

\begin{figure}
\begin{center}
\includegraphics[width=0.5\textwidth]{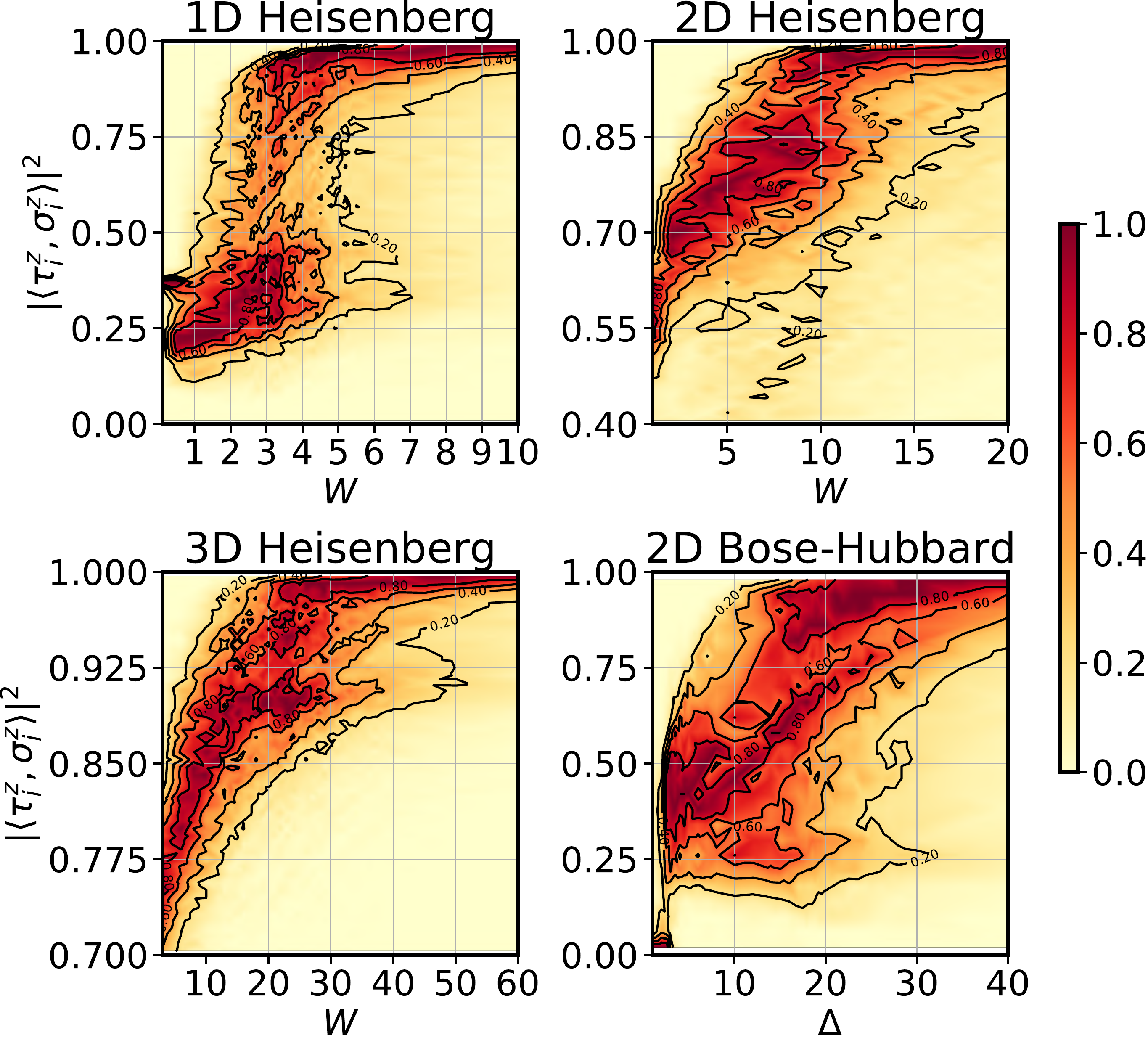}
\end{center}
\caption{Interpolated histograms of $|\langle \tau^z_i, \sigma^z_i \rangle|^2$ at different disorder strengths. The histograms are made of 50 evenly spaced bins (25 for 2D Bose-Hubbard) and are normalized so that at a fixed disorder strength the maximum of the histogram is at a value of 1. The black lines are contour lines corresponding to normalized histogram values of 0.2, 0.4, 0.6, and 0.8.} \label{fig:fig3}
\end{figure}

\begin{figure}
\begin{center}
\includegraphics[width=0.45\textwidth]{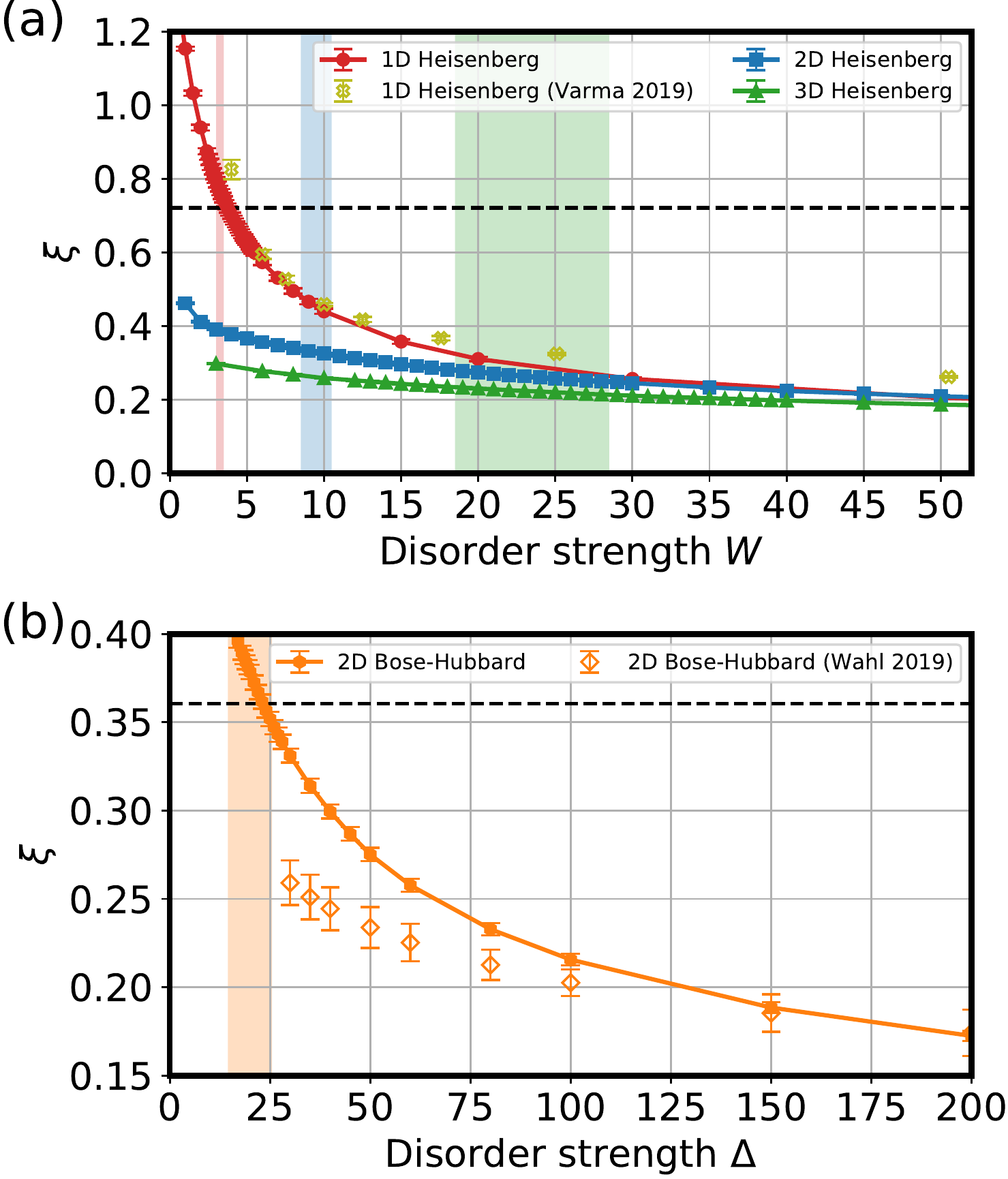}
\end{center}
\caption{The average correlation lengths of our $\tau^z_i$ operators versus disorder strength. For comparison, we show average correlation lengths of $\ell$-bits obtained by Ref.~\onlinecite{Varma2019} (Varma 2019) for the 1D model and by Ref.~\onlinecite{Wahl2019} (Wahl 2019) for the 2D Bose-Hubbard model. Horizontal dashed lines are drawn at (a) $\xi = 1/\ln(4)$ and (b) $\xi = 1/\ln(4^2)$; shading indicates our estimates of the transition regions (see Fig.~\ref{fig:fig3} and supplement).} \label{fig:fig4}
\end{figure}

\emph{Results and discussion.}--- Using our algorithm, we obtain $\tau^z_i$ operators for 1600 random realizations of the disordered Heisenberg models of Eq.~(\ref{eq:Hheisenberg}) and for 800 realizations of the disordered hard-core Bose-Hubbard model of Eq.~(\ref{eq:Hbosehubbard}) \footnote{We use the same set of (scaled) disorder patterns for all $W$ for a fixed model but different disorder patterns for different models.}. In this section, we present some statistical properties of the (normalized) $\tau^z_i$ operators that our algorithm finds after the final iteration of basis expansions (see supplement for earlier iterations).  

At high disorder, we find $\tau^z_i$ operators that are largely binary and nearly commute with the Hamiltonian for all four models studied (see Fig.~\ref{fig:fig2}). This is anticipated in an MBL phase where quasilocal operators should be well represented by a small local basis of operators. However, the algorithm's ability to find good $\ell$-bits 
becomes 
1--2 orders of magnitude
worse with respect to both the commutator norm $\norm{[H, \tau^z_i]}^2$ and binarity $\norm{(\tau^z_i)^2-I}^2$ with decreasing disorder strength.  
We also compare the rate of convergence as a function of basis size (see Figs.~S23-24 in supplement); while the errors decrease with basis size, they fall off slowly. Improving the rate of convergence is an interesting area for future improvement of the algorithm.

An important statistical quantity that we consider is the overlap $|\langle \tau^z_i, \sigma^z_i \rangle|^2$ \footnote{Note that when we compute this quantity we use the $\sigma^z_i$ on the site $i$ with the largest weight (see Eq.~(\ref{eq:weight})) rather than the $\sigma^z_i$ used to initialize the basis $B$. In general, the $\tau^z_i$ operators discovered with our method can ``drift'' away from their initial site, though this tends to only become significant at low disorder strength (see supplement).} (see Fig.~\ref{fig:fig3} for their distributions). At high disorder, most $\tau^z_i$ operators are localized so that $|\langle \tau^z_i, \sigma^z_i \rangle|^2 \approx 1$, with the distribution exhibiting a quickly decaying tail away from this value. At low disorder, there are almost no operators with $|\langle \tau^z_i, \sigma^z_i \rangle|^2 \approx 1$; instead most operators have an overlap with a non-zero value significantly below one. For all the models studied, we find a rapid change in the probability distribution of these operator overlaps over a narrow region of disorder; within this region we see hints of bimodality \cite{kjall2014many,Yu2016,Villalonga2018} of the probability distribution. We would anticipate that this rapid change signals a ``transition.''

We find in 1D that the location of this transition region is in good agreement with the accepted location of the MBL-ergodic transition in the range $3 \lesssim W \lesssim 3.5$ \cite{Pal2010,DeLuca2013,Luitz2015,Serbyn2016,Yu2016,Wahl2017,Chandran2015,OBrien2016,Mierzejewski2018,Pancotti2018,Goihl2018,Kulshreshta2018}.
Moreover, the transition region of $14.5 \lesssim \Delta \lesssim 25.5$ in the 2D hard-core Bose-Hubbard model is consistent with the critical disorder strength of $\Delta_c^{tn} \approx 19$ estimated by Ref.~\onlinecite{Wahl2019}.
The rapid changes in the probability distributions of $|\langle \tau^z_i, \sigma^z_i \rangle|^2$ in the 2D and 3D Heisenberg models and their high overlap at large disorder then suggests that similar MBL transitions exist in these models as well.
These transitions happen around $8.5 \lesssim W \lesssim 10.5$ and $18.5 \lesssim W \lesssim 28.5$, respectively. See supplement for details on the estimation of the approximate location of the transition regions.

We note that in 1D, the two peaks of $|\langle \tau^z_i, \sigma^z_i \rangle|^2$ in the transition region are more separated than in higher dimensions.  We believe this is due to limitations of the basis size; in 1D, as the basis size $|B|$ grows the separation between the peaks also grows (see supplement) and we expect the same to hold for other models.

Another quantity we use to characterize  $\tau^z_i$  is the correlation length, shown in Fig.~\ref{fig:fig4}. We obtain correlation lengths by fitting the function $\tilde{w}_\vec{r} = e^{-\norm{\vec{r} - \vec{r}_i}/\xi}/(\sum_{\vec{r}'} e^{-\norm{\vec{r}' - \vec{r}_i}/\xi})$ to the weight $w_{\vec{r}}$ of Eq.~(\ref{eq:weight}) for the $\tau^z_i$ centered at site $\vec{r}_i$ using a non-linear least-squares fit 
\footnote{Note that the summation in the denominator of $\tilde{w}_{\vec{r}}$ is only over the positions $\vec{r}'$ where $w_{\vec{r}'} \neq 0$ and $\vec{r}_i \equiv \textrm{argmax}_{\vec{r}} w_{\vec{r}}$.}. We should note that while this fitting procedure gave sensible results for all models, other reasonable ways of fitting these approximate $\ell$-bits were less robust. 
For a wide range of disorder strengths, our 1D Heisenberg model correlation lengths agree with those obtained by Ref.~\onlinecite{Varma2019} (see supplement for additional correlation length comparisons). 
For large disorder strengths, our 2D Bose-Hubbard correlation lengths agree with those obtained by Ref.~\onlinecite{Wahl2019} using shallow 2D tensor networks, but take on larger values at low disorder strength. As shown in Fig.~\ref{fig:fig2}(b), our $\ell$-bits have significantly lower commutator norms, so might be able to more accurately capture the $\tau^z_i$ operators near the transition. As expected theoretically, none of the correlation lengths diverge at the ``transition.'' Interestingly, we empirically find that $\xi \approx 1/\ln(4^d)$, where $d$ is the spatial dimension, near the transition region.  While the $d=1$ value agrees with some theoretical predictions \cite{Varma2019}, we are not aware of expected values of correlation lengths at the transition region in higher $d$ and these values in larger dimensions might be coincidental.

Finally, we note that for the 2D Bose-Hubbard model we see a sharp change in the histogram of $|\langle \tau^z_i, \sigma^z_i \rangle|^2$ at $\Delta \approx 3$ (see Fig.~\ref{fig:fig3}) somewhat close to the $\Delta_c^{exp} \approx 5.5(4)$ value obtained experimentally by Ref.~\onlinecite{Choi2016}. Near this disorder strength the binarity of our $\ell$-bits increases sharply and so this behavior could simply be attributed to a breakdown of our algorithm (see supplement); nonetheless, we cannot rule out that the algorithm breaking down near this low $\Delta$ is somehow related to the results seen in the experimental systems.

\emph{Outlook.}--- We present an algorithm for constructing high-quality approximations of quasilocal binary integrals of motion and use it to study MBL in four different models. This algorithm works by adaptively building a basis of operators in which to construct the quasilocal integrals of motion ($\ell$-bits). Using this algorithm, we find the first theoretical evidence for MBL in three dimensions.

Our algorithm is well suited for studying $\ell$-bits in more general settings than has previously been possible. 
For example, it can be used to construct approximate $\ell$-bits for models on complicated lattice geometries, for fermionic models (in which Majorana strings can be used instead of Pauli strings; see Ref.~\onlinecite{Chertkov2020}), or for models with potential MBL-MBL transitions \cite{Pekker2014}. 
Moreover, using the strategy of Ref.~\onlinecite{Inglis2016}, the $\ell$-bits constructed with this algorithm could be used to push highly excited states into the ground state.  Our algorithm can also be applied beyond MBL to construct localized zero modes in interacting topological systems \cite{Katsura2015,Chertkov2020} or (with slight adjustment) to construct unitary operators that commute with given Hamiltonians or symmetries.

\begin{acknowledgments}

\emph{Acknowledgments.}--- We acknowledge useful discussions with Ryan Levy, Greg Hamilton, and David Pekker. We thank Steve Simon, Arijeet Pal, Thorsten Wahl, David Huse, and David Luitz for a careful reading of and comments on the manuscript. We acknowledge support from the Department of Energy grant DOE de-sc0020165. This research is part of the Blue Waters sustained-petascale computing project, which is supported by the National Science Foundation (Awards No. OCI-0725070 and No. ACI-1238993) and the state of Illinois. Blue Waters is a joint effort of the University of Illinois at Urbana-Champaign and its National Center for Supercomputing Applications.

\end{acknowledgments}

\bibliography{refs,refs_extra}

\begin{thebibliography}{79}%
\makeatletter
\providecommand \@ifxundefined [1]{%
 \@ifx{#1\undefined}
}%
\providecommand \@ifnum [1]{%
 \ifnum #1\expandafter \@firstoftwo
 \else \expandafter \@secondoftwo
 \fi
}%
\providecommand \@ifx [1]{%
 \ifx #1\expandafter \@firstoftwo
 \else \expandafter \@secondoftwo
 \fi
}%
\providecommand \natexlab [1]{#1}%
\providecommand \enquote  [1]{``#1''}%
\providecommand \bibnamefont  [1]{#1}%
\providecommand \bibfnamefont [1]{#1}%
\providecommand \citenamefont [1]{#1}%
\providecommand \href@noop [0]{\@secondoftwo}%
\providecommand \href [0]{\begingroup \@sanitize@url \@href}%
\providecommand \@href[1]{\@@startlink{#1}\@@href}%
\providecommand \@@href[1]{\endgroup#1\@@endlink}%
\providecommand \@sanitize@url [0]{\catcode `\\12\catcode `\$12\catcode
  `\&12\catcode `\#12\catcode `\^12\catcode `\_12\catcode `\%12\relax}%
\providecommand \@@startlink[1]{}%
\providecommand \@@endlink[0]{}%
\providecommand \url  [0]{\begingroup\@sanitize@url \@url }%
\providecommand \@url [1]{\endgroup\@href {#1}{\urlprefix }}%
\providecommand \urlprefix  [0]{URL }%
\providecommand \Eprint [0]{\href }%
\providecommand \doibase [0]{https://doi.org/}%
\providecommand \selectlanguage [0]{\@gobble}%
\providecommand \bibinfo  [0]{\@secondoftwo}%
\providecommand \bibfield  [0]{\@secondoftwo}%
\providecommand \translation [1]{[#1]}%
\providecommand \BibitemOpen [0]{}%
\providecommand \bibitemStop [0]{}%
\providecommand \bibitemNoStop [0]{.\EOS\space}%
\providecommand \EOS [0]{\spacefactor3000\relax}%
\providecommand \BibitemShut  [1]{\csname bibitem#1\endcsname}%
\let\auto@bib@innerbib\@empty
\bibitem [{\citenamefont {Anderson}(1958)}]{Anderson1958}%
  \BibitemOpen
  \bibfield  {author} {\bibinfo {author} {\bibfnamefont {P.~W.}\ \bibnamefont
  {Anderson}},\ }\bibfield  {title} {\bibinfo {title} {{Absence of Diffusion in
  Certain Random Lattices}},\ }\href {https://doi.org/10.1103/PhysRev.109.1492}
  {\bibfield  {journal} {\bibinfo  {journal} {Phys. Rev.}\ }\textbf {\bibinfo
  {volume} {109}},\ \bibinfo {pages} {1492} (\bibinfo {year}
  {1958})}\BibitemShut {NoStop}%
\bibitem [{\citenamefont {Fleishman}\ and\ \citenamefont
  {Anderson}(1980)}]{fleishman1980interactions}%
  \BibitemOpen
  \bibfield  {author} {\bibinfo {author} {\bibfnamefont {L.}~\bibnamefont
  {Fleishman}}\ and\ \bibinfo {author} {\bibfnamefont {P.~W.}\ \bibnamefont
  {Anderson}},\ }\bibfield  {title} {\bibinfo {title} {{Interactions and the
  Anderson transition}},\ }\href {https://doi.org/10.1103/PhysRevB.21.2366}
  {\bibfield  {journal} {\bibinfo  {journal} {Physical Review B}\ }\textbf
  {\bibinfo {volume} {21}},\ \bibinfo {pages} {2366} (\bibinfo {year}
  {1980})}\BibitemShut {NoStop}%
\bibitem [{\citenamefont {Gornyi}\ \emph {et~al.}(2005)\citenamefont {Gornyi},
  \citenamefont {Mirlin},\ and\ \citenamefont
  {Polyakov}}]{gornyi2005interacting}%
  \BibitemOpen
  \bibfield  {author} {\bibinfo {author} {\bibfnamefont {I.~V.}\ \bibnamefont
  {Gornyi}}, \bibinfo {author} {\bibfnamefont {A.~D.}\ \bibnamefont {Mirlin}},\
  and\ \bibinfo {author} {\bibfnamefont {D.~G.}\ \bibnamefont {Polyakov}},\
  }\bibfield  {title} {\bibinfo {title} {{Interacting electrons in disordered
  wires: Anderson localization and low-T transport}},\ }\href
  {https://doi.org/10.1103/PhysRevLett.95.206603} {\bibfield  {journal}
  {\bibinfo  {journal} {Phys. Rev. Lett.}\ }\textbf {\bibinfo {volume} {95}},\
  \bibinfo {pages} {206603} (\bibinfo {year} {2005})}\BibitemShut {NoStop}%
\bibitem [{\citenamefont {Basko}\ \emph {et~al.}(2006)\citenamefont {Basko},
  \citenamefont {Aleiner},\ and\ \citenamefont {Altshuler}}]{basko2006metal}%
  \BibitemOpen
  \bibfield  {author} {\bibinfo {author} {\bibfnamefont {D.~M.}\ \bibnamefont
  {Basko}}, \bibinfo {author} {\bibfnamefont {I.~L.}\ \bibnamefont {Aleiner}},\
  and\ \bibinfo {author} {\bibfnamefont {B.~L.}\ \bibnamefont {Altshuler}},\
  }\bibfield  {title} {\bibinfo {title} {{Metal--insulator transition in a
  weakly interacting many-electron system with localized single-particle
  states}},\ }\href {https://doi.org/10.1016/j.aop.2005.11.014} {\bibfield
  {journal} {\bibinfo  {journal} {Ann. Phys. (N. Y.)}\ }\textbf {\bibinfo
  {volume} {321}},\ \bibinfo {pages} {1126} (\bibinfo {year}
  {2006})}\BibitemShut {NoStop}%
\bibitem [{\citenamefont {Nandkishore}\ and\ \citenamefont
  {Huse}(2015)}]{Nandkishore2015}%
  \BibitemOpen
  \bibfield  {author} {\bibinfo {author} {\bibfnamefont {R.}~\bibnamefont
  {Nandkishore}}\ and\ \bibinfo {author} {\bibfnamefont {D.~A.}\ \bibnamefont
  {Huse}},\ }\bibfield  {title} {\bibinfo {title} {Many-body localization and
  thermalization in quantum statistical mechanics},\ }\href
  {https://doi.org/10.1146/annurev-conmatphys-031214-014726} {\bibfield
  {journal} {\bibinfo  {journal} {Annu. Rev. Condens. Matter Phys.}\ }\textbf
  {\bibinfo {volume} {6}},\ \bibinfo {pages} {15} (\bibinfo {year}
  {2015})}\BibitemShut {NoStop}%
\bibitem [{\citenamefont {Abanin}\ and\ \citenamefont
  {Papi\ifmmode~\acute{c}\else \'{c}\fi{}}(2017)}]{Abanin2017}%
  \BibitemOpen
  \bibfield  {author} {\bibinfo {author} {\bibfnamefont {D.~A.}\ \bibnamefont
  {Abanin}}\ and\ \bibinfo {author} {\bibfnamefont {Z.}~\bibnamefont
  {Papi\ifmmode~\acute{c}\else \'{c}\fi{}}},\ }\bibfield  {title} {\bibinfo
  {title} {Recent progress in many-body localization},\ }\href
  {https://doi.org/10.1002/andp.201700169} {\bibfield  {journal} {\bibinfo
  {journal} {Ann. Phys. (Berl.)}\ }\textbf {\bibinfo {volume} {529}},\ \bibinfo
  {pages} {1700169} (\bibinfo {year} {2017})}\BibitemShut {NoStop}%
\bibitem [{\citenamefont {Abanin}\ \emph {et~al.}(2019)\citenamefont {Abanin},
  \citenamefont {Altman}, \citenamefont {Bloch},\ and\ \citenamefont
  {Serbyn}}]{Abanin2019}%
  \BibitemOpen
  \bibfield  {author} {\bibinfo {author} {\bibfnamefont {D.~A.}\ \bibnamefont
  {Abanin}}, \bibinfo {author} {\bibfnamefont {E.}~\bibnamefont {Altman}},
  \bibinfo {author} {\bibfnamefont {I.}~\bibnamefont {Bloch}},\ and\ \bibinfo
  {author} {\bibfnamefont {M.}~\bibnamefont {Serbyn}},\ }\bibfield  {title}
  {\bibinfo {title} {Colloquium: Many-body localization, thermalization, and
  entanglement},\ }\href {https://doi.org/10.1103/RevModPhys.91.021001}
  {\bibfield  {journal} {\bibinfo  {journal} {Rev. Mod. Phys.}\ }\textbf
  {\bibinfo {volume} {91}},\ \bibinfo {pages} {021001} (\bibinfo {year}
  {2019})}\BibitemShut {NoStop}%
\bibitem [{Note1()}]{Note1}%
  \BibitemOpen
  \bibinfo {note} {In the MBL literature, a ``quasilocal'' operator refers to
  an operator that has compact support over a finite region and exponentially
  decaying tails beyond that region. In other contexts, such as when discussing
  Anderson localization, such operators would be called local or localized
  instead.}\BibitemShut {Stop}%
\bibitem [{\citenamefont {Serbyn}\ \emph {et~al.}(2013)\citenamefont {Serbyn},
  \citenamefont {Papi\ifmmode~\acute{c}\else \'{c}\fi{}},\ and\ \citenamefont
  {Abanin}}]{Serbyn2013}%
  \BibitemOpen
  \bibfield  {author} {\bibinfo {author} {\bibfnamefont {M.}~\bibnamefont
  {Serbyn}}, \bibinfo {author} {\bibfnamefont {Z.}~\bibnamefont
  {Papi\ifmmode~\acute{c}\else \'{c}\fi{}}},\ and\ \bibinfo {author}
  {\bibfnamefont {D.~A.}\ \bibnamefont {Abanin}},\ }\bibfield  {title}
  {\bibinfo {title} {Local conservation laws and the structure of the many-body
  localized states},\ }\href {https://doi.org/10.1103/PhysRevLett.111.127201}
  {\bibfield  {journal} {\bibinfo  {journal} {Phys. Rev. Lett.}\ }\textbf
  {\bibinfo {volume} {111}},\ \bibinfo {pages} {127201(R)} (\bibinfo {year}
  {2013})}\BibitemShut {NoStop}%
\bibitem [{\citenamefont {Huse}\ \emph {et~al.}(2014)\citenamefont {Huse},
  \citenamefont {Nandkishore},\ and\ \citenamefont {Oganesyan}}]{Huse2014}%
  \BibitemOpen
  \bibfield  {author} {\bibinfo {author} {\bibfnamefont {D.~A.}\ \bibnamefont
  {Huse}}, \bibinfo {author} {\bibfnamefont {R.}~\bibnamefont {Nandkishore}},\
  and\ \bibinfo {author} {\bibfnamefont {V.}~\bibnamefont {Oganesyan}},\
  }\bibfield  {title} {\bibinfo {title} {Phenomenology of fully
  many-body-localized systems},\ }\href
  {https://doi.org/10.1103/PhysRevB.90.174202} {\bibfield  {journal} {\bibinfo
  {journal} {Phys. Rev. B}\ }\textbf {\bibinfo {volume} {90}},\ \bibinfo
  {pages} {174202} (\bibinfo {year} {2014})}\BibitemShut {NoStop}%
\bibitem [{\citenamefont {Imbrie}\ \emph {et~al.}(2017)\citenamefont {Imbrie},
  \citenamefont {Ros},\ and\ \citenamefont {Scardicchio}}]{Imbrie2017}%
  \BibitemOpen
  \bibfield  {author} {\bibinfo {author} {\bibfnamefont {J.~Z.}\ \bibnamefont
  {Imbrie}}, \bibinfo {author} {\bibfnamefont {V.}~\bibnamefont {Ros}},\ and\
  \bibinfo {author} {\bibfnamefont {A.}~\bibnamefont {Scardicchio}},\
  }\bibfield  {title} {\bibinfo {title} {Local integrals of motion in many-body
  localized systems},\ }\href {https://doi.org/10.1002/andp.201600278}
  {\bibfield  {journal} {\bibinfo  {journal} {Ann. Phys. (Berl.)}\ }\textbf
  {\bibinfo {volume} {529}},\ \bibinfo {pages} {1600278} (\bibinfo {year}
  {2017})}\BibitemShut {NoStop}%
\bibitem [{\citenamefont {Luitz}\ \emph {et~al.}(2015)\citenamefont {Luitz},
  \citenamefont {Laflorencie},\ and\ \citenamefont {Alet}}]{Luitz2015}%
  \BibitemOpen
  \bibfield  {author} {\bibinfo {author} {\bibfnamefont {D.~J.}\ \bibnamefont
  {Luitz}}, \bibinfo {author} {\bibfnamefont {N.}~\bibnamefont {Laflorencie}},\
  and\ \bibinfo {author} {\bibfnamefont {F.}~\bibnamefont {Alet}},\ }\bibfield
  {title} {\bibinfo {title} {Many-body localization edge in the random-field
  heisenberg chain},\ }\href {https://doi.org/10.1103/PhysRevB.91.081103}
  {\bibfield  {journal} {\bibinfo  {journal} {Phys. Rev. B}\ }\textbf {\bibinfo
  {volume} {91}},\ \bibinfo {pages} {081103(R)} (\bibinfo {year}
  {2015})}\BibitemShut {NoStop}%
\bibitem [{\citenamefont {Villalonga}\ \emph {et~al.}(2018)\citenamefont
  {Villalonga}, \citenamefont {Yu}, \citenamefont {Luitz},\ and\ \citenamefont
  {Clark}}]{Villalonga2018}%
  \BibitemOpen
  \bibfield  {author} {\bibinfo {author} {\bibfnamefont {B.}~\bibnamefont
  {Villalonga}}, \bibinfo {author} {\bibfnamefont {X.}~\bibnamefont {Yu}},
  \bibinfo {author} {\bibfnamefont {D.~J.}\ \bibnamefont {Luitz}},\ and\
  \bibinfo {author} {\bibfnamefont {B.~K.}\ \bibnamefont {Clark}},\ }\bibfield
  {title} {\bibinfo {title} {Exploring one-particle orbitals in large many-body
  localized systems},\ }\href {https://doi.org/10.1103/PhysRevB.97.104406}
  {\bibfield  {journal} {\bibinfo  {journal} {Phys. Rev. B}\ }\textbf {\bibinfo
  {volume} {97}},\ \bibinfo {pages} {104406} (\bibinfo {year}
  {2018})}\BibitemShut {NoStop}%
\bibitem [{\citenamefont {Serbyn}\ and\ \citenamefont
  {Moore}(2016)}]{Serbyn2016}%
  \BibitemOpen
  \bibfield  {author} {\bibinfo {author} {\bibfnamefont {M.}~\bibnamefont
  {Serbyn}}\ and\ \bibinfo {author} {\bibfnamefont {J.~E.}\ \bibnamefont
  {Moore}},\ }\bibfield  {title} {\bibinfo {title} {Spectral statistics across
  the many-body localization transition},\ }\href
  {https://doi.org/10.1103/PhysRevB.93.041424} {\bibfield  {journal} {\bibinfo
  {journal} {Phys. Rev. B}\ }\textbf {\bibinfo {volume} {93}},\ \bibinfo
  {pages} {041424(R)} (\bibinfo {year} {2016})}\BibitemShut {NoStop}%
\bibitem [{\citenamefont {Bauer}\ and\ \citenamefont
  {Nayak}(2013)}]{bauer2013area}%
  \BibitemOpen
  \bibfield  {author} {\bibinfo {author} {\bibfnamefont {B.}~\bibnamefont
  {Bauer}}\ and\ \bibinfo {author} {\bibfnamefont {C.}~\bibnamefont {Nayak}},\
  }\bibfield  {title} {\bibinfo {title} {{Area laws in a many-body localized
  state and its implications for topological order}},\ }\href
  {https://doi.org/10.1088/1742-5468/2013/09/P09005} {\bibfield  {journal}
  {\bibinfo  {journal} {J. Stat. Mech.}\ }\textbf {\bibinfo {volume} {2013}},\
  \bibinfo {pages} {P09005} (\bibinfo {year} {2013})}\BibitemShut {NoStop}%
\bibitem [{\citenamefont {Kj{\"a}ll}\ \emph {et~al.}(2014)\citenamefont
  {Kj{\"a}ll}, \citenamefont {Bardarson},\ and\ \citenamefont
  {Pollmann}}]{kjall2014many}%
  \BibitemOpen
  \bibfield  {author} {\bibinfo {author} {\bibfnamefont {J.~A.}\ \bibnamefont
  {Kj{\"a}ll}}, \bibinfo {author} {\bibfnamefont {J.~H.}\ \bibnamefont
  {Bardarson}},\ and\ \bibinfo {author} {\bibfnamefont {F.}~\bibnamefont
  {Pollmann}},\ }\bibfield  {title} {\bibinfo {title} {{Many-body localization
  in a disordered quantum Ising chain}},\ }\href
  {https://doi.org/10.1103/PhysRevLett.113.107204} {\bibfield  {journal}
  {\bibinfo  {journal} {Phys. Rev. Lett.}\ }\textbf {\bibinfo {volume} {113}},\
  \bibinfo {pages} {107204} (\bibinfo {year} {2014})}\BibitemShut {NoStop}%
\bibitem [{\citenamefont {Yu}\ \emph {et~al.}(2016)\citenamefont {Yu},
  \citenamefont {Luitz},\ and\ \citenamefont {Clark}}]{Yu2016}%
  \BibitemOpen
  \bibfield  {author} {\bibinfo {author} {\bibfnamefont {X.}~\bibnamefont
  {Yu}}, \bibinfo {author} {\bibfnamefont {D.~J.}\ \bibnamefont {Luitz}},\ and\
  \bibinfo {author} {\bibfnamefont {B.~K.}\ \bibnamefont {Clark}},\ }\bibfield
  {title} {\bibinfo {title} {Bimodal entanglement entropy distribution in the
  many-body localization transition},\ }\href
  {https://doi.org/10.1103/PhysRevB.94.184202} {\bibfield  {journal} {\bibinfo
  {journal} {Phys. Rev. B}\ }\textbf {\bibinfo {volume} {94}},\ \bibinfo
  {pages} {184202} (\bibinfo {year} {2016})}\BibitemShut {NoStop}%
\bibitem [{\citenamefont {Yu}\ \emph {et~al.}(2017)\citenamefont {Yu},
  \citenamefont {Pekker},\ and\ \citenamefont {Clark}}]{Yu2017}%
  \BibitemOpen
  \bibfield  {author} {\bibinfo {author} {\bibfnamefont {X.}~\bibnamefont
  {Yu}}, \bibinfo {author} {\bibfnamefont {D.}~\bibnamefont {Pekker}},\ and\
  \bibinfo {author} {\bibfnamefont {B.~K.}\ \bibnamefont {Clark}},\ }\bibfield
  {title} {\bibinfo {title} {{Finding Matrix Product State Representations of
  Highly Excited Eigenstates of Many-Body Localized Hamiltonians}},\ }\href
  {https://doi.org/10.1103/PhysRevLett.118.017201} {\bibfield  {journal}
  {\bibinfo  {journal} {Phys. Rev. Lett.}\ }\textbf {\bibinfo {volume} {118}},\
  \bibinfo {pages} {017201} (\bibinfo {year} {2017})}\BibitemShut {NoStop}%
\bibitem [{\citenamefont {Luitz}(2016)}]{luitz2016long}%
  \BibitemOpen
  \bibfield  {author} {\bibinfo {author} {\bibfnamefont {D.~J.}\ \bibnamefont
  {Luitz}},\ }\bibfield  {title} {\bibinfo {title} {Long tail distributions
  near the many-body localization transition},\ }\href
  {https://doi.org/10.1103/PhysRevB.93.134201} {\bibfield  {journal} {\bibinfo
  {journal} {Phys. Rev. B}\ }\textbf {\bibinfo {volume} {93}},\ \bibinfo
  {pages} {134201} (\bibinfo {year} {2016})}\BibitemShut {NoStop}%
\bibitem [{\citenamefont {Khemani}\ \emph {et~al.}(2016)\citenamefont
  {Khemani}, \citenamefont {Pollmann},\ and\ \citenamefont
  {Sondhi}}]{Khemani2016}%
  \BibitemOpen
  \bibfield  {author} {\bibinfo {author} {\bibfnamefont {V.}~\bibnamefont
  {Khemani}}, \bibinfo {author} {\bibfnamefont {F.}~\bibnamefont {Pollmann}},\
  and\ \bibinfo {author} {\bibfnamefont {S.~L.}\ \bibnamefont {Sondhi}},\
  }\bibfield  {title} {\bibinfo {title} {{Obtaining Highly Excited Eigenstates
  of Many-Body Localized Hamiltonians by the Density Matrix Renormalization
  Group Approach}},\ }\href {https://doi.org/10.1103/PhysRevLett.116.247204}
  {\bibfield  {journal} {\bibinfo  {journal} {Phys. Rev. Lett.}\ }\textbf
  {\bibinfo {volume} {116}},\ \bibinfo {pages} {247204} (\bibinfo {year}
  {2016})}\BibitemShut {NoStop}%
\bibitem [{\citenamefont {Lim}\ and\ \citenamefont
  {Sheng}(2016)}]{lim2016many}%
  \BibitemOpen
  \bibfield  {author} {\bibinfo {author} {\bibfnamefont {S.~P.}\ \bibnamefont
  {Lim}}\ and\ \bibinfo {author} {\bibfnamefont {D.~N.}\ \bibnamefont
  {Sheng}},\ }\bibfield  {title} {\bibinfo {title} {{Many-body localization and
  transition by density matrix renormalization group and exact diagonalization
  studies}},\ }\href {https://doi.org/10.1103/PhysRevB.94.045111} {\bibfield
  {journal} {\bibinfo  {journal} {Phys. Rev. B}\ }\textbf {\bibinfo {volume}
  {94}},\ \bibinfo {pages} {045111} (\bibinfo {year} {2016})}\BibitemShut
  {NoStop}%
\bibitem [{\citenamefont {Panda}\ \emph {et~al.}(2020)\citenamefont {Panda},
  \citenamefont {Scardicchio}, \citenamefont {Schulz}, \citenamefont {Taylor},\
  and\ \citenamefont {{\v{Z}}nidari{\v{c}}}}]{panda2020can}%
  \BibitemOpen
  \bibfield  {author} {\bibinfo {author} {\bibfnamefont {R.~K.}\ \bibnamefont
  {Panda}}, \bibinfo {author} {\bibfnamefont {A.}~\bibnamefont {Scardicchio}},
  \bibinfo {author} {\bibfnamefont {M.}~\bibnamefont {Schulz}}, \bibinfo
  {author} {\bibfnamefont {S.~R.}\ \bibnamefont {Taylor}},\ and\ \bibinfo
  {author} {\bibfnamefont {M.}~\bibnamefont {{\v{Z}}nidari{\v{c}}}},\
  }\bibfield  {title} {\bibinfo {title} {Can we study the many-body
  localisation transition?},\ }\href@noop {} {\bibfield  {journal} {\bibinfo
  {journal} {EPL (Europhysics Letters)}\ }\textbf {\bibinfo {volume} {128}},\
  \bibinfo {pages} {67003} (\bibinfo {year} {2020})}\BibitemShut {NoStop}%
\bibitem [{\citenamefont {Choi}\ \emph {et~al.}(2016)\citenamefont {Choi},
  \citenamefont {Hild}, \citenamefont {Zeiher}, \citenamefont {Schau{\ss}},
  \citenamefont {Rubio-Abadal}, \citenamefont {Yefsah}, \citenamefont
  {Khemani}, \citenamefont {Huse}, \citenamefont {Bloch},\ and\ \citenamefont
  {Gross}}]{Choi2016}%
  \BibitemOpen
  \bibfield  {author} {\bibinfo {author} {\bibfnamefont {J.-Y.}\ \bibnamefont
  {Choi}}, \bibinfo {author} {\bibfnamefont {S.}~\bibnamefont {Hild}}, \bibinfo
  {author} {\bibfnamefont {J.}~\bibnamefont {Zeiher}}, \bibinfo {author}
  {\bibfnamefont {P.}~\bibnamefont {Schau{\ss}}}, \bibinfo {author}
  {\bibfnamefont {A.}~\bibnamefont {Rubio-Abadal}}, \bibinfo {author}
  {\bibfnamefont {T.}~\bibnamefont {Yefsah}}, \bibinfo {author} {\bibfnamefont
  {V.}~\bibnamefont {Khemani}}, \bibinfo {author} {\bibfnamefont {D.~A.}\
  \bibnamefont {Huse}}, \bibinfo {author} {\bibfnamefont {I.}~\bibnamefont
  {Bloch}},\ and\ \bibinfo {author} {\bibfnamefont {C.}~\bibnamefont {Gross}},\
  }\bibfield  {title} {\bibinfo {title} {Exploring the many-body localization
  transition in two dimensions},\ }\href
  {https://doi.org/10.1126/science.aaf8834} {\bibfield  {journal} {\bibinfo
  {journal} {Science}\ }\textbf {\bibinfo {volume} {352}},\ \bibinfo {pages}
  {1547} (\bibinfo {year} {2016})}\BibitemShut {NoStop}%
\bibitem [{\citenamefont {Bordia}\ \emph {et~al.}(2017)\citenamefont {Bordia},
  \citenamefont {L\"uschen}, \citenamefont {Scherg}, \citenamefont
  {Gopalakrishnan}, \citenamefont {Knap}, \citenamefont {Schneider},\ and\
  \citenamefont {Bloch}}]{Bordia2017}%
  \BibitemOpen
  \bibfield  {author} {\bibinfo {author} {\bibfnamefont {P.}~\bibnamefont
  {Bordia}}, \bibinfo {author} {\bibfnamefont {H.}~\bibnamefont {L\"uschen}},
  \bibinfo {author} {\bibfnamefont {S.}~\bibnamefont {Scherg}}, \bibinfo
  {author} {\bibfnamefont {S.}~\bibnamefont {Gopalakrishnan}}, \bibinfo
  {author} {\bibfnamefont {M.}~\bibnamefont {Knap}}, \bibinfo {author}
  {\bibfnamefont {U.}~\bibnamefont {Schneider}},\ and\ \bibinfo {author}
  {\bibfnamefont {I.}~\bibnamefont {Bloch}},\ }\bibfield  {title} {\bibinfo
  {title} {Probing slow relaxation and many-body localization in
  two-dimensional quasiperiodic systems},\ }\href
  {https://doi.org/10.1103/PhysRevX.7.041047} {\bibfield  {journal} {\bibinfo
  {journal} {Phys. Rev. X}\ }\textbf {\bibinfo {volume} {7}},\ \bibinfo {pages}
  {041047} (\bibinfo {year} {2017})}\BibitemShut {NoStop}%
\bibitem [{\citenamefont {Kondov}\ \emph {et~al.}(2015)\citenamefont {Kondov},
  \citenamefont {McGehee}, \citenamefont {Xu},\ and\ \citenamefont
  {DeMarco}}]{Kondov2015}%
  \BibitemOpen
  \bibfield  {author} {\bibinfo {author} {\bibfnamefont {S.~S.}\ \bibnamefont
  {Kondov}}, \bibinfo {author} {\bibfnamefont {W.~R.}\ \bibnamefont {McGehee}},
  \bibinfo {author} {\bibfnamefont {W.}~\bibnamefont {Xu}},\ and\ \bibinfo
  {author} {\bibfnamefont {B.}~\bibnamefont {DeMarco}},\ }\bibfield  {title}
  {\bibinfo {title} {Disorder-induced localization in a strongly correlated
  atomic hubbard gas},\ }\href {https://doi.org/10.1103/PhysRevLett.114.083002}
  {\bibfield  {journal} {\bibinfo  {journal} {Phys. Rev. Lett.}\ }\textbf
  {\bibinfo {volume} {114}},\ \bibinfo {pages} {083002} (\bibinfo {year}
  {2015})}\BibitemShut {NoStop}%
\bibitem [{\citenamefont {De~Roeck}\ and\ \citenamefont
  {Huveneers}(2017)}]{DeRoeck2017a}%
  \BibitemOpen
  \bibfield  {author} {\bibinfo {author} {\bibfnamefont {W.}~\bibnamefont
  {De~Roeck}}\ and\ \bibinfo {author} {\bibfnamefont {F.}~\bibnamefont
  {Huveneers}},\ }\bibfield  {title} {\bibinfo {title} {Stability and
  instability towards delocalization in many-body localization systems},\
  }\href {https://doi.org/10.1103/PhysRevB.95.155129} {\bibfield  {journal}
  {\bibinfo  {journal} {Phys. Rev. B}\ }\textbf {\bibinfo {volume} {95}},\
  \bibinfo {pages} {155129} (\bibinfo {year} {2017})}\BibitemShut {NoStop}%
\bibitem [{\citenamefont {De~Roeck}\ and\ \citenamefont
  {Imbrie}(2017)}]{DeRoeck2017b}%
  \BibitemOpen
  \bibfield  {author} {\bibinfo {author} {\bibfnamefont {W.}~\bibnamefont
  {De~Roeck}}\ and\ \bibinfo {author} {\bibfnamefont {J.~Z.}\ \bibnamefont
  {Imbrie}},\ }\bibfield  {title} {\bibinfo {title} {Many-body localization:
  stability and instability},\ }\href {https://doi.org/10.1098/rsta.2016.0422}
  {\bibfield  {journal} {\bibinfo  {journal} {Philos. Trans. R. Soc. A}\
  }\textbf {\bibinfo {volume} {375}},\ \bibinfo {pages} {20160422} (\bibinfo
  {year} {2017})}\BibitemShut {NoStop}%
\bibitem [{\citenamefont {Chandran}\ \emph {et~al.}(2016)\citenamefont
  {Chandran}, \citenamefont {Pal}, \citenamefont {Laumann},\ and\ \citenamefont
  {Scardicchio}}]{Chandran2016}%
  \BibitemOpen
  \bibfield  {author} {\bibinfo {author} {\bibfnamefont {A.}~\bibnamefont
  {Chandran}}, \bibinfo {author} {\bibfnamefont {A.}~\bibnamefont {Pal}},
  \bibinfo {author} {\bibfnamefont {C.~R.}\ \bibnamefont {Laumann}},\ and\
  \bibinfo {author} {\bibfnamefont {A.}~\bibnamefont {Scardicchio}},\
  }\bibfield  {title} {\bibinfo {title} {Many-body localization beyond
  eigenstates in all dimensions},\ }\href
  {https://doi.org/10.1103/PhysRevB.94.144203} {\bibfield  {journal} {\bibinfo
  {journal} {Phys. Rev. B}\ }\textbf {\bibinfo {volume} {94}},\ \bibinfo
  {pages} {144203} (\bibinfo {year} {2016})}\BibitemShut {NoStop}%
\bibitem [{\citenamefont {Agarwal}\ \emph {et~al.}(2017)\citenamefont
  {Agarwal}, \citenamefont {Altman}, \citenamefont {Demler}, \citenamefont
  {Gopalakrishnan}, \citenamefont {Huse},\ and\ \citenamefont
  {Knap}}]{Agarwal2017}%
  \BibitemOpen
  \bibfield  {author} {\bibinfo {author} {\bibfnamefont {K.}~\bibnamefont
  {Agarwal}}, \bibinfo {author} {\bibfnamefont {E.}~\bibnamefont {Altman}},
  \bibinfo {author} {\bibfnamefont {E.}~\bibnamefont {Demler}}, \bibinfo
  {author} {\bibfnamefont {S.}~\bibnamefont {Gopalakrishnan}}, \bibinfo
  {author} {\bibfnamefont {D.~A.}\ \bibnamefont {Huse}},\ and\ \bibinfo
  {author} {\bibfnamefont {M.}~\bibnamefont {Knap}},\ }\bibfield  {title}
  {\bibinfo {title} {{Rare-region effects and dynamics near the many-body
  localization transition}},\ }\href {https://doi.org/10.1002/andp.201600326}
  {\bibfield  {journal} {\bibinfo  {journal} {Ann. Phys. (Berl.)}\ }\textbf
  {\bibinfo {volume} {529}},\ \bibinfo {pages} {1600326} (\bibinfo {year}
  {2017})}\BibitemShut {NoStop}%
\bibitem [{\citenamefont {Gopalakrishnan}\ and\ \citenamefont
  {Huse}(2019)}]{Gopalakrishnan2019}%
  \BibitemOpen
  \bibfield  {author} {\bibinfo {author} {\bibfnamefont {S.}~\bibnamefont
  {Gopalakrishnan}}\ and\ \bibinfo {author} {\bibfnamefont {D.~A.}\
  \bibnamefont {Huse}},\ }\bibfield  {title} {\bibinfo {title} {Instability of
  many-body localized systems as a phase transition in a nonstandard
  thermodynamic limit},\ }\href {https://doi.org/10.1103/PhysRevB.99.134305}
  {\bibfield  {journal} {\bibinfo  {journal} {Phys. Rev. B}\ }\textbf {\bibinfo
  {volume} {99}},\ \bibinfo {pages} {134305} (\bibinfo {year}
  {2019})}\BibitemShut {NoStop}%
\bibitem [{\citenamefont {Lev}\ and\ \citenamefont
  {Reichman}(2016)}]{BarLev2016}%
  \BibitemOpen
  \bibfield  {author} {\bibinfo {author} {\bibfnamefont {Y.~B.}\ \bibnamefont
  {Lev}}\ and\ \bibinfo {author} {\bibfnamefont {D.~R.}\ \bibnamefont
  {Reichman}},\ }\bibfield  {title} {\bibinfo {title} {{Slow dynamics in a
  two-dimensional Anderson-Hubbard model}},\ }\href
  {https://doi.org/10.1209/0295-5075/113/46001} {\bibfield  {journal} {\bibinfo
   {journal} {{EPL}}\ }\textbf {\bibinfo {volume} {113}},\ \bibinfo {pages}
  {46001} (\bibinfo {year} {2016})}\BibitemShut {NoStop}%
\bibitem [{\citenamefont {Inglis}\ and\ \citenamefont
  {Pollet}(2016)}]{Inglis2016}%
  \BibitemOpen
  \bibfield  {author} {\bibinfo {author} {\bibfnamefont {S.}~\bibnamefont
  {Inglis}}\ and\ \bibinfo {author} {\bibfnamefont {L.}~\bibnamefont
  {Pollet}},\ }\bibfield  {title} {\bibinfo {title} {Accessing many-body
  localized states through the generalized gibbs ensemble},\ }\href
  {https://doi.org/10.1103/PhysRevLett.117.120402} {\bibfield  {journal}
  {\bibinfo  {journal} {Phys. Rev. Lett.}\ }\textbf {\bibinfo {volume} {117}},\
  \bibinfo {pages} {120402} (\bibinfo {year} {2016})}\BibitemShut {NoStop}%
\bibitem [{\citenamefont {Thomson}\ and\ \citenamefont
  {Schir\'o}(2018)}]{Thomson2018}%
  \BibitemOpen
  \bibfield  {author} {\bibinfo {author} {\bibfnamefont {S.~J.}\ \bibnamefont
  {Thomson}}\ and\ \bibinfo {author} {\bibfnamefont {M.}~\bibnamefont
  {Schir\'o}},\ }\bibfield  {title} {\bibinfo {title} {Time evolution of
  many-body localized systems with the flow equation approach},\ }\href
  {https://doi.org/10.1103/PhysRevB.97.060201} {\bibfield  {journal} {\bibinfo
  {journal} {Phys. Rev. B}\ }\textbf {\bibinfo {volume} {97}},\ \bibinfo
  {pages} {060201(R)} (\bibinfo {year} {2018})}\BibitemShut {NoStop}%
\bibitem [{\citenamefont {{Kennes}}(2018)}]{Kennes2018}%
  \BibitemOpen
  \bibfield  {author} {\bibinfo {author} {\bibfnamefont {D.~M.}\ \bibnamefont
  {{Kennes}}},\ }\bibfield  {title} {\bibinfo {title} {{Many-Body Localization
  in Two Dimensions from Projected Entangled-Pair States}},\ }\href@noop {} {\
  (\bibinfo {year} {2018})},\ \Eprint {https://arxiv.org/abs/1811.04126}
  {arXiv:1811.04126} \BibitemShut {NoStop}%
\bibitem [{\citenamefont {Wahl}\ \emph {et~al.}(2019)\citenamefont {Wahl},
  \citenamefont {Pal},\ and\ \citenamefont {Simon}}]{Wahl2019}%
  \BibitemOpen
  \bibfield  {author} {\bibinfo {author} {\bibfnamefont {T.}~\bibnamefont
  {Wahl}}, \bibinfo {author} {\bibfnamefont {A.}~\bibnamefont {Pal}},\ and\
  \bibinfo {author} {\bibfnamefont {S.}~\bibnamefont {Simon}},\ }\bibfield
  {title} {\bibinfo {title} {Signatures of the many-body localized regime in
  two dimensions},\ }\href {https://doi.org/10.1038/s41567-018-0339-x}
  {\bibfield  {journal} {\bibinfo  {journal} {Nat. Phys}\ }\textbf {\bibinfo
  {volume} {15}},\ \bibinfo {pages} {164} (\bibinfo {year} {2019})}\BibitemShut
  {NoStop}%
\bibitem [{\citenamefont {{Gei{\ss}ler}}\ and\ \citenamefont
  {{Pupillo}}(2019)}]{Geissler2019}%
  \BibitemOpen
  \bibfield  {author} {\bibinfo {author} {\bibfnamefont {A.}~\bibnamefont
  {{Gei{\ss}ler}}}\ and\ \bibinfo {author} {\bibfnamefont {G.}~\bibnamefont
  {{Pupillo}}},\ }\bibfield  {title} {\bibinfo {title} {{Many-body localization
  in the two dimensional Bose-Hubbard model}},\ }\href@noop {} {\  (\bibinfo
  {year} {2019})},\ \Eprint {https://arxiv.org/abs/1909.09247}
  {arXiv:1909.09247} \BibitemShut {NoStop}%
\bibitem [{\citenamefont {De~Tomasi}\ \emph {et~al.}(2019)\citenamefont
  {De~Tomasi}, \citenamefont {Pollmann},\ and\ \citenamefont
  {Heyl}}]{DeTomasi2019}%
  \BibitemOpen
  \bibfield  {author} {\bibinfo {author} {\bibfnamefont {G.}~\bibnamefont
  {De~Tomasi}}, \bibinfo {author} {\bibfnamefont {F.}~\bibnamefont
  {Pollmann}},\ and\ \bibinfo {author} {\bibfnamefont {M.}~\bibnamefont
  {Heyl}},\ }\bibfield  {title} {\bibinfo {title} {Efficiently solving the
  dynamics of many-body localized systems at strong disorder},\ }\href
  {https://doi.org/10.1103/PhysRevB.99.241114} {\bibfield  {journal} {\bibinfo
  {journal} {Phys. Rev. B}\ }\textbf {\bibinfo {volume} {99}},\ \bibinfo
  {pages} {241114(R)} (\bibinfo {year} {2019})}\BibitemShut {NoStop}%
\bibitem [{\citenamefont {Th\'{e}veniaut}\ \emph {et~al.}(2019)\citenamefont
  {Th\'{e}veniaut}, \citenamefont {Lan},\ and\ \citenamefont
  {Alet}}]{Theveniaut2019}%
  \BibitemOpen
  \bibfield  {author} {\bibinfo {author} {\bibfnamefont {H.}~\bibnamefont
  {Th\'{e}veniaut}}, \bibinfo {author} {\bibfnamefont {Z.}~\bibnamefont
  {Lan}},\ and\ \bibinfo {author} {\bibfnamefont {F.}~\bibnamefont {Alet}},\
  }\bibfield  {title} {\bibinfo {title} {Many-body localization transition in a
  two-dimensional disordered quantum dimer model},\ }\href@noop {} {\
  (\bibinfo {year} {2019})},\ \Eprint {https://arxiv.org/abs/1902.04091}
  {arXiv:1902.04091} \BibitemShut {NoStop}%
\bibitem [{\citenamefont {Kshetrimayum}\ \emph {et~al.}(2019)\citenamefont
  {Kshetrimayum}, \citenamefont {Goihl},\ and\ \citenamefont
  {Eisert}}]{Kshetrimayum2019}%
  \BibitemOpen
  \bibfield  {author} {\bibinfo {author} {\bibfnamefont {A.}~\bibnamefont
  {Kshetrimayum}}, \bibinfo {author} {\bibfnamefont {M.}~\bibnamefont
  {Goihl}},\ and\ \bibinfo {author} {\bibfnamefont {J.}~\bibnamefont
  {Eisert}},\ }\bibfield  {title} {\bibinfo {title} {Time evolution of
  many-body localized systems in two spatial dimensions},\ }\href@noop {} {\
  (\bibinfo {year} {2019})},\ \Eprint {https://arxiv.org/abs/1910.11359}
  {arXiv:1910.11359} \BibitemShut {NoStop}%
\bibitem [{\citenamefont {Pietracaprina}\ and\ \citenamefont
  {Alet}(2020)}]{Pietracaprina2020}%
  \BibitemOpen
  \bibfield  {author} {\bibinfo {author} {\bibfnamefont {F.}~\bibnamefont
  {Pietracaprina}}\ and\ \bibinfo {author} {\bibfnamefont {F.}~\bibnamefont
  {Alet}},\ }\bibfield  {title} {\bibinfo {title} {Probing many-body
  localization in a disordered quantum dimer model on the honeycomb lattice},\
  }\href@noop {} {\  (\bibinfo {year} {2020})},\ \Eprint
  {https://arxiv.org/abs/2005.10233} {arXiv:2005.10233} \BibitemShut {NoStop}%
\bibitem [{\citenamefont {Doggen}\ \emph {et~al.}(2020)\citenamefont {Doggen},
  \citenamefont {Gornyi}, \citenamefont {Mirlin},\ and\ \citenamefont
  {Polyakov}}]{Doggen2020}%
  \BibitemOpen
  \bibfield  {author} {\bibinfo {author} {\bibfnamefont {E.~V.~H.}\
  \bibnamefont {Doggen}}, \bibinfo {author} {\bibfnamefont {I.~V.}\
  \bibnamefont {Gornyi}}, \bibinfo {author} {\bibfnamefont {A.~D.}\
  \bibnamefont {Mirlin}},\ and\ \bibinfo {author} {\bibfnamefont {D.~G.}\
  \bibnamefont {Polyakov}},\ }\bibfield  {title} {\bibinfo {title} {Slow
  many-body delocalization beyond one dimension},\ }\href@noop {} {\  (\bibinfo
  {year} {2020})},\ \Eprint {https://arxiv.org/abs/2002.07635}
  {arXiv:2002.07635} \BibitemShut {NoStop}%
\bibitem [{\citenamefont {Pekker}\ \emph {et~al.}(2017)\citenamefont {Pekker},
  \citenamefont {Clark}, \citenamefont {Oganesyan},\ and\ \citenamefont
  {Refael}}]{Pekker2017b}%
  \BibitemOpen
  \bibfield  {author} {\bibinfo {author} {\bibfnamefont {D.}~\bibnamefont
  {Pekker}}, \bibinfo {author} {\bibfnamefont {B.~K.}\ \bibnamefont {Clark}},
  \bibinfo {author} {\bibfnamefont {V.}~\bibnamefont {Oganesyan}},\ and\
  \bibinfo {author} {\bibfnamefont {G.}~\bibnamefont {Refael}},\ }\bibfield
  {title} {\bibinfo {title} {Fixed points of wegner-wilson flows and many-body
  localization},\ }\href {https://doi.org/10.1103/PhysRevLett.119.075701}
  {\bibfield  {journal} {\bibinfo  {journal} {Phys. Rev. Lett.}\ }\textbf
  {\bibinfo {volume} {119}},\ \bibinfo {pages} {075701} (\bibinfo {year}
  {2017})}\BibitemShut {NoStop}%
\bibitem [{\citenamefont {Kulshreshtha}\ \emph {et~al.}(2018)\citenamefont
  {Kulshreshtha}, \citenamefont {Pal}, \citenamefont {Wahl},\ and\
  \citenamefont {Simon}}]{Kulshreshta2018}%
  \BibitemOpen
  \bibfield  {author} {\bibinfo {author} {\bibfnamefont {A.~K.}\ \bibnamefont
  {Kulshreshtha}}, \bibinfo {author} {\bibfnamefont {A.}~\bibnamefont {Pal}},
  \bibinfo {author} {\bibfnamefont {T.~B.}\ \bibnamefont {Wahl}},\ and\
  \bibinfo {author} {\bibfnamefont {S.~H.}\ \bibnamefont {Simon}},\ }\bibfield
  {title} {\bibinfo {title} {Behavior of l-bits near the many-body localization
  transition},\ }\href {https://doi.org/10.1103/PhysRevB.98.184201} {\bibfield
  {journal} {\bibinfo  {journal} {Phys. Rev. B}\ }\textbf {\bibinfo {volume}
  {98}},\ \bibinfo {pages} {184201} (\bibinfo {year} {2018})}\BibitemShut
  {NoStop}%
\bibitem [{\citenamefont {Goihl}\ \emph {et~al.}(2018)\citenamefont {Goihl},
  \citenamefont {Gluza}, \citenamefont {Krumnow},\ and\ \citenamefont
  {Eisert}}]{Goihl2018}%
  \BibitemOpen
  \bibfield  {author} {\bibinfo {author} {\bibfnamefont {M.}~\bibnamefont
  {Goihl}}, \bibinfo {author} {\bibfnamefont {M.}~\bibnamefont {Gluza}},
  \bibinfo {author} {\bibfnamefont {C.}~\bibnamefont {Krumnow}},\ and\ \bibinfo
  {author} {\bibfnamefont {J.}~\bibnamefont {Eisert}},\ }\bibfield  {title}
  {\bibinfo {title} {Construction of exact constants of motion and effective
  models for many-body localized systems},\ }\href
  {https://doi.org/10.1103/PhysRevB.97.134202} {\bibfield  {journal} {\bibinfo
  {journal} {Phys. Rev. B}\ }\textbf {\bibinfo {volume} {97}},\ \bibinfo
  {pages} {134202} (\bibinfo {year} {2018})}\BibitemShut {NoStop}%
\bibitem [{\citenamefont {{Yu}}\ \emph {et~al.}(2019)\citenamefont {{Yu}},
  \citenamefont {{Pekker}},\ and\ \citenamefont {{Clark}}}]{Yu2019}%
  \BibitemOpen
  \bibfield  {author} {\bibinfo {author} {\bibfnamefont {X.}~\bibnamefont
  {{Yu}}}, \bibinfo {author} {\bibfnamefont {D.}~\bibnamefont {{Pekker}}},\
  and\ \bibinfo {author} {\bibfnamefont {B.~K.}\ \bibnamefont {{Clark}}},\
  }\bibfield  {title} {\bibinfo {title} {{Bulk Geometry of the Many Body
  Localized Phase from Wilson-Wegner Flow}},\ }\href@noop {} {\  (\bibinfo
  {year} {2019})},\ \Eprint {https://arxiv.org/abs/1909.11097}
  {arXiv:1909.11097} \BibitemShut {NoStop}%
\bibitem [{\citenamefont {Varma}\ \emph {et~al.}(2019)\citenamefont {Varma},
  \citenamefont {Raj}, \citenamefont {Gopalakrishnan}, \citenamefont
  {Oganesyan},\ and\ \citenamefont {Pekker}}]{Varma2019}%
  \BibitemOpen
  \bibfield  {author} {\bibinfo {author} {\bibfnamefont {V.~K.}\ \bibnamefont
  {Varma}}, \bibinfo {author} {\bibfnamefont {A.}~\bibnamefont {Raj}}, \bibinfo
  {author} {\bibfnamefont {S.}~\bibnamefont {Gopalakrishnan}}, \bibinfo
  {author} {\bibfnamefont {V.}~\bibnamefont {Oganesyan}},\ and\ \bibinfo
  {author} {\bibfnamefont {D.}~\bibnamefont {Pekker}},\ }\bibfield  {title}
  {\bibinfo {title} {Length scales in the many-body localized phase and their
  spectral signatures},\ }\href {https://doi.org/10.1103/PhysRevB.100.115136}
  {\bibfield  {journal} {\bibinfo  {journal} {Phys. Rev. B}\ }\textbf {\bibinfo
  {volume} {100}},\ \bibinfo {pages} {115136} (\bibinfo {year}
  {2019})}\BibitemShut {NoStop}%
\bibitem [{\citenamefont {Peng}\ \emph {et~al.}(2019)\citenamefont {Peng},
  \citenamefont {Li}, \citenamefont {Yan}, \citenamefont {Wei},\ and\
  \citenamefont {Cappellaro}}]{Peng2019}%
  \BibitemOpen
  \bibfield  {author} {\bibinfo {author} {\bibfnamefont {P.}~\bibnamefont
  {Peng}}, \bibinfo {author} {\bibfnamefont {Z.}~\bibnamefont {Li}}, \bibinfo
  {author} {\bibfnamefont {H.}~\bibnamefont {Yan}}, \bibinfo {author}
  {\bibfnamefont {K.~X.}\ \bibnamefont {Wei}},\ and\ \bibinfo {author}
  {\bibfnamefont {P.}~\bibnamefont {Cappellaro}},\ }\bibfield  {title}
  {\bibinfo {title} {Comparing many-body localization lengths via
  nonperturbative construction of local integrals of motion},\ }\href
  {https://doi.org/10.1103/PhysRevB.100.214203} {\bibfield  {journal} {\bibinfo
   {journal} {Phys. Rev. B}\ }\textbf {\bibinfo {volume} {100}},\ \bibinfo
  {pages} {214203} (\bibinfo {year} {2019})}\BibitemShut {NoStop}%
\bibitem [{\citenamefont {Kelly}\ \emph {et~al.}(2020)\citenamefont {Kelly},
  \citenamefont {Nandkishore},\ and\ \citenamefont {Marino}}]{Kelly2020}%
  \BibitemOpen
  \bibfield  {author} {\bibinfo {author} {\bibfnamefont {S.~P.}\ \bibnamefont
  {Kelly}}, \bibinfo {author} {\bibfnamefont {R.}~\bibnamefont {Nandkishore}},\
  and\ \bibinfo {author} {\bibfnamefont {J.}~\bibnamefont {Marino}},\
  }\bibfield  {title} {\bibinfo {title} {Exploring many-body localization in
  quantum systems coupled to an environment via wegner-wilson flows},\ }\href
  {https://doi.org/https://doi.org/10.1016/j.nuclphysb.2019.114886} {\bibfield
  {journal} {\bibinfo  {journal} {Nucl. Phys. B}\ }\textbf {\bibinfo {volume}
  {951}},\ \bibinfo {pages} {114886} (\bibinfo {year} {2020})}\BibitemShut
  {NoStop}%
\bibitem [{\citenamefont {Kim}\ \emph {et~al.}(2015)\citenamefont {Kim},
  \citenamefont {Ba\~nuls}, \citenamefont {Cirac}, \citenamefont {Hastings},\
  and\ \citenamefont {Huse}}]{Kim2015}%
  \BibitemOpen
  \bibfield  {author} {\bibinfo {author} {\bibfnamefont {H.}~\bibnamefont
  {Kim}}, \bibinfo {author} {\bibfnamefont {M.~C.}\ \bibnamefont {Ba\~nuls}},
  \bibinfo {author} {\bibfnamefont {J.~I.}\ \bibnamefont {Cirac}}, \bibinfo
  {author} {\bibfnamefont {M.~B.}\ \bibnamefont {Hastings}},\ and\ \bibinfo
  {author} {\bibfnamefont {D.~A.}\ \bibnamefont {Huse}},\ }\bibfield  {title}
  {\bibinfo {title} {Slowest local operators in quantum spin chains},\ }\href
  {https://doi.org/10.1103/PhysRevE.92.012128} {\bibfield  {journal} {\bibinfo
  {journal} {Phys. Rev. E}\ }\textbf {\bibinfo {volume} {92}},\ \bibinfo
  {pages} {012128} (\bibinfo {year} {2015})}\BibitemShut {NoStop}%
\bibitem [{\citenamefont {Chandran}\ \emph {et~al.}(2015)\citenamefont
  {Chandran}, \citenamefont {Kim}, \citenamefont {Vidal},\ and\ \citenamefont
  {Abanin}}]{Chandran2015}%
  \BibitemOpen
  \bibfield  {author} {\bibinfo {author} {\bibfnamefont {A.}~\bibnamefont
  {Chandran}}, \bibinfo {author} {\bibfnamefont {I.~H.}\ \bibnamefont {Kim}},
  \bibinfo {author} {\bibfnamefont {G.}~\bibnamefont {Vidal}},\ and\ \bibinfo
  {author} {\bibfnamefont {D.~A.}\ \bibnamefont {Abanin}},\ }\bibfield  {title}
  {\bibinfo {title} {Constructing local integrals of motion in the many-body
  localized phase},\ }\href {https://doi.org/10.1103/PhysRevB.91.085425}
  {\bibfield  {journal} {\bibinfo  {journal} {Phys. Rev. B}\ }\textbf {\bibinfo
  {volume} {91}},\ \bibinfo {pages} {085425} (\bibinfo {year}
  {2015})}\BibitemShut {NoStop}%
\bibitem [{\citenamefont {O'Brien}\ \emph {et~al.}(2016)\citenamefont
  {O'Brien}, \citenamefont {Abanin}, \citenamefont {Vidal},\ and\ \citenamefont
  {Papi\ifmmode~\acute{c}\else \'{c}\fi{}}}]{OBrien2016}%
  \BibitemOpen
  \bibfield  {author} {\bibinfo {author} {\bibfnamefont {T.~E.}\ \bibnamefont
  {O'Brien}}, \bibinfo {author} {\bibfnamefont {D.~A.}\ \bibnamefont {Abanin}},
  \bibinfo {author} {\bibfnamefont {G.}~\bibnamefont {Vidal}},\ and\ \bibinfo
  {author} {\bibfnamefont {Z.}~\bibnamefont {Papi\ifmmode~\acute{c}\else
  \'{c}\fi{}}},\ }\bibfield  {title} {\bibinfo {title} {Explicit construction
  of local conserved operators in disordered many-body systems},\ }\href
  {https://doi.org/10.1103/PhysRevB.94.144208} {\bibfield  {journal} {\bibinfo
  {journal} {Phys. Rev. B}\ }\textbf {\bibinfo {volume} {94}},\ \bibinfo
  {pages} {144208} (\bibinfo {year} {2016})}\BibitemShut {NoStop}%
\bibitem [{\citenamefont {Lin}\ and\ \citenamefont
  {Motrunich}(2017)}]{Lin2017}%
  \BibitemOpen
  \bibfield  {author} {\bibinfo {author} {\bibfnamefont {C.-J.}\ \bibnamefont
  {Lin}}\ and\ \bibinfo {author} {\bibfnamefont {O.~I.}\ \bibnamefont
  {Motrunich}},\ }\bibfield  {title} {\bibinfo {title} {Explicit construction
  of quasiconserved local operator of translationally invariant nonintegrable
  quantum spin chain in prethermalization},\ }\href
  {https://doi.org/10.1103/PhysRevB.96.214301} {\bibfield  {journal} {\bibinfo
  {journal} {Phys. Rev. B}\ }\textbf {\bibinfo {volume} {96}},\ \bibinfo
  {pages} {214301} (\bibinfo {year} {2017})}\BibitemShut {NoStop}%
\bibitem [{\citenamefont {Mierzejewski}\ \emph {et~al.}(2018)\citenamefont
  {Mierzejewski}, \citenamefont {Kozarzewski},\ and\ \citenamefont
  {Prelov\ifmmode~\check{s}\else \v{s}\fi{}ek}}]{Mierzejewski2018}%
  \BibitemOpen
  \bibfield  {author} {\bibinfo {author} {\bibfnamefont {M.}~\bibnamefont
  {Mierzejewski}}, \bibinfo {author} {\bibfnamefont {M.}~\bibnamefont
  {Kozarzewski}},\ and\ \bibinfo {author} {\bibfnamefont {P.}~\bibnamefont
  {Prelov\ifmmode~\check{s}\else \v{s}\fi{}ek}},\ }\bibfield  {title} {\bibinfo
  {title} {Counting local integrals of motion in disordered spinless-fermion
  and hubbard chains},\ }\href {https://doi.org/10.1103/PhysRevB.97.064204}
  {\bibfield  {journal} {\bibinfo  {journal} {Phys. Rev. B}\ }\textbf {\bibinfo
  {volume} {97}},\ \bibinfo {pages} {064204} (\bibinfo {year}
  {2018})}\BibitemShut {NoStop}%
\bibitem [{\citenamefont {Pancotti}\ \emph {et~al.}(2018)\citenamefont
  {Pancotti}, \citenamefont {Knap}, \citenamefont {Huse}, \citenamefont
  {Cirac},\ and\ \citenamefont {Ba\~nuls}}]{Pancotti2018}%
  \BibitemOpen
  \bibfield  {author} {\bibinfo {author} {\bibfnamefont {N.}~\bibnamefont
  {Pancotti}}, \bibinfo {author} {\bibfnamefont {M.}~\bibnamefont {Knap}},
  \bibinfo {author} {\bibfnamefont {D.~A.}\ \bibnamefont {Huse}}, \bibinfo
  {author} {\bibfnamefont {J.~I.}\ \bibnamefont {Cirac}},\ and\ \bibinfo
  {author} {\bibfnamefont {M.~C.}\ \bibnamefont {Ba\~nuls}},\ }\bibfield
  {title} {\bibinfo {title} {Almost conserved operators in nearly many-body
  localized systems},\ }\href {https://doi.org/10.1103/PhysRevB.97.094206}
  {\bibfield  {journal} {\bibinfo  {journal} {Phys. Rev. B}\ }\textbf {\bibinfo
  {volume} {97}},\ \bibinfo {pages} {094206} (\bibinfo {year}
  {2018})}\BibitemShut {NoStop}%
\bibitem [{\citenamefont {Oganesyan}\ and\ \citenamefont
  {Huse}(2007)}]{Oganesyan2007}%
  \BibitemOpen
  \bibfield  {author} {\bibinfo {author} {\bibfnamefont {V.}~\bibnamefont
  {Oganesyan}}\ and\ \bibinfo {author} {\bibfnamefont {D.~A.}\ \bibnamefont
  {Huse}},\ }\bibfield  {title} {\bibinfo {title} {Localization of interacting
  fermions at high temperature},\ }\href
  {https://doi.org/10.1103/PhysRevB.75.155111} {\bibfield  {journal} {\bibinfo
  {journal} {Phys. Rev. B}\ }\textbf {\bibinfo {volume} {75}},\ \bibinfo
  {pages} {155111} (\bibinfo {year} {2007})}\BibitemShut {NoStop}%
\bibitem [{\citenamefont {Pal}\ and\ \citenamefont {Huse}(2010)}]{Pal2010}%
  \BibitemOpen
  \bibfield  {author} {\bibinfo {author} {\bibfnamefont {A.}~\bibnamefont
  {Pal}}\ and\ \bibinfo {author} {\bibfnamefont {D.~A.}\ \bibnamefont {Huse}},\
  }\bibfield  {title} {\bibinfo {title} {Many-body localization phase
  transition},\ }\href {https://doi.org/10.1103/PhysRevB.82.174411} {\bibfield
  {journal} {\bibinfo  {journal} {Phys. Rev. B}\ }\textbf {\bibinfo {volume}
  {82}},\ \bibinfo {pages} {174411} (\bibinfo {year} {2010})}\BibitemShut
  {NoStop}%
\bibitem [{\citenamefont {\ifmmode \check{Z}\else
  \v{Z}\fi{}nidari\ifmmode~\check{c}\else \v{c}\fi{}}\ \emph
  {et~al.}(2008)\citenamefont {\ifmmode \check{Z}\else
  \v{Z}\fi{}nidari\ifmmode~\check{c}\else \v{c}\fi{}}, \citenamefont {Prosen},\
  and\ \citenamefont {Prelov\ifmmode~\check{s}\else
  \v{s}\fi{}ek}}]{Znidaric2008}%
  \BibitemOpen
  \bibfield  {author} {\bibinfo {author} {\bibfnamefont {M.}~\bibnamefont
  {\ifmmode \check{Z}\else \v{Z}\fi{}nidari\ifmmode~\check{c}\else
  \v{c}\fi{}}}, \bibinfo {author} {\bibfnamefont {T.}~\bibnamefont {Prosen}},\
  and\ \bibinfo {author} {\bibfnamefont {P.}~\bibnamefont
  {Prelov\ifmmode~\check{s}\else \v{s}\fi{}ek}},\ }\bibfield  {title} {\bibinfo
  {title} {{Many-body localization in the Heisenberg $XXZ$ magnet in a random
  field}},\ }\href {https://doi.org/10.1103/PhysRevB.77.064426} {\bibfield
  {journal} {\bibinfo  {journal} {Phys. Rev. B}\ }\textbf {\bibinfo {volume}
  {77}},\ \bibinfo {pages} {064426} (\bibinfo {year} {2008})}\BibitemShut
  {NoStop}%
\bibitem [{\citenamefont {Bardarson}\ \emph {et~al.}(2012)\citenamefont
  {Bardarson}, \citenamefont {Pollmann},\ and\ \citenamefont
  {Moore}}]{Bardarson2012}%
  \BibitemOpen
  \bibfield  {author} {\bibinfo {author} {\bibfnamefont {J.~H.}\ \bibnamefont
  {Bardarson}}, \bibinfo {author} {\bibfnamefont {F.}~\bibnamefont
  {Pollmann}},\ and\ \bibinfo {author} {\bibfnamefont {J.~E.}\ \bibnamefont
  {Moore}},\ }\bibfield  {title} {\bibinfo {title} {Unbounded growth of
  entanglement in models of many-body localization},\ }\href
  {https://doi.org/10.1103/PhysRevLett.109.017202} {\bibfield  {journal}
  {\bibinfo  {journal} {Phys. Rev. Lett.}\ }\textbf {\bibinfo {volume} {109}},\
  \bibinfo {pages} {017202} (\bibinfo {year} {2012})}\BibitemShut {NoStop}%
\bibitem [{\citenamefont {Luca}\ and\ \citenamefont
  {Scardicchio}(2013)}]{DeLuca2013}%
  \BibitemOpen
  \bibfield  {author} {\bibinfo {author} {\bibfnamefont {A.~D.}\ \bibnamefont
  {Luca}}\ and\ \bibinfo {author} {\bibfnamefont {A.}~\bibnamefont
  {Scardicchio}},\ }\bibfield  {title} {\bibinfo {title} {Ergodicity breaking
  in a model showing many-body localization},\ }\href
  {https://doi.org/10.1209/0295-5075/101/37003} {\bibfield  {journal} {\bibinfo
   {journal} {{EPL}}\ }\textbf {\bibinfo {volume} {101}},\ \bibinfo {pages}
  {37003} (\bibinfo {year} {2013})}\BibitemShut {NoStop}%
\bibitem [{\citenamefont {Pollmann}\ \emph {et~al.}(2016)\citenamefont
  {Pollmann}, \citenamefont {Khemani}, \citenamefont {Cirac},\ and\
  \citenamefont {Sondhi}}]{Pollman2016}%
  \BibitemOpen
  \bibfield  {author} {\bibinfo {author} {\bibfnamefont {F.}~\bibnamefont
  {Pollmann}}, \bibinfo {author} {\bibfnamefont {V.}~\bibnamefont {Khemani}},
  \bibinfo {author} {\bibfnamefont {J.~I.}\ \bibnamefont {Cirac}},\ and\
  \bibinfo {author} {\bibfnamefont {S.~L.}\ \bibnamefont {Sondhi}},\ }\bibfield
   {title} {\bibinfo {title} {{Efficient variational diagonalization of fully
  many-body localized Hamiltonians}},\ }\href
  {https://doi.org/10.1103/PhysRevB.94.041116} {\bibfield  {journal} {\bibinfo
  {journal} {Phys. Rev. B}\ }\textbf {\bibinfo {volume} {94}},\ \bibinfo
  {pages} {041116(R)} (\bibinfo {year} {2016})}\BibitemShut {NoStop}%
\bibitem [{\citenamefont {Pekker}\ and\ \citenamefont
  {Clark}(2017)}]{Pekker2017a}%
  \BibitemOpen
  \bibfield  {author} {\bibinfo {author} {\bibfnamefont {D.}~\bibnamefont
  {Pekker}}\ and\ \bibinfo {author} {\bibfnamefont {B.~K.}\ \bibnamefont
  {Clark}},\ }\bibfield  {title} {\bibinfo {title} {Encoding the structure of
  many-body localization with matrix product operators},\ }\href
  {https://doi.org/10.1103/PhysRevB.95.035116} {\bibfield  {journal} {\bibinfo
  {journal} {Phys. Rev. B}\ }\textbf {\bibinfo {volume} {95}},\ \bibinfo
  {pages} {035116} (\bibinfo {year} {2017})}\BibitemShut {NoStop}%
\bibitem [{\citenamefont {Wahl}\ \emph {et~al.}(2017)\citenamefont {Wahl},
  \citenamefont {Pal},\ and\ \citenamefont {Simon}}]{Wahl2017}%
  \BibitemOpen
  \bibfield  {author} {\bibinfo {author} {\bibfnamefont {T.~B.}\ \bibnamefont
  {Wahl}}, \bibinfo {author} {\bibfnamefont {A.}~\bibnamefont {Pal}},\ and\
  \bibinfo {author} {\bibfnamefont {S.~H.}\ \bibnamefont {Simon}},\ }\bibfield
  {title} {\bibinfo {title} {Efficient representation of fully many-body
  localized systems using tensor networks},\ }\href
  {https://doi.org/10.1103/PhysRevX.7.021018} {\bibfield  {journal} {\bibinfo
  {journal} {Phys. Rev. X}\ }\textbf {\bibinfo {volume} {7}},\ \bibinfo {pages}
  {021018} (\bibinfo {year} {2017})}\BibitemShut {NoStop}%
\bibitem [{Note2()}]{Note2}%
  \BibitemOpen
  \bibinfo {note} {See Supplemental Material for additional details on the
  methods used and for additional data obtained in this work. The supplement
  includes Refs.~\protect \rev@citealp
  {scipy2020,Canovi2011,Rademaker2017}.}\BibitemShut {Stop}%
\bibitem [{\citenamefont {Bender}\ and\ \citenamefont
  {Davidson}(1969)}]{Bender1969}%
  \BibitemOpen
  \bibfield  {author} {\bibinfo {author} {\bibfnamefont {C.~F.}\ \bibnamefont
  {Bender}}\ and\ \bibinfo {author} {\bibfnamefont {E.~R.}\ \bibnamefont
  {Davidson}},\ }\bibfield  {title} {\bibinfo {title} {{Studies in
  Configuration Interaction: The First-Row Diatomic Hydrides}},\ }\href
  {https://doi.org/10.1103/PhysRev.183.23} {\bibfield  {journal} {\bibinfo
  {journal} {Phys. Rev.}\ }\textbf {\bibinfo {volume} {183}},\ \bibinfo {pages}
  {23} (\bibinfo {year} {1969})}\BibitemShut {NoStop}%
\bibitem [{\citenamefont {Whitten}\ and\ \citenamefont
  {Hackmeyer}(1969)}]{Whitten1969}%
  \BibitemOpen
  \bibfield  {author} {\bibinfo {author} {\bibfnamefont {J.~L.}\ \bibnamefont
  {Whitten}}\ and\ \bibinfo {author} {\bibfnamefont {M.}~\bibnamefont
  {Hackmeyer}},\ }\bibfield  {title} {\bibinfo {title} {{Configuration
  Interaction Studies of Ground and Excited States of Polyatomic Molecules. I.
  The CI Formulation and Studies of Formaldehyde}},\ }\href
  {https://doi.org/10.1063/1.1671985} {\bibfield  {journal} {\bibinfo
  {journal} {J. Chem. Phys.}\ }\textbf {\bibinfo {volume} {51}},\ \bibinfo
  {pages} {5584} (\bibinfo {year} {1969})}\BibitemShut {NoStop}%
\bibitem [{\citenamefont {Holmes}\ \emph {et~al.}(2016)\citenamefont {Holmes},
  \citenamefont {Tubman},\ and\ \citenamefont {Umrigar}}]{Holmes2016}%
  \BibitemOpen
  \bibfield  {author} {\bibinfo {author} {\bibfnamefont {A.~A.}\ \bibnamefont
  {Holmes}}, \bibinfo {author} {\bibfnamefont {N.~A.}\ \bibnamefont {Tubman}},\
  and\ \bibinfo {author} {\bibfnamefont {C.~J.}\ \bibnamefont {Umrigar}},\
  }\bibfield  {title} {\bibinfo {title} {{Heat-Bath Configuration Interaction:
  An Efficient Selected Configuration Interaction Algorithm Inspired by
  Heat-Bath Sampling}},\ }\href {https://doi.org/10.1021/acs.jctc.6b00407}
  {\bibfield  {journal} {\bibinfo  {journal} {J. Chem. Theory Comput.}\
  }\textbf {\bibinfo {volume} {12}},\ \bibinfo {pages} {3674} (\bibinfo {year}
  {2016})}\BibitemShut {NoStop}%
\bibitem [{\citenamefont {Tubman}\ \emph {et~al.}(2016)\citenamefont {Tubman},
  \citenamefont {Lee}, \citenamefont {Takeshita}, \citenamefont {Head-Gordon},\
  and\ \citenamefont {Whaley}}]{Tubman2016}%
  \BibitemOpen
  \bibfield  {author} {\bibinfo {author} {\bibfnamefont {N.~M.}\ \bibnamefont
  {Tubman}}, \bibinfo {author} {\bibfnamefont {J.}~\bibnamefont {Lee}},
  \bibinfo {author} {\bibfnamefont {T.~Y.}\ \bibnamefont {Takeshita}}, \bibinfo
  {author} {\bibfnamefont {M.}~\bibnamefont {Head-Gordon}},\ and\ \bibinfo
  {author} {\bibfnamefont {K.~B.}\ \bibnamefont {Whaley}},\ }\bibfield  {title}
  {\bibinfo {title} {A deterministic alternative to the full configuration
  interaction quantum monte carlo method},\ }\href
  {https://doi.org/10.1063/1.4955109} {\bibfield  {journal} {\bibinfo
  {journal} {J. Chem. Phys.}\ }\textbf {\bibinfo {volume} {145}},\ \bibinfo
  {pages} {044112} (\bibinfo {year} {2016})}\BibitemShut {NoStop}%
\bibitem [{Note3()}]{Note3}%
  \BibitemOpen
  \bibinfo {note} {In order to save memory and time in our calculations, we
  modified step (a) so that only the largest 2000 terms of $[H, \tau ^z_i]$
  were kept before computing $[H, [H, \tau ^z_i]]$.}\BibitemShut {Stop}%
\bibitem [{\citenamefont {Chertkov}(2020)}]{bioms}%
  \BibitemOpen
  \bibfield  {author} {\bibinfo {author} {\bibfnamefont {E.}~\bibnamefont
  {Chertkov}},\ }\href@noop {} {\bibinfo {title} {{BIOMS: Binary Integrals of
  Motion}}},\ \bibinfo {howpublished}
  {\url{https://github.com/ClarkResearchGroup/bioms}} (\bibinfo {year}
  {2020})\BibitemShut {NoStop}%
\bibitem [{\citenamefont {Chertkov}(2019)}]{qosy}%
  \BibitemOpen
  \bibfield  {author} {\bibinfo {author} {\bibfnamefont {E.}~\bibnamefont
  {Chertkov}},\ }\href@noop {} {\bibinfo {title} {{Qosy: Quantum Operators from
  Symmetry}}},\ \bibinfo {howpublished}
  {\url{https://github.com/ClarkResearchGroup/qosy}} (\bibinfo {year}
  {2019})\BibitemShut {NoStop}%
\bibitem [{Note4()}]{Note4}%
  \BibitemOpen
  \bibinfo {note} {We use the same set of (scaled) disorder patterns for all
  $W$ for a fixed model but different disorder patterns for different
  models.}\BibitemShut {Stop}%
\bibitem [{Note5()}]{Note5}%
  \BibitemOpen
  \bibinfo {note} {Note that when we compute this quantity we use the $\sigma
  ^z_i$ on the site $i$ with the largest weight (see Eq.~(\ref {eq:weight}))
  rather than the $\sigma ^z_i$ used to initialize the basis $B$. In general,
  the $\tau ^z_i$ operators discovered with our method can ``drift'' away from
  their initial site, though this tends to only become significant at low
  disorder strength (see supplement).}\BibitemShut {Stop}%
\bibitem [{Note6()}]{Note6}%
  \BibitemOpen
  \bibinfo {note} {Note that the summation in the denominator of $\protect
  \tilde {w}_{\protect \mathbf {r}}$ is only over the positions $\protect
  \mathbf {r}'$ where $w_{\protect \mathbf {r}'} \protect \neq 0$ and $\protect
  \mathbf {r}_i \equiv \protect \textrm {argmax}_{\protect \mathbf {r}}
  w_{\protect \mathbf {r}}$.}\BibitemShut {Stop}%
\bibitem [{\citenamefont {Chertkov}\ \emph {et~al.}(2020)\citenamefont
  {Chertkov}, \citenamefont {Villalonga},\ and\ \citenamefont
  {Clark}}]{Chertkov2020}%
  \BibitemOpen
  \bibfield  {author} {\bibinfo {author} {\bibfnamefont {E.}~\bibnamefont
  {Chertkov}}, \bibinfo {author} {\bibfnamefont {B.}~\bibnamefont
  {Villalonga}},\ and\ \bibinfo {author} {\bibfnamefont {B.~K.}\ \bibnamefont
  {Clark}},\ }\bibfield  {title} {\bibinfo {title} {{Engineering topological
  models with a general-purpose symmetry-to-Hamiltonian approach}},\ }\href
  {https://doi.org/10.1103/PhysRevResearch.2.023348} {\bibfield  {journal}
  {\bibinfo  {journal} {Phys. Rev. Research}\ }\textbf {\bibinfo {volume}
  {2}},\ \bibinfo {pages} {023348} (\bibinfo {year} {2020})}\BibitemShut
  {NoStop}%
\bibitem [{\citenamefont {Pekker}\ \emph {et~al.}(2014)\citenamefont {Pekker},
  \citenamefont {Refael}, \citenamefont {Altman}, \citenamefont {Demler},\ and\
  \citenamefont {Oganesyan}}]{Pekker2014}%
  \BibitemOpen
  \bibfield  {author} {\bibinfo {author} {\bibfnamefont {D.}~\bibnamefont
  {Pekker}}, \bibinfo {author} {\bibfnamefont {G.}~\bibnamefont {Refael}},
  \bibinfo {author} {\bibfnamefont {E.}~\bibnamefont {Altman}}, \bibinfo
  {author} {\bibfnamefont {E.}~\bibnamefont {Demler}},\ and\ \bibinfo {author}
  {\bibfnamefont {V.}~\bibnamefont {Oganesyan}},\ }\bibfield  {title} {\bibinfo
  {title} {Hilbert-glass transition: New universality of temperature-tuned
  many-body dynamical quantum criticality},\ }\href
  {https://doi.org/10.1103/PhysRevX.4.011052} {\bibfield  {journal} {\bibinfo
  {journal} {Phys. Rev. X}\ }\textbf {\bibinfo {volume} {4}},\ \bibinfo {pages}
  {011052} (\bibinfo {year} {2014})}\BibitemShut {NoStop}%
\bibitem [{\citenamefont {{Katsura}}\ \emph {et~al.}(2015)\citenamefont
  {{Katsura}}, \citenamefont {{Schuricht}},\ and\ \citenamefont
  {{Takahashi}}}]{Katsura2015}%
  \BibitemOpen
  \bibfield  {author} {\bibinfo {author} {\bibfnamefont {H.}~\bibnamefont
  {{Katsura}}}, \bibinfo {author} {\bibfnamefont {D.}~\bibnamefont
  {{Schuricht}}},\ and\ \bibinfo {author} {\bibfnamefont {M.}~\bibnamefont
  {{Takahashi}}},\ }\bibfield  {title} {\bibinfo {title} {{Exact Ground States
  and Topological Order in Interacting Kitaev/Majorana Chains}},\ }\href
  {https://doi.org/10.1103/PhysRevB.92.115137} {\bibfield  {journal} {\bibinfo
  {journal} {Phys. Rev. B}\ }\textbf {\bibinfo {volume} {92}},\ \bibinfo
  {pages} {115137} (\bibinfo {year} {2015})}\BibitemShut {NoStop}%
\bibitem [{\citenamefont {Virtanen}\ \emph {et~al.}(2020)\citenamefont
  {Virtanen} \emph {et~al.}}]{scipy2020}%
  \BibitemOpen
  \bibfield  {author} {\bibinfo {author} {\bibfnamefont {P.}~\bibnamefont
  {Virtanen}} \emph {et~al.},\ }\bibfield  {title} {\bibinfo {title} {{SciPy
  1.0: Fundamental Algorithms for Scientific Computing in Python}},\ }\href
  {https://doi.org/https://doi.org/10.1038/s41592-019-0686-2} {\bibfield
  {journal} {\bibinfo  {journal} {Nature Methods}\ }\textbf {\bibinfo {volume}
  {17}},\ \bibinfo {pages} {261} (\bibinfo {year} {2020})}\BibitemShut
  {NoStop}%
\bibitem [{\citenamefont {Canovi}\ \emph {et~al.}(2011)\citenamefont {Canovi},
  \citenamefont {Rossini}, \citenamefont {Fazio}, \citenamefont {Santoro},\
  and\ \citenamefont {Silva}}]{Canovi2011}%
  \BibitemOpen
  \bibfield  {author} {\bibinfo {author} {\bibfnamefont {E.}~\bibnamefont
  {Canovi}}, \bibinfo {author} {\bibfnamefont {D.}~\bibnamefont {Rossini}},
  \bibinfo {author} {\bibfnamefont {R.}~\bibnamefont {Fazio}}, \bibinfo
  {author} {\bibfnamefont {G.~E.}\ \bibnamefont {Santoro}},\ and\ \bibinfo
  {author} {\bibfnamefont {A.}~\bibnamefont {Silva}},\ }\bibfield  {title}
  {\bibinfo {title} {Quantum quenches, thermalization, and many-body
  localization},\ }\href {https://doi.org/10.1103/PhysRevB.83.094431}
  {\bibfield  {journal} {\bibinfo  {journal} {Phys. Rev. B}\ }\textbf {\bibinfo
  {volume} {83}},\ \bibinfo {pages} {094431} (\bibinfo {year}
  {2011})}\BibitemShut {NoStop}%
\bibitem [{\citenamefont {Rademaker}\ \emph {et~al.}(2017)\citenamefont
  {Rademaker}, \citenamefont {Ortu{\~{n}}o},\ and\ \citenamefont
  {Somoza}}]{Rademaker2017}%
  \BibitemOpen
  \bibfield  {author} {\bibinfo {author} {\bibfnamefont {L.}~\bibnamefont
  {Rademaker}}, \bibinfo {author} {\bibfnamefont {M.}~\bibnamefont
  {Ortu{\~{n}}o}},\ and\ \bibinfo {author} {\bibfnamefont {A.~M.}\ \bibnamefont
  {Somoza}},\ }\bibfield  {title} {\bibinfo {title} {{Many-body localization
  from the perspective of Integrals of Motion}},\ }\href
  {https://doi.org/10.1002/andp.201600322} {\bibfield  {journal} {\bibinfo
  {journal} {Ann. Phys. (Berl.)}\ }\textbf {\bibinfo {volume} {529}},\ \bibinfo
  {pages} {1600322} (\bibinfo {year} {2017})}\BibitemShut {NoStop}%
\end{thebibliography}%


\begin{thebibliography}{20}%
\makeatletter
\providecommand \@ifxundefined [1]{%
 \@ifx{#1\undefined}
}%
\providecommand \@ifnum [1]{%
 \ifnum #1\expandafter \@firstoftwo
 \else \expandafter \@secondoftwo
 \fi
}%
\providecommand \@ifx [1]{%
 \ifx #1\expandafter \@firstoftwo
 \else \expandafter \@secondoftwo
 \fi
}%
\providecommand \natexlab [1]{#1}%
\providecommand \enquote  [1]{``#1''}%
\providecommand \bibnamefont  [1]{#1}%
\providecommand \bibfnamefont [1]{#1}%
\providecommand \citenamefont [1]{#1}%
\providecommand \href@noop [0]{\@secondoftwo}%
\providecommand \href [0]{\begingroup \@sanitize@url \@href}%
\providecommand \@href[1]{\@@startlink{#1}\@@href}%
\providecommand \@@href[1]{\endgroup#1\@@endlink}%
\providecommand \@sanitize@url [0]{\catcode `\\12\catcode `\$12\catcode
  `\&12\catcode `\#12\catcode `\^12\catcode `\_12\catcode `\%12\relax}%
\providecommand \@@startlink[1]{}%
\providecommand \@@endlink[0]{}%
\providecommand \url  [0]{\begingroup\@sanitize@url \@url }%
\providecommand \@url [1]{\endgroup\@href {#1}{\urlprefix }}%
\providecommand \urlprefix  [0]{URL }%
\providecommand \Eprint [0]{\href }%
\providecommand \doibase [0]{https://doi.org/}%
\providecommand \selectlanguage [0]{\@gobble}%
\providecommand \bibinfo  [0]{\@secondoftwo}%
\providecommand \bibfield  [0]{\@secondoftwo}%
\providecommand \translation [1]{[#1]}%
\providecommand \BibitemOpen [0]{}%
\providecommand \bibitemStop [0]{}%
\providecommand \bibitemNoStop [0]{.\EOS\space}%
\providecommand \EOS [0]{\spacefactor3000\relax}%
\providecommand \BibitemShut  [1]{\csname bibitem#1\endcsname}%
\let\auto@bib@innerbib\@empty
\bibitem [{\citenamefont {Chertkov}\ \emph {et~al.}(2020)\citenamefont
  {Chertkov}, \citenamefont {Villalonga},\ and\ \citenamefont
  {Clark}}]{Chertkov2020}%
  \BibitemOpen
  \bibfield  {author} {\bibinfo {author} {\bibfnamefont {E.}~\bibnamefont
  {Chertkov}}, \bibinfo {author} {\bibfnamefont {B.}~\bibnamefont
  {Villalonga}},\ and\ \bibinfo {author} {\bibfnamefont {B.~K.}\ \bibnamefont
  {Clark}},\ }\bibfield  {title} {\bibinfo {title} {{Engineering topological
  models with a general-purpose symmetry-to-Hamiltonian approach}},\ }\href
  {https://doi.org/10.1103/PhysRevResearch.2.023348} {\bibfield  {journal}
  {\bibinfo  {journal} {Phys. Rev. Research}\ }\textbf {\bibinfo {volume}
  {2}},\ \bibinfo {pages} {023348} (\bibinfo {year} {2020})}\BibitemShut
  {NoStop}%
\bibitem [{\citenamefont {Chertkov}(2020)}]{bioms}%
  \BibitemOpen
  \bibfield  {author} {\bibinfo {author} {\bibfnamefont {E.}~\bibnamefont
  {Chertkov}},\ }\href@noop {} {\bibinfo {title} {{BIOMS: Binary Integrals of
  Motion}}},\ \bibinfo {howpublished}
  {\url{https://github.com/ClarkResearchGroup/bioms}} (\bibinfo {year}
  {2020})\BibitemShut {NoStop}%
\bibitem [{\citenamefont {Chertkov}(2019)}]{qosy}%
  \BibitemOpen
  \bibfield  {author} {\bibinfo {author} {\bibfnamefont {E.}~\bibnamefont
  {Chertkov}},\ }\href@noop {} {\bibinfo {title} {{Qosy: Quantum Operators from
  Symmetry}}},\ \bibinfo {howpublished}
  {\url{https://github.com/ClarkResearchGroup/qosy}} (\bibinfo {year}
  {2019})\BibitemShut {NoStop}%
\bibitem [{\citenamefont {Virtanen}\ \emph {et~al.}(2020)\citenamefont
  {Virtanen} \emph {et~al.}}]{scipy2020}%
  \BibitemOpen
  \bibfield  {author} {\bibinfo {author} {\bibfnamefont {P.}~\bibnamefont
  {Virtanen}} \emph {et~al.},\ }\bibfield  {title} {\bibinfo {title} {{SciPy
  1.0: Fundamental Algorithms for Scientific Computing in Python}},\ }\href
  {https://doi.org/https://doi.org/10.1038/s41592-019-0686-2} {\bibfield
  {journal} {\bibinfo  {journal} {Nature Methods}\ }\textbf {\bibinfo {volume}
  {17}},\ \bibinfo {pages} {261} (\bibinfo {year} {2020})}\BibitemShut
  {NoStop}%
\bibitem [{\citenamefont {Huse}\ \emph {et~al.}(2014)\citenamefont {Huse},
  \citenamefont {Nandkishore},\ and\ \citenamefont {Oganesyan}}]{Huse2014}%
  \BibitemOpen
  \bibfield  {author} {\bibinfo {author} {\bibfnamefont {D.~A.}\ \bibnamefont
  {Huse}}, \bibinfo {author} {\bibfnamefont {R.}~\bibnamefont {Nandkishore}},\
  and\ \bibinfo {author} {\bibfnamefont {V.}~\bibnamefont {Oganesyan}},\
  }\bibfield  {title} {\bibinfo {title} {Phenomenology of fully
  many-body-localized systems},\ }\href
  {https://doi.org/10.1103/PhysRevB.90.174202} {\bibfield  {journal} {\bibinfo
  {journal} {Phys. Rev. B}\ }\textbf {\bibinfo {volume} {90}},\ \bibinfo
  {pages} {174202} (\bibinfo {year} {2014})}\BibitemShut {NoStop}%
\bibitem [{\citenamefont {Villalonga}\ \emph {et~al.}(2018)\citenamefont
  {Villalonga}, \citenamefont {Yu}, \citenamefont {Luitz},\ and\ \citenamefont
  {Clark}}]{Villalonga2018}%
  \BibitemOpen
  \bibfield  {author} {\bibinfo {author} {\bibfnamefont {B.}~\bibnamefont
  {Villalonga}}, \bibinfo {author} {\bibfnamefont {X.}~\bibnamefont {Yu}},
  \bibinfo {author} {\bibfnamefont {D.~J.}\ \bibnamefont {Luitz}},\ and\
  \bibinfo {author} {\bibfnamefont {B.~K.}\ \bibnamefont {Clark}},\ }\bibfield
  {title} {\bibinfo {title} {Exploring one-particle orbitals in large many-body
  localized systems},\ }\href {https://doi.org/10.1103/PhysRevB.97.104406}
  {\bibfield  {journal} {\bibinfo  {journal} {Phys. Rev. B}\ }\textbf {\bibinfo
  {volume} {97}},\ \bibinfo {pages} {104406} (\bibinfo {year}
  {2018})}\BibitemShut {NoStop}%
\bibitem [{\citenamefont {Canovi}\ \emph {et~al.}(2011)\citenamefont {Canovi},
  \citenamefont {Rossini}, \citenamefont {Fazio}, \citenamefont {Santoro},\
  and\ \citenamefont {Silva}}]{Canovi2011}%
  \BibitemOpen
  \bibfield  {author} {\bibinfo {author} {\bibfnamefont {E.}~\bibnamefont
  {Canovi}}, \bibinfo {author} {\bibfnamefont {D.}~\bibnamefont {Rossini}},
  \bibinfo {author} {\bibfnamefont {R.}~\bibnamefont {Fazio}}, \bibinfo
  {author} {\bibfnamefont {G.~E.}\ \bibnamefont {Santoro}},\ and\ \bibinfo
  {author} {\bibfnamefont {A.}~\bibnamefont {Silva}},\ }\bibfield  {title}
  {\bibinfo {title} {Quantum quenches, thermalization, and many-body
  localization},\ }\href {https://doi.org/10.1103/PhysRevB.83.094431}
  {\bibfield  {journal} {\bibinfo  {journal} {Phys. Rev. B}\ }\textbf {\bibinfo
  {volume} {83}},\ \bibinfo {pages} {094431} (\bibinfo {year}
  {2011})}\BibitemShut {NoStop}%
\bibitem [{\citenamefont {Serbyn}\ \emph {et~al.}(2013)\citenamefont {Serbyn},
  \citenamefont {Papi\ifmmode~\acute{c}\else \'{c}\fi{}},\ and\ \citenamefont
  {Abanin}}]{Serbyn2013}%
  \BibitemOpen
  \bibfield  {author} {\bibinfo {author} {\bibfnamefont {M.}~\bibnamefont
  {Serbyn}}, \bibinfo {author} {\bibfnamefont {Z.}~\bibnamefont
  {Papi\ifmmode~\acute{c}\else \'{c}\fi{}}},\ and\ \bibinfo {author}
  {\bibfnamefont {D.~A.}\ \bibnamefont {Abanin}},\ }\bibfield  {title}
  {\bibinfo {title} {Local conservation laws and the structure of the many-body
  localized states},\ }\href {https://doi.org/10.1103/PhysRevLett.111.127201}
  {\bibfield  {journal} {\bibinfo  {journal} {Phys. Rev. Lett.}\ }\textbf
  {\bibinfo {volume} {111}},\ \bibinfo {pages} {127201(R)} (\bibinfo {year}
  {2013})}\BibitemShut {NoStop}%
\bibitem [{\citenamefont {Lin}\ and\ \citenamefont
  {Motrunich}(2017)}]{Lin2017}%
  \BibitemOpen
  \bibfield  {author} {\bibinfo {author} {\bibfnamefont {C.-J.}\ \bibnamefont
  {Lin}}\ and\ \bibinfo {author} {\bibfnamefont {O.~I.}\ \bibnamefont
  {Motrunich}},\ }\bibfield  {title} {\bibinfo {title} {Explicit construction
  of quasiconserved local operator of translationally invariant nonintegrable
  quantum spin chain in prethermalization},\ }\href
  {https://doi.org/10.1103/PhysRevB.96.214301} {\bibfield  {journal} {\bibinfo
  {journal} {Phys. Rev. B}\ }\textbf {\bibinfo {volume} {96}},\ \bibinfo
  {pages} {214301} (\bibinfo {year} {2017})}\BibitemShut {NoStop}%
\bibitem [{\citenamefont {Rademaker}\ \emph {et~al.}(2017)\citenamefont
  {Rademaker}, \citenamefont {Ortu{\~{n}}o},\ and\ \citenamefont
  {Somoza}}]{Rademaker2017}%
  \BibitemOpen
  \bibfield  {author} {\bibinfo {author} {\bibfnamefont {L.}~\bibnamefont
  {Rademaker}}, \bibinfo {author} {\bibfnamefont {M.}~\bibnamefont
  {Ortu{\~{n}}o}},\ and\ \bibinfo {author} {\bibfnamefont {A.~M.}\ \bibnamefont
  {Somoza}},\ }\bibfield  {title} {\bibinfo {title} {{Many-body localization
  from the perspective of Integrals of Motion}},\ }\href
  {https://doi.org/10.1002/andp.201600322} {\bibfield  {journal} {\bibinfo
  {journal} {Ann. Phys. (Berl.)}\ }\textbf {\bibinfo {volume} {529}},\ \bibinfo
  {pages} {1600322} (\bibinfo {year} {2017})}\BibitemShut {NoStop}%
\bibitem [{\citenamefont {Thomson}\ and\ \citenamefont
  {Schir\'o}(2018)}]{Thomson2018}%
  \BibitemOpen
  \bibfield  {author} {\bibinfo {author} {\bibfnamefont {S.~J.}\ \bibnamefont
  {Thomson}}\ and\ \bibinfo {author} {\bibfnamefont {M.}~\bibnamefont
  {Schir\'o}},\ }\bibfield  {title} {\bibinfo {title} {Time evolution of
  many-body localized systems with the flow equation approach},\ }\href
  {https://doi.org/10.1103/PhysRevB.97.060201} {\bibfield  {journal} {\bibinfo
  {journal} {Phys. Rev. B}\ }\textbf {\bibinfo {volume} {97}},\ \bibinfo
  {pages} {060201(R)} (\bibinfo {year} {2018})}\BibitemShut {NoStop}%
\bibitem [{\citenamefont {Kulshreshtha}\ \emph {et~al.}(2018)\citenamefont
  {Kulshreshtha}, \citenamefont {Pal}, \citenamefont {Wahl},\ and\
  \citenamefont {Simon}}]{Kulshreshta2018}%
  \BibitemOpen
  \bibfield  {author} {\bibinfo {author} {\bibfnamefont {A.~K.}\ \bibnamefont
  {Kulshreshtha}}, \bibinfo {author} {\bibfnamefont {A.}~\bibnamefont {Pal}},
  \bibinfo {author} {\bibfnamefont {T.~B.}\ \bibnamefont {Wahl}},\ and\
  \bibinfo {author} {\bibfnamefont {S.~H.}\ \bibnamefont {Simon}},\ }\bibfield
  {title} {\bibinfo {title} {Behavior of l-bits near the many-body localization
  transition},\ }\href {https://doi.org/10.1103/PhysRevB.98.184201} {\bibfield
  {journal} {\bibinfo  {journal} {Phys. Rev. B}\ }\textbf {\bibinfo {volume}
  {98}},\ \bibinfo {pages} {184201} (\bibinfo {year} {2018})}\BibitemShut
  {NoStop}%
\bibitem [{\citenamefont {Pancotti}\ \emph {et~al.}(2018)\citenamefont
  {Pancotti}, \citenamefont {Knap}, \citenamefont {Huse}, \citenamefont
  {Cirac},\ and\ \citenamefont {Ba\~nuls}}]{Pancotti2018}%
  \BibitemOpen
  \bibfield  {author} {\bibinfo {author} {\bibfnamefont {N.}~\bibnamefont
  {Pancotti}}, \bibinfo {author} {\bibfnamefont {M.}~\bibnamefont {Knap}},
  \bibinfo {author} {\bibfnamefont {D.~A.}\ \bibnamefont {Huse}}, \bibinfo
  {author} {\bibfnamefont {J.~I.}\ \bibnamefont {Cirac}},\ and\ \bibinfo
  {author} {\bibfnamefont {M.~C.}\ \bibnamefont {Ba\~nuls}},\ }\bibfield
  {title} {\bibinfo {title} {Almost conserved operators in nearly many-body
  localized systems},\ }\href {https://doi.org/10.1103/PhysRevB.97.094206}
  {\bibfield  {journal} {\bibinfo  {journal} {Phys. Rev. B}\ }\textbf {\bibinfo
  {volume} {97}},\ \bibinfo {pages} {094206} (\bibinfo {year}
  {2018})}\BibitemShut {NoStop}%
\bibitem [{\citenamefont {Peng}\ \emph {et~al.}(2019)\citenamefont {Peng},
  \citenamefont {Li}, \citenamefont {Yan}, \citenamefont {Wei},\ and\
  \citenamefont {Cappellaro}}]{Peng2019}%
  \BibitemOpen
  \bibfield  {author} {\bibinfo {author} {\bibfnamefont {P.}~\bibnamefont
  {Peng}}, \bibinfo {author} {\bibfnamefont {Z.}~\bibnamefont {Li}}, \bibinfo
  {author} {\bibfnamefont {H.}~\bibnamefont {Yan}}, \bibinfo {author}
  {\bibfnamefont {K.~X.}\ \bibnamefont {Wei}},\ and\ \bibinfo {author}
  {\bibfnamefont {P.}~\bibnamefont {Cappellaro}},\ }\bibfield  {title}
  {\bibinfo {title} {Comparing many-body localization lengths via
  nonperturbative construction of local integrals of motion},\ }\href
  {https://doi.org/10.1103/PhysRevB.100.214203} {\bibfield  {journal} {\bibinfo
   {journal} {Phys. Rev. B}\ }\textbf {\bibinfo {volume} {100}},\ \bibinfo
  {pages} {214203} (\bibinfo {year} {2019})}\BibitemShut {NoStop}%
\bibitem [{\citenamefont {Wahl}\ \emph {et~al.}(2019)\citenamefont {Wahl},
  \citenamefont {Pal},\ and\ \citenamefont {Simon}}]{Wahl2019}%
  \BibitemOpen
  \bibfield  {author} {\bibinfo {author} {\bibfnamefont {T.}~\bibnamefont
  {Wahl}}, \bibinfo {author} {\bibfnamefont {A.}~\bibnamefont {Pal}},\ and\
  \bibinfo {author} {\bibfnamefont {S.}~\bibnamefont {Simon}},\ }\bibfield
  {title} {\bibinfo {title} {Signatures of the many-body localized regime in
  two dimensions},\ }\href {https://doi.org/10.1038/s41567-018-0339-x}
  {\bibfield  {journal} {\bibinfo  {journal} {Nat. Phys}\ }\textbf {\bibinfo
  {volume} {15}},\ \bibinfo {pages} {164} (\bibinfo {year} {2019})}\BibitemShut
  {NoStop}%
\bibitem [{\citenamefont {Kim}\ \emph {et~al.}(2015)\citenamefont {Kim},
  \citenamefont {Ba\~nuls}, \citenamefont {Cirac}, \citenamefont {Hastings},\
  and\ \citenamefont {Huse}}]{Kim2015}%
  \BibitemOpen
  \bibfield  {author} {\bibinfo {author} {\bibfnamefont {H.}~\bibnamefont
  {Kim}}, \bibinfo {author} {\bibfnamefont {M.~C.}\ \bibnamefont {Ba\~nuls}},
  \bibinfo {author} {\bibfnamefont {J.~I.}\ \bibnamefont {Cirac}}, \bibinfo
  {author} {\bibfnamefont {M.~B.}\ \bibnamefont {Hastings}},\ and\ \bibinfo
  {author} {\bibfnamefont {D.~A.}\ \bibnamefont {Huse}},\ }\bibfield  {title}
  {\bibinfo {title} {Slowest local operators in quantum spin chains},\ }\href
  {https://doi.org/10.1103/PhysRevE.92.012128} {\bibfield  {journal} {\bibinfo
  {journal} {Phys. Rev. E}\ }\textbf {\bibinfo {volume} {92}},\ \bibinfo
  {pages} {012128} (\bibinfo {year} {2015})}\BibitemShut {NoStop}%
\bibitem [{\citenamefont {Varma}\ \emph {et~al.}(2019)\citenamefont {Varma},
  \citenamefont {Raj}, \citenamefont {Gopalakrishnan}, \citenamefont
  {Oganesyan},\ and\ \citenamefont {Pekker}}]{Varma2019}%
  \BibitemOpen
  \bibfield  {author} {\bibinfo {author} {\bibfnamefont {V.~K.}\ \bibnamefont
  {Varma}}, \bibinfo {author} {\bibfnamefont {A.}~\bibnamefont {Raj}}, \bibinfo
  {author} {\bibfnamefont {S.}~\bibnamefont {Gopalakrishnan}}, \bibinfo
  {author} {\bibfnamefont {V.}~\bibnamefont {Oganesyan}},\ and\ \bibinfo
  {author} {\bibfnamefont {D.}~\bibnamefont {Pekker}},\ }\bibfield  {title}
  {\bibinfo {title} {Length scales in the many-body localized phase and their
  spectral signatures},\ }\href {https://doi.org/10.1103/PhysRevB.100.115136}
  {\bibfield  {journal} {\bibinfo  {journal} {Phys. Rev. B}\ }\textbf {\bibinfo
  {volume} {100}},\ \bibinfo {pages} {115136} (\bibinfo {year}
  {2019})}\BibitemShut {NoStop}%
\bibitem [{\citenamefont {Choi}\ \emph {et~al.}(2016)\citenamefont {Choi},
  \citenamefont {Hild}, \citenamefont {Zeiher}, \citenamefont {Schau{\ss}},
  \citenamefont {Rubio-Abadal}, \citenamefont {Yefsah}, \citenamefont
  {Khemani}, \citenamefont {Huse}, \citenamefont {Bloch},\ and\ \citenamefont
  {Gross}}]{Choi2016}%
  \BibitemOpen
  \bibfield  {author} {\bibinfo {author} {\bibfnamefont {J.-Y.}\ \bibnamefont
  {Choi}}, \bibinfo {author} {\bibfnamefont {S.}~\bibnamefont {Hild}}, \bibinfo
  {author} {\bibfnamefont {J.}~\bibnamefont {Zeiher}}, \bibinfo {author}
  {\bibfnamefont {P.}~\bibnamefont {Schau{\ss}}}, \bibinfo {author}
  {\bibfnamefont {A.}~\bibnamefont {Rubio-Abadal}}, \bibinfo {author}
  {\bibfnamefont {T.}~\bibnamefont {Yefsah}}, \bibinfo {author} {\bibfnamefont
  {V.}~\bibnamefont {Khemani}}, \bibinfo {author} {\bibfnamefont {D.~A.}\
  \bibnamefont {Huse}}, \bibinfo {author} {\bibfnamefont {I.}~\bibnamefont
  {Bloch}},\ and\ \bibinfo {author} {\bibfnamefont {C.}~\bibnamefont {Gross}},\
  }\bibfield  {title} {\bibinfo {title} {Exploring the many-body localization
  transition in two dimensions},\ }\href
  {https://doi.org/10.1126/science.aaf8834} {\bibfield  {journal} {\bibinfo
  {journal} {Science}\ }\textbf {\bibinfo {volume} {352}},\ \bibinfo {pages}
  {1547} (\bibinfo {year} {2016})}\BibitemShut {NoStop}%
\bibitem [{\citenamefont {Kj{\"a}ll}\ \emph {et~al.}(2014)\citenamefont
  {Kj{\"a}ll}, \citenamefont {Bardarson},\ and\ \citenamefont
  {Pollmann}}]{kjall2014many}%
  \BibitemOpen
  \bibfield  {author} {\bibinfo {author} {\bibfnamefont {J.~A.}\ \bibnamefont
  {Kj{\"a}ll}}, \bibinfo {author} {\bibfnamefont {J.~H.}\ \bibnamefont
  {Bardarson}},\ and\ \bibinfo {author} {\bibfnamefont {F.}~\bibnamefont
  {Pollmann}},\ }\bibfield  {title} {\bibinfo {title} {{Many-body localization
  in a disordered quantum Ising chain}},\ }\href
  {https://doi.org/10.1103/PhysRevLett.113.107204} {\bibfield  {journal}
  {\bibinfo  {journal} {Phys. Rev. Lett.}\ }\textbf {\bibinfo {volume} {113}},\
  \bibinfo {pages} {107204} (\bibinfo {year} {2014})}\BibitemShut {NoStop}%
\bibitem [{\citenamefont {Yu}\ \emph {et~al.}(2016)\citenamefont {Yu},
  \citenamefont {Luitz},\ and\ \citenamefont {Clark}}]{Yu2016}%
  \BibitemOpen
  \bibfield  {author} {\bibinfo {author} {\bibfnamefont {X.}~\bibnamefont
  {Yu}}, \bibinfo {author} {\bibfnamefont {D.~J.}\ \bibnamefont {Luitz}},\ and\
  \bibinfo {author} {\bibfnamefont {B.~K.}\ \bibnamefont {Clark}},\ }\bibfield
  {title} {\bibinfo {title} {Bimodal entanglement entropy distribution in the
  many-body localization transition},\ }\href
  {https://doi.org/10.1103/PhysRevB.94.184202} {\bibfield  {journal} {\bibinfo
  {journal} {Phys. Rev. B}\ }\textbf {\bibinfo {volume} {94}},\ \bibinfo
  {pages} {184202} (\bibinfo {year} {2016})}\BibitemShut {NoStop}%
\end{thebibliography}%

\end{document}


\begin{center}
\textbf{Supplemental Material: Numerical evidence for many-body localization in two and three dimensions}
\end{center}

\section{Hard-core Bose-Hubbard model as a spin-1/2 model}

If the bosons in the Bose-Hubbard model of Eq.~(2) are hard-core bosons, then they can be represented with Pauli matrices. In particular, the creation and annihilation operators can be represented as $a_i^\dagger = \sigma_i^{+} = (\sigma_i^x + i \sigma_i^y)/2$ and $a_i = \sigma_i^{-} = (\sigma_i^x - i \sigma_i^y)/2$ and the number operator as $n_i = a^\dagger_i a_i = (I + \sigma^z_i)/2 = n_i^2$. Substituting these expressions into Eq.~(2), we obtain the spin-$1/2$ Hamiltonian
\begin{align}
H = -\frac{J}{2}\sum_{\langle ij \rangle} \left(\sigma^x_i \sigma^x_j + \sigma^y_i \sigma^y_j\right) + \frac{1}{2}\sum_i \delta_i \sigma^z_i \label{eq:Hxx}
\end{align}
plus a term proportional to the identity operator, which we ignore because it commutes with all operators. This is a two-dimensional $XX$ model with random magnetic fields. We use Eq.~(\ref{eq:Hxx}) in our hard-core Bose-Hubbard calculations.

\section{Algorithmic details}

Here we describe the details of the gradient descent calculation performed to optimize the objective function in~Eq.~(6). As described in Ref.~\onlinecite{Chertkov2020}, for an operator $O = \sum_a c_a \mathcal{O}_a$, the norm of the commutator with a Hamiltonian $H = \sum_a J_a \mathcal{O}_a$ can be expressed as the quadratic form
\begin{align}
Z_C \equiv \norm{[H, O]}^2 = \sum_{ab} c_a (C_H)_{ab} c_b \label{eq:comnorm}
\end{align}
where $(C_H)_{ab} = \tr{[H, \mathcal{O}_a]^\dagger [H, \mathcal{O}_b]}/\tr{I}$ is the commutant matrix, which is Hermitian and positive semi-definite. Finding the (normalized) operator $O$ that minimizes this quantity amounts to finding the smallest eigenvalue eigenvector of $C_H$. Since the Pauli strings $\mathcal{O}_a$ are orthonormal with respect to the Frobenius inner product, the commutant matrix can be written as $C_H = L_H^\dagger L_H$, where $(L_H)_{ca} = \sum_b J_b f_{ba}^c$ is the Liouvillian matrix and $f_{ab}^c$ are structure constants that describe how Pauli strings commute ($[\mathcal{O}_a, \mathcal{O}_b] = \sum_c f_{ab}^c \mathcal{O}_c$). For Pauli strings, the tensor $f_{ab}^c$ and matrix $L_H$ are sparse and efficient to compute numerically \cite{Chertkov2020}.

The ``binarity'' $\norm{O^2 - I}^2$ of the operator $O$ can be written as
\begin{align}
Z_B &\equiv \norm{O^2 - I}^2 = \norm{\frac{1}{2}\{O, O\} - I}^2 \nonumber \\
&= \norm{\frac{1}{2}\sum_{ab} c_a c_b \{\mathcal{O}_a, \mathcal{O}_b \} - I}^2 \nonumber \\
&= \norm{\sum_c\left(\frac{1}{2}\sum_{ab} c_a c_b \bar{f}_{ab}^c - \delta_{c,0}\right) \mathcal{O}_c}^2 \nonumber \\
&= 1 - \sum_{ab} c_a c_b \bar{f}_{ab}^0 + \frac{1}{4}\sum_{ab{a'}{b'}c} c_a c_b c_{a'} c_{b'} \bar{f}_{ab}^c \bar{f}_{{a'}{b'}}^c ,
\end{align}
where $\bar{f}_{ab}^c$ are real constants that describe how Pauli strings anti-commute ($\{\mathcal{O}_a, \mathcal{O}_b\} = \sum_c \bar{f}_{ab}^c \mathcal{O}_c$) and the $c=0$ index corresponds to the identity operator so that $\mathcal{O}_0 = I$. Note that $Z_B$ is \emph{quartic} in the $\{c_a\}$ parameters, making it difficult to minimize in general. 

To minimize the non-linear objective $\alpha Z_C + \beta Z_B$, we perform gradient descent using Newton's method, which requires calculation of gradients and Hessians. To avoid the numerical cost and stability issues of using finite-difference derivatives, we use exact expressions for these quantities. The gradient and Hessian of the commutator norm can be expressed as
\begin{align}
\partial_a Z_C \equiv \frac{\partial Z_C}{\partial c_a} = \sum_b \left(C_H + C_H^\dagger \right)_{ab} c_b \label{eq:gradZC}
\end{align}
and
\begin{align}
\partial_a \partial_b Z_C = \left(C_H + C_H^\dagger \right)_{ab}. \label{eq:hessZC}
\end{align}
The gradient and Hessian of the binarity can be expressed as
\begin{align}
\partial_a Z_B = \sum_{bc} c_b (\bar{L}_O)_{bc} (\bar{L}_O)_{ca} - 2 (\bar{L}_O)_{a,0} \label{eq:gradZB}
\end{align}
and
\begin{align}
\partial_a \partial_b Z_B = \sum_c \left[2(\bar{L}_O)_{ca} (\bar{L}_O)_{cb} + \sum_d (\bar{L}_O)_{cd} c_d \bar{f}_{ab}^c \right] - 2 f_{ab}^0, \label{eq:hessZB}
\end{align}
where $(\bar{L}_O)_{ca} \equiv \sum_b c_b \bar{f}_{ba}^c$ is itself a function of the $\{c_a\}$ parameters. We checked Eqs.~(\ref{eq:gradZC})--(\ref{eq:hessZB}) numerically against finite-difference derivatives and Hessians and found that they agreed.

Our open-source Python code for performing this optimization is available online \cite{bioms}. To compute commutators and anti-commutators between Pauli strings, we used the quantum operators from symmetry (QOSY) python package \cite{qosy,Chertkov2020}. The gradient descent was performed using Newton's method with conjugate-gradient iteration as implemented in the \texttt{scipy.optimize.minimize} function in the scipy library \cite{scipy2020}, with the desired relative error required for convergence set to \texttt{xtol}$=10^{-6}$. We used python version 3.5.5, scipy version 1.0.0, and numpy version 1.14.2.

During the calculations, we stored the commutators and anti-commutators between Pauli strings, i.e., the constants $f_{ab}^c$ and $\bar{f}_{ab}^c$, into python dictionaries (hash tables). In our calculations, memory was a main bottleneck, which prevented us from working with larger basis sizes $|B|$. To gather reliable statistics, we performed many independent optimizations on different random realizations of the Hamiltonians in Eqs.~(1)~and~(2) by running the optimization in parallel over many processes and nodes of the Blue Waters supercomputer at the National Center for Supercomputing Applications at the University of Illinois at Urbana-Champaign. A typical slow single disorder realization for a low-disorder, three-dimensional calculation (for all 11 basis expansions) takes approximately 1 core-hour and about 7 gigabytes of memory on a Blue Waters XE node. Running the full suite of realizations/models in this work takes approximately 10,000 node-hours on Blue Waters. 

\section{Future extensions}

In future work, it would be useful to consider alternative more memory-efficient numerical minimization procedures that would allow us to go to larger basis sizes and/or optimize more disordered realizations in parallel. It would also be interesting to develop other basis initializations or basis expansion procedures that are more efficient or lead to better minimization of the objective Eq.~(6). One could also explore whether basis reduction techniques for pruning the basis of Pauli strings could improve performance and memory usage. It would also be interesting to use this method to find many mutually commuting binary integrals of motion at once, which can be done by modifying the objective in Eq.~(6) to include multiple $\tau^z_i$ operators, but we generally expect this to be computationally expensive. 

Another interesting extension would be to modify the second term in objective function in Eq.~(6) from a binarity to a \emph{unitarity}, i.e., $\norm{U^\dagger U - I}^2$. This would allow the algorithm to find unitaries that commute with the given Hamiltonian.

\section{Statistical properties of the approximate integrals of motion}

\subsection{Quantities considered}

In this work, we analyzed a variety of quantities to understand the properties of the approximate $\ell$-bits that our algorithm produced as output.
In summary, given an approximate $\ell$-bit, we consider its commutator norm, binarity, normalization, overlap with a single-site $\tau^z_i$ operator, range ($r$), locality ($k$), correlation length ($\xi$), and the $IPR$.

The commutator norm $\norm{[H, \tau^z_i]}^2$ and binarity $\norm{(\tau^z_i)^2 - I}^2$ make up the objective function in Eq.~(6) that we minimize with our algorithm. Note that the minimization of Eq.~(6) is performed without any constraint on the normalization of $\tau^z_i$; a perfectly binary operator is normalized, and so minimizing the binarity takes care of keeping operators nearly normalized. Therefore, it is interesting to consider $\norm{\tau^z_i}$ as a metric for the performance of the algorithm. A strong deviation from $\norm{\tau^z_i}\approx 1$ potentially indicates a failure of our algorithm. Note that for consistency the commutator norms, binarities, and other quantities depicted in all of the figures in the main paper and the supplemental material use \emph{normalized} $\tau^z_i$ operators. 

An important quantity that we discuss in the main paper is the overlap of the $\tau^z_i$ operator with $\sigma^z_i$. As shown in Fig.~3 in the main text, the probability distribution of $|\langle \tau^z_i, \sigma^z_i \rangle |^2$ changes significantly as one decreases disorder strength. In particular, the distribution rapidly changes near the ergodic-MBL transition for the 1D Heisenberg model and behaves similarly for the other models considered.

Another quantity related to the spatial localization of a $\tau^z_i$ operator is the ``range''
\begin{align}
r = \frac{1}{\sum_b |c_b|^2} \left(\sum_a |c_a|^2 \max_{\vec{r},\vec{r}' \in R_a}\norm{\vec{r} - \vec{r}'} \right), \label{eq:avgr}
\end{align}
where $R_a$ is the set of lattice coordinates that Pauli string $\mathcal{O}_a$ acts on. The range is a weighted average over the spatial extents of the operator's Pauli strings and is similar to other definitions for the range of an $\ell$-bit used in studies of 1D MBL systems \cite{Huse2014,Villalonga2018}. A similar quantity that we consider is the ``locality'' $k$ of a $\tau^z_i$ operator, which is defined as
\begin{align}
k = \frac{\sum_a |c_a|^2 k_a}{\sum_b |c_b|^2}, \label{eq:avgk}
\end{align}
where $k_a = |R_a|$ is the number of sites a Pauli string $\mathcal{O}_a$ acts on (i.e., $\mathcal{O}_a$ is $k_a$-local), regardless of their position in the lattice. The locality measures the average contribution of non-local Pauli strings to the operator.

In Eq.~(5) in the main text, we introduced the weights $w_\vec{r}$ of a $\tau^z_i$ operator, which can be interpreted as the spatial probability distribution of $\tau^z_i$ over the lattice positions $\vec{r}$. We fit the weights $w_\vec{r}$ of each $\tau^z_i$ operator to an exponential decay with a correlation length $\xi$. In addition, we also use the weights $w_\vec{r}$ to define the operator inverse participation ratio
\begin{align}
IPR = \frac{1}{\sum_{\vec{r}} w_{\vec{r}}^2}, \label{eq:ipr}
\end{align}
a quantity analogous to inverse participation ratios considered in other MBL studies \cite{Canovi2011,Serbyn2013,Lin2017}. For a $\tau^z_i$ operator localized on site $i$, $w_{\vec{r}} \approx \delta_{\vec{r},\vec{r}_i}$ and so $IPR \approx 1$. For a $\tau^z_i$ operator delocalized evenly over $N$ sites, $w_{\vec{r}}$ is approximately $1/N$ on those sites so that $IPR \approx 1/(N \cdot 1/N^2) = N$.

\subsection{Distributions}

\begin{figure}[H]
\begin{center}
\begin{tabular}{cc}
\includegraphics[height=0.4\textheight]{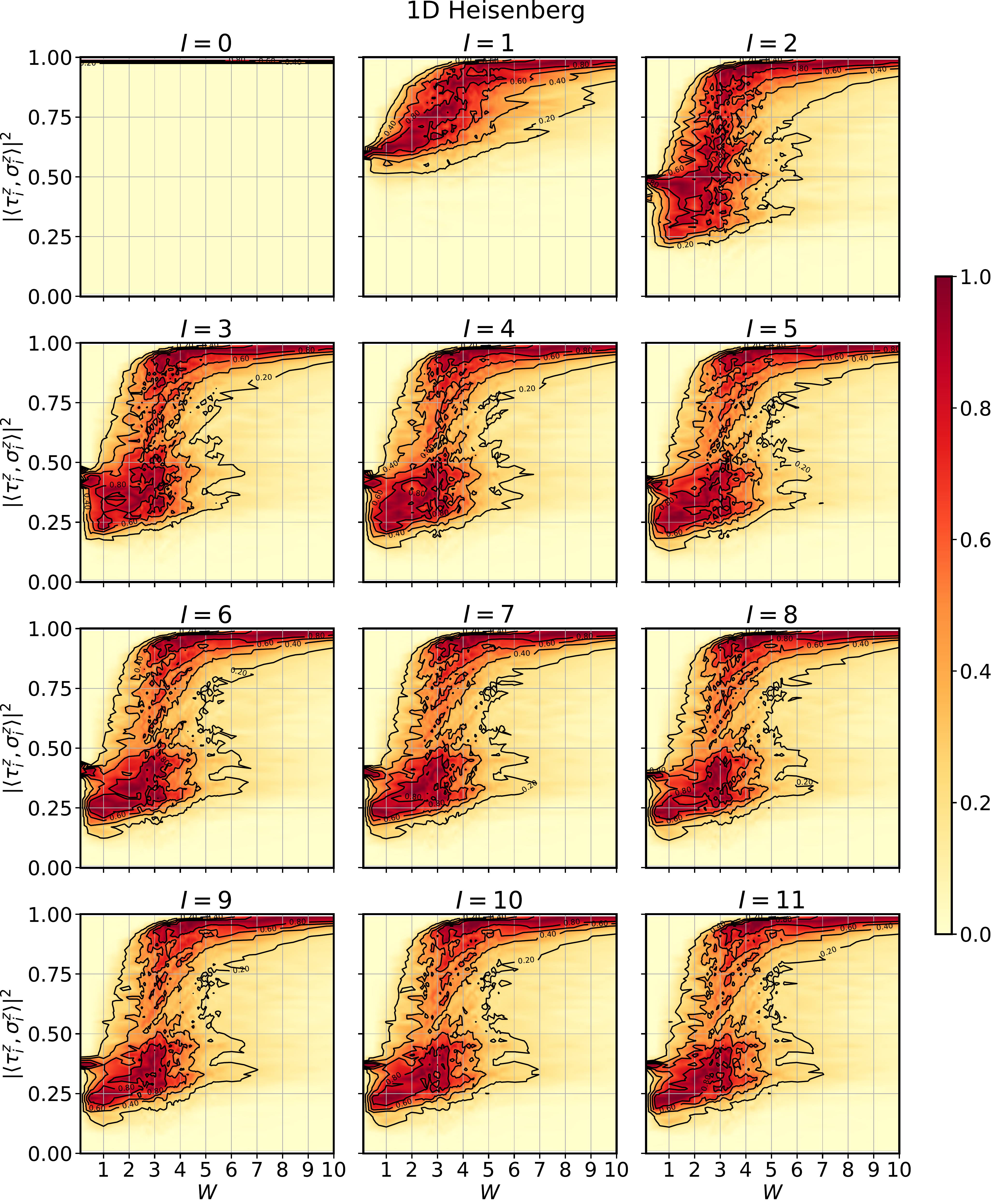} & \includegraphics[height=0.4\textheight]{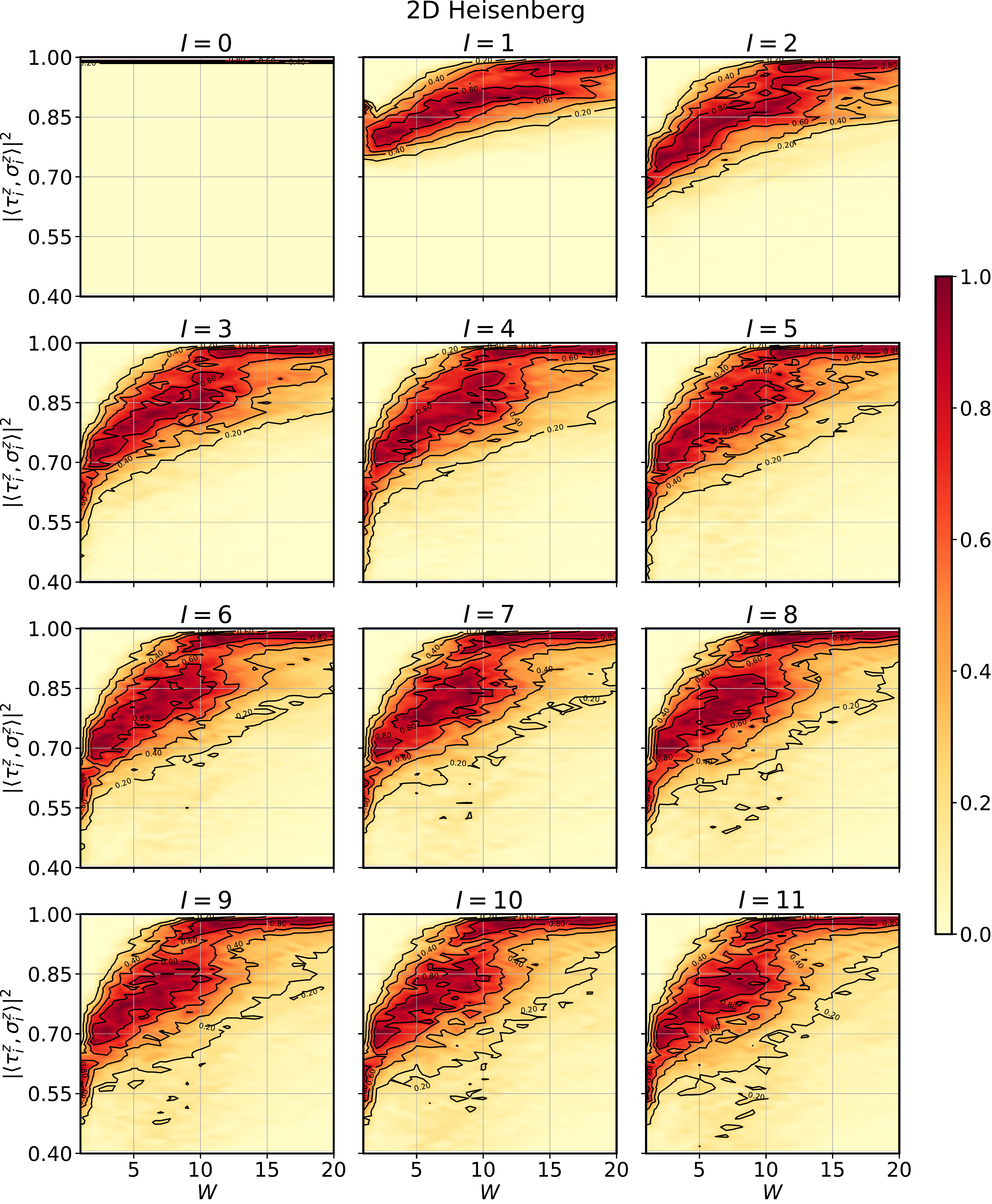} \\
\includegraphics[height=0.4\textheight]{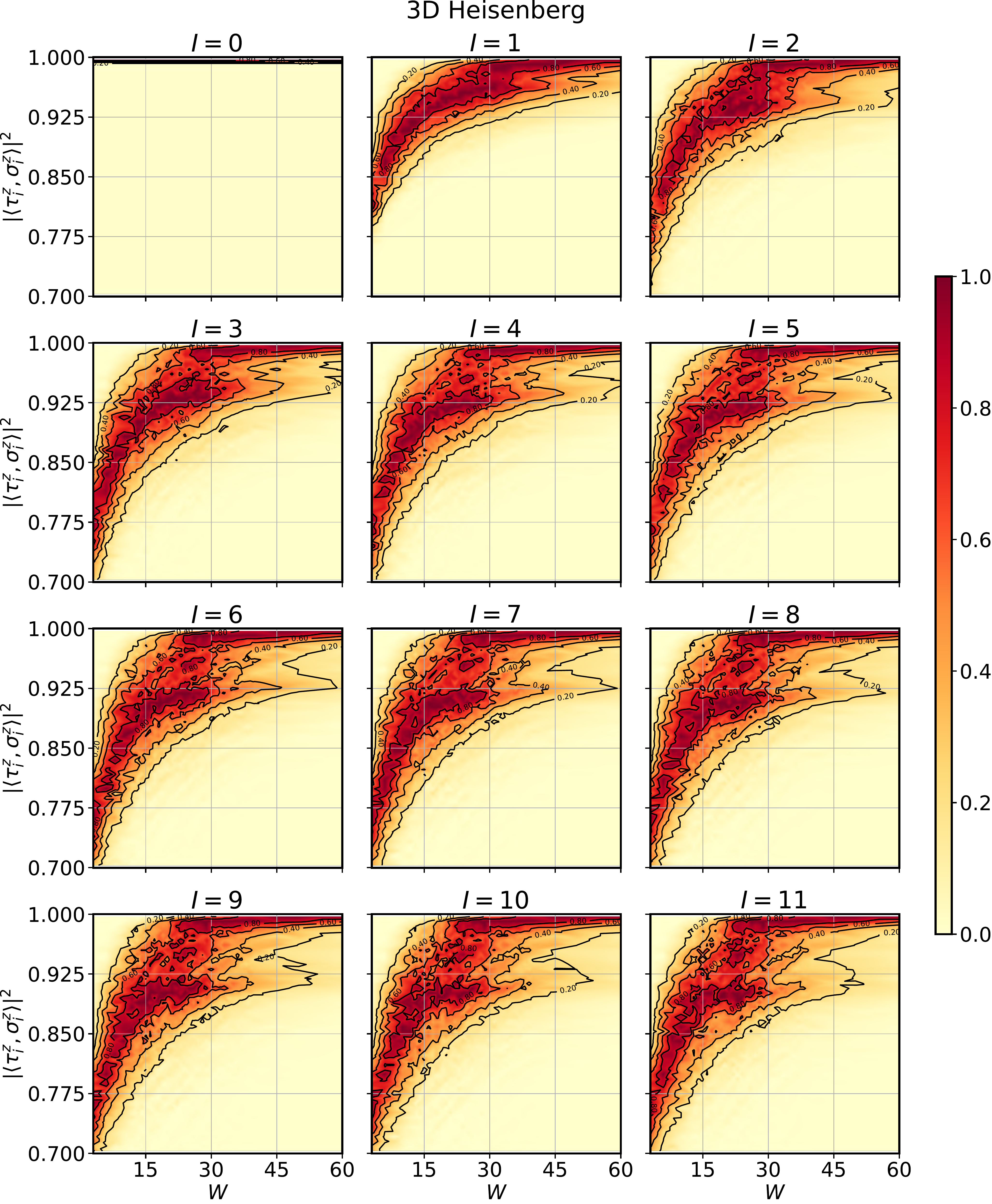} & \includegraphics[height=0.4\textheight]{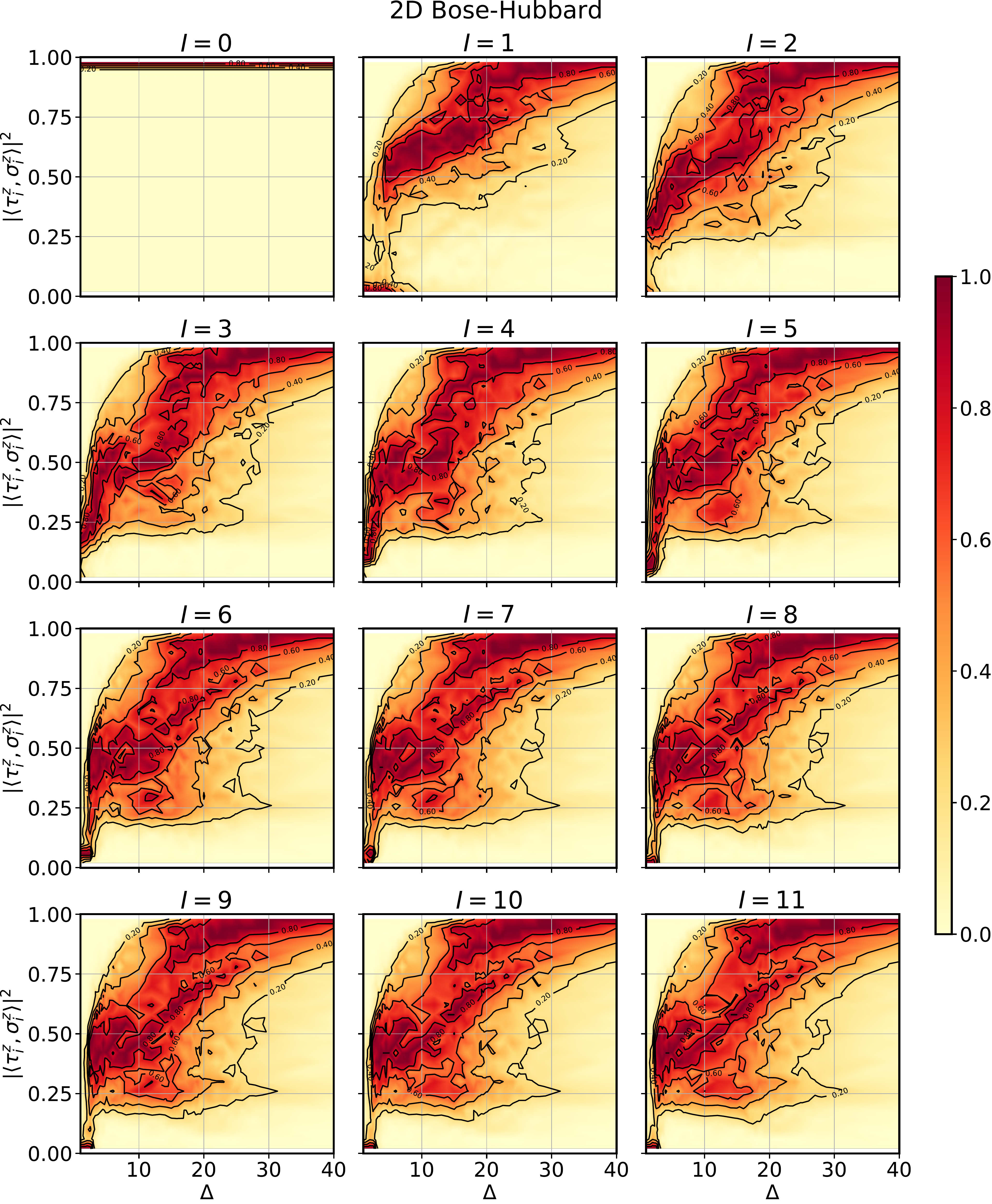}
\end{tabular}
\end{center}
\caption{Interpolated histograms of $|\langle \tau^z_i, \sigma^z_i \rangle|^2$ overlaps at different disorder strengths for the four models studied at different iterations $I$ of the basis expansion.} \label{fig:p_zoverlap} 
\end{figure}

\begin{figure}[H]
\begin{center}
\begin{tabular}{cc}
\includegraphics[height=0.4\textheight]{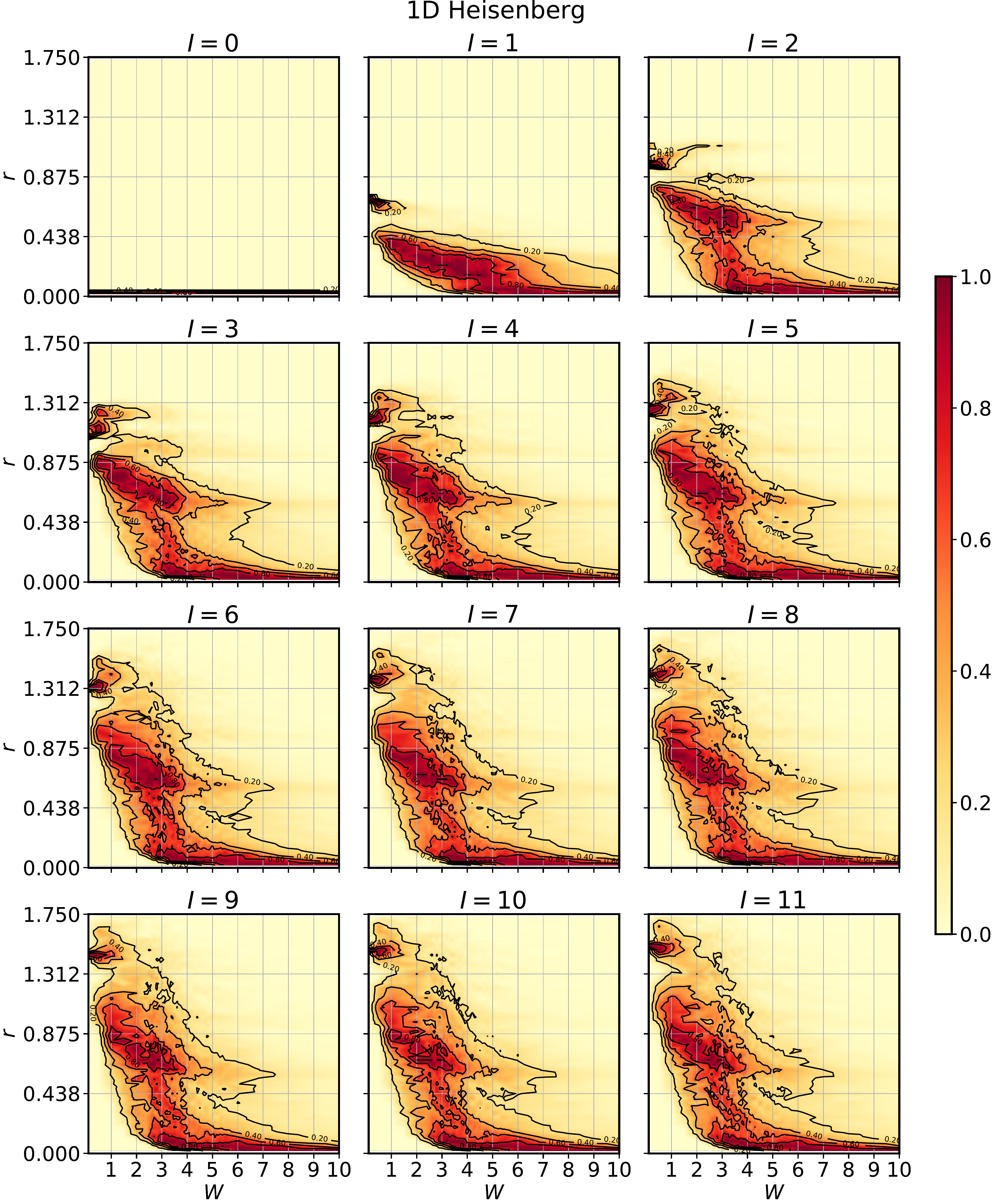} & \includegraphics[height=0.4\textheight]{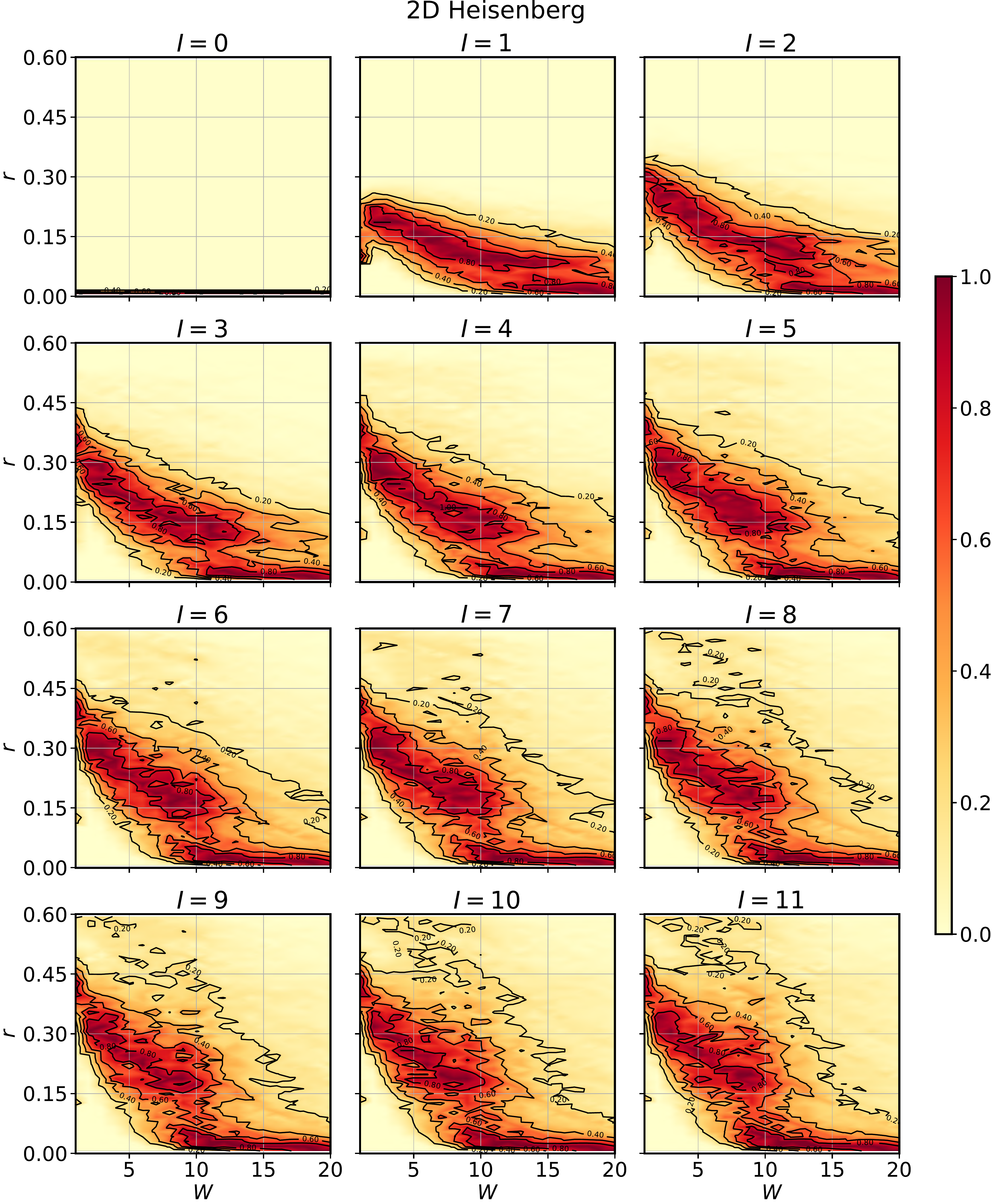} \\
\includegraphics[height=0.4\textheight]{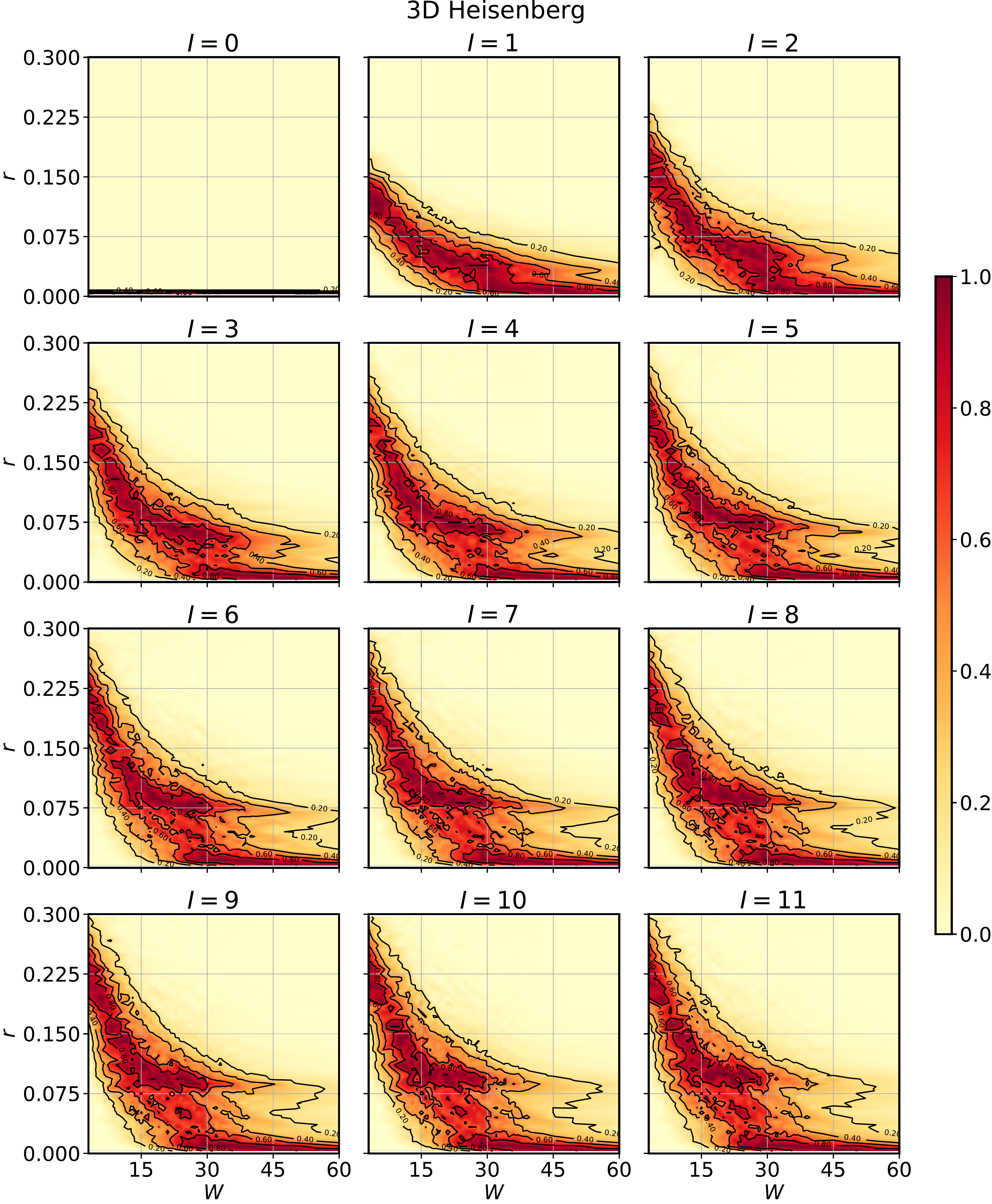} & \includegraphics[height=0.4\textheight]{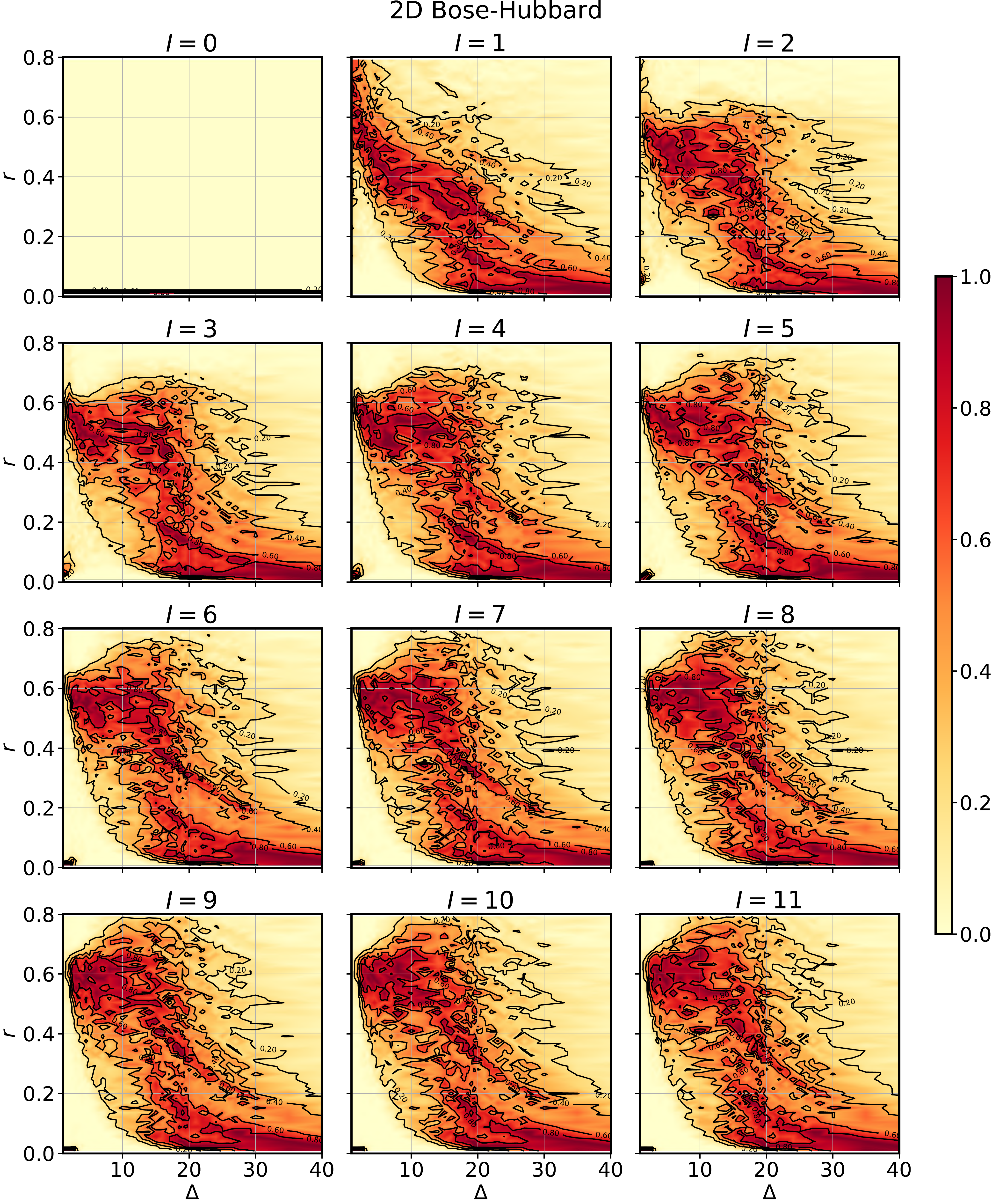}
\end{tabular}
\end{center}
\caption{Interpolated histograms of the range $r$ at different disorder strengths for the four models studied at different iterations $I$ of the basis expansion.} \label{fig:p_range} 
\end{figure}

\begin{figure}[H]
\begin{center}
\begin{tabular}{cc}
\includegraphics[height=0.4\textheight]{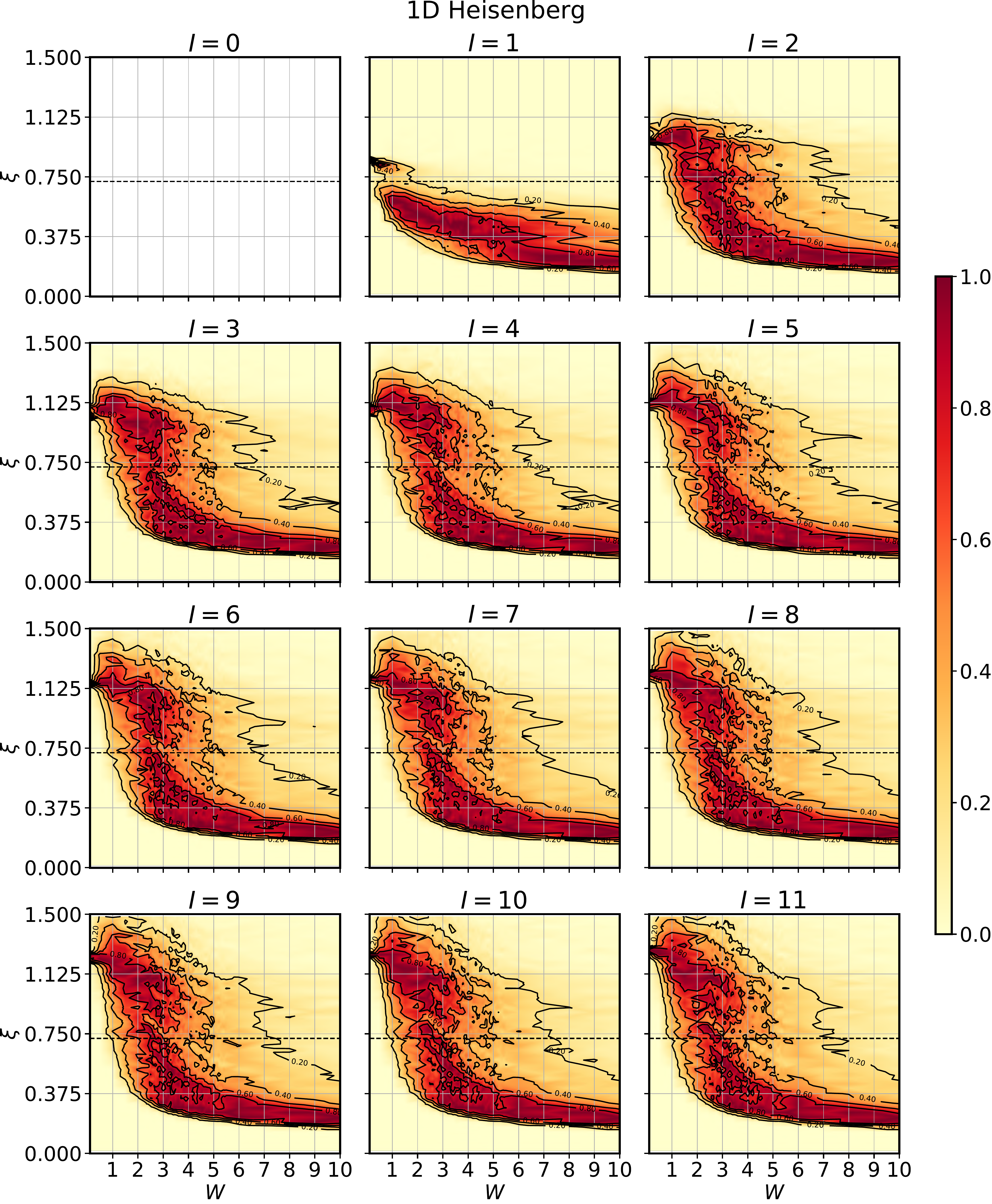} & \includegraphics[height=0.4\textheight]{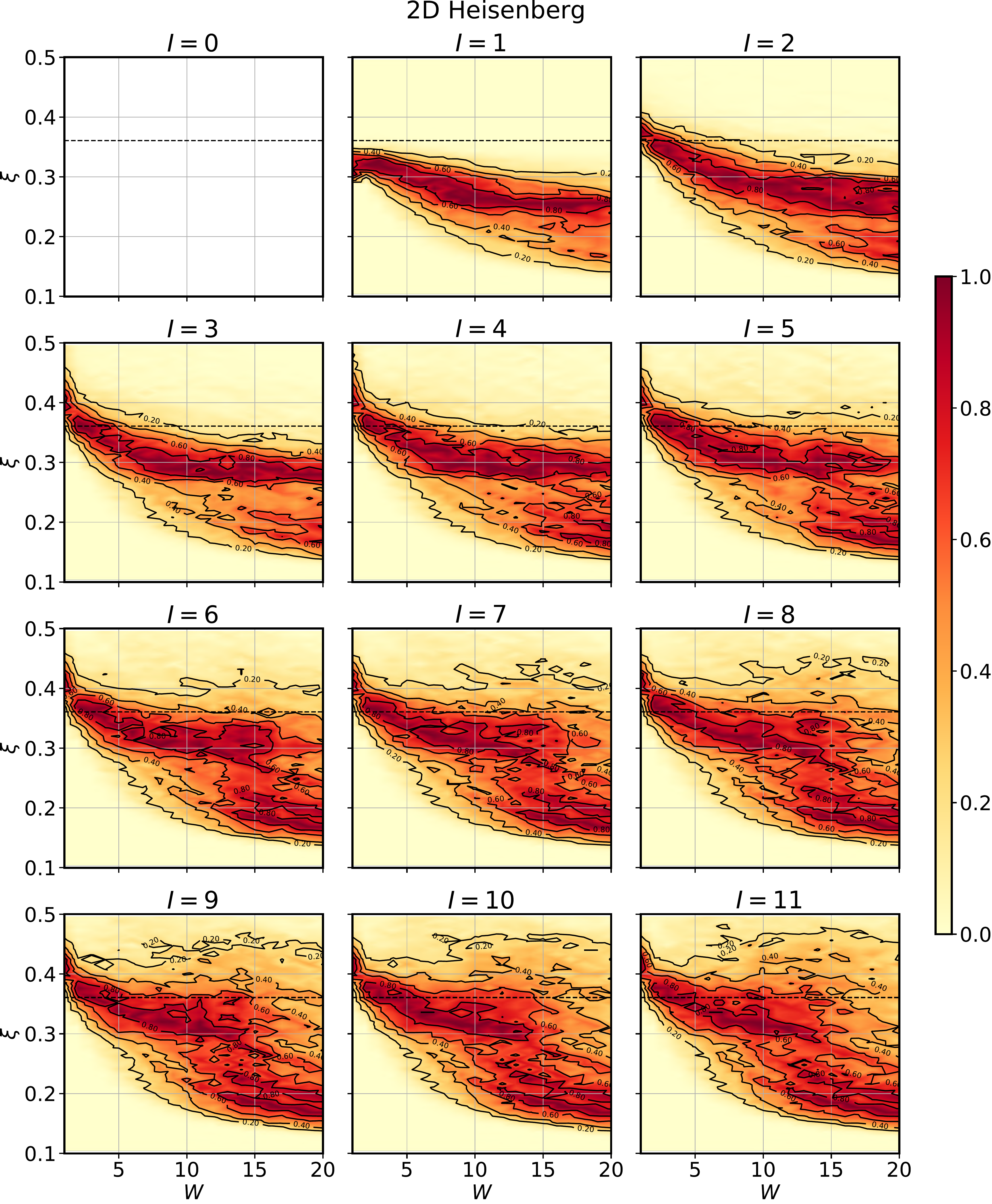} \\
\includegraphics[height=0.4\textheight]{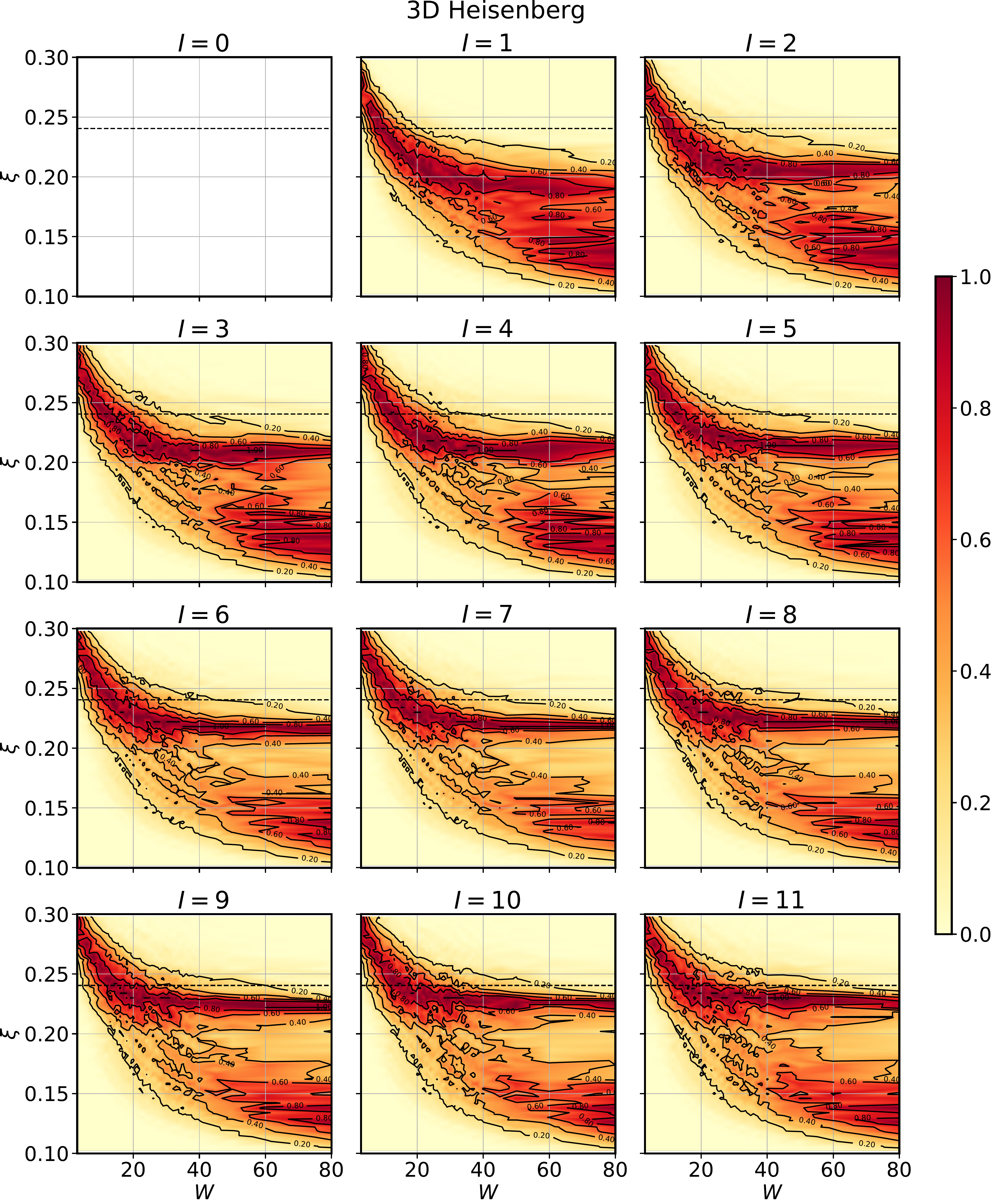} & \includegraphics[height=0.4\textheight]{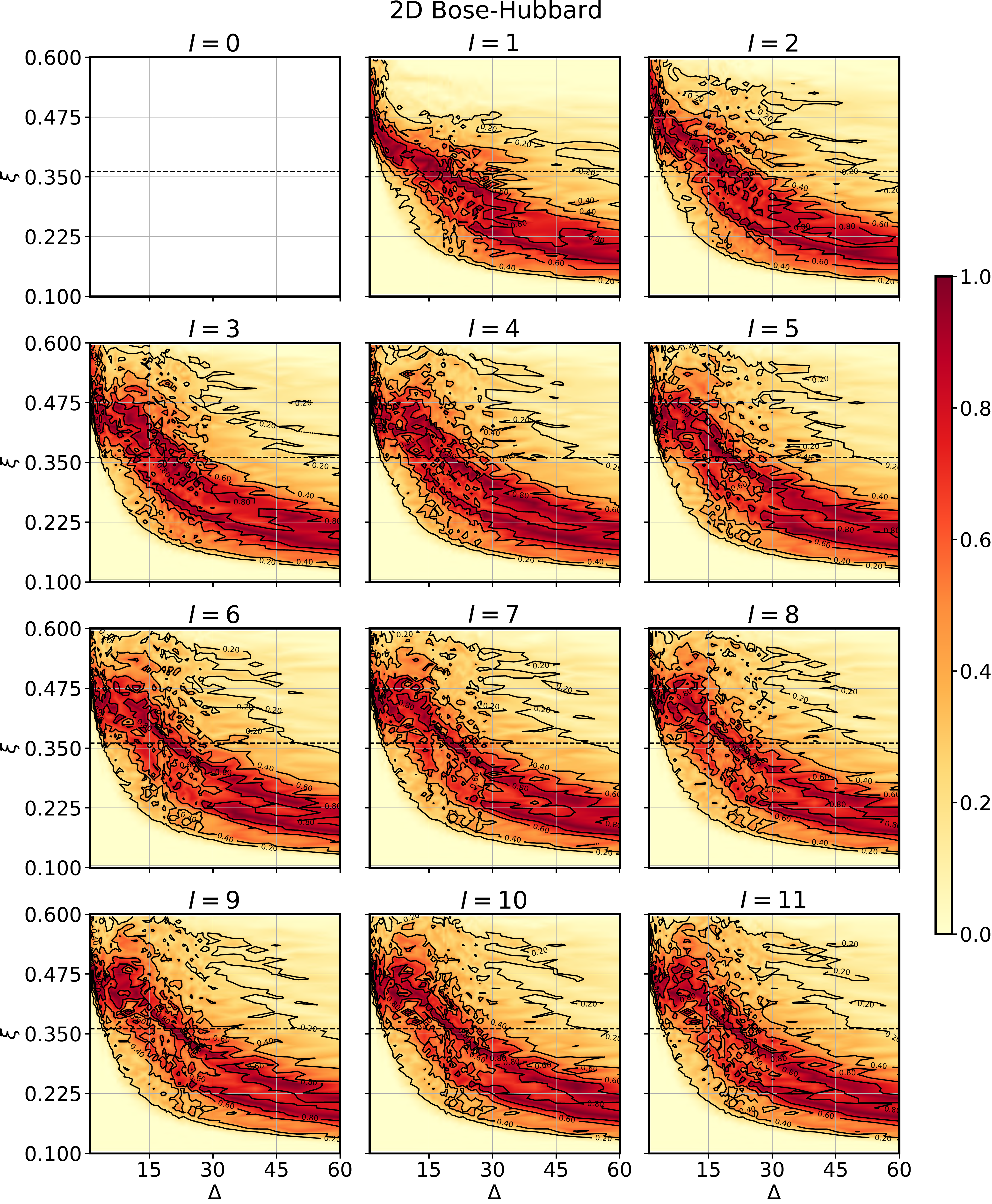}
\end{tabular}
\end{center}
\caption{Interpolated histograms of the correlation length $\xi$ at different disorder strengths for the four models studied at different iterations $I$ of the basis expansion. For reference, $1/\ln(4^d)$,  where $d$ is the spatial dimension, is marked with a horizontal dashed line.} \label{fig:p_corr_length} 
\end{figure}

Figs.~\ref{fig:p_zoverlap}--\ref{fig:p_corr_length} show interpolated histograms of $|\langle \tau^z_i, \sigma^z_i \rangle|^2$, $r$, and $\xi$ versus disorder strength for the four models studied and for different iterations of the algorithm. The histograms are normalized so that at a fixed disorder strength the maximum of the histogram is one.

\subsection{Averages, medians, and fluctuations}

\begin{figure}[H]
\begin{center}
\includegraphics[height=0.4\textheight]{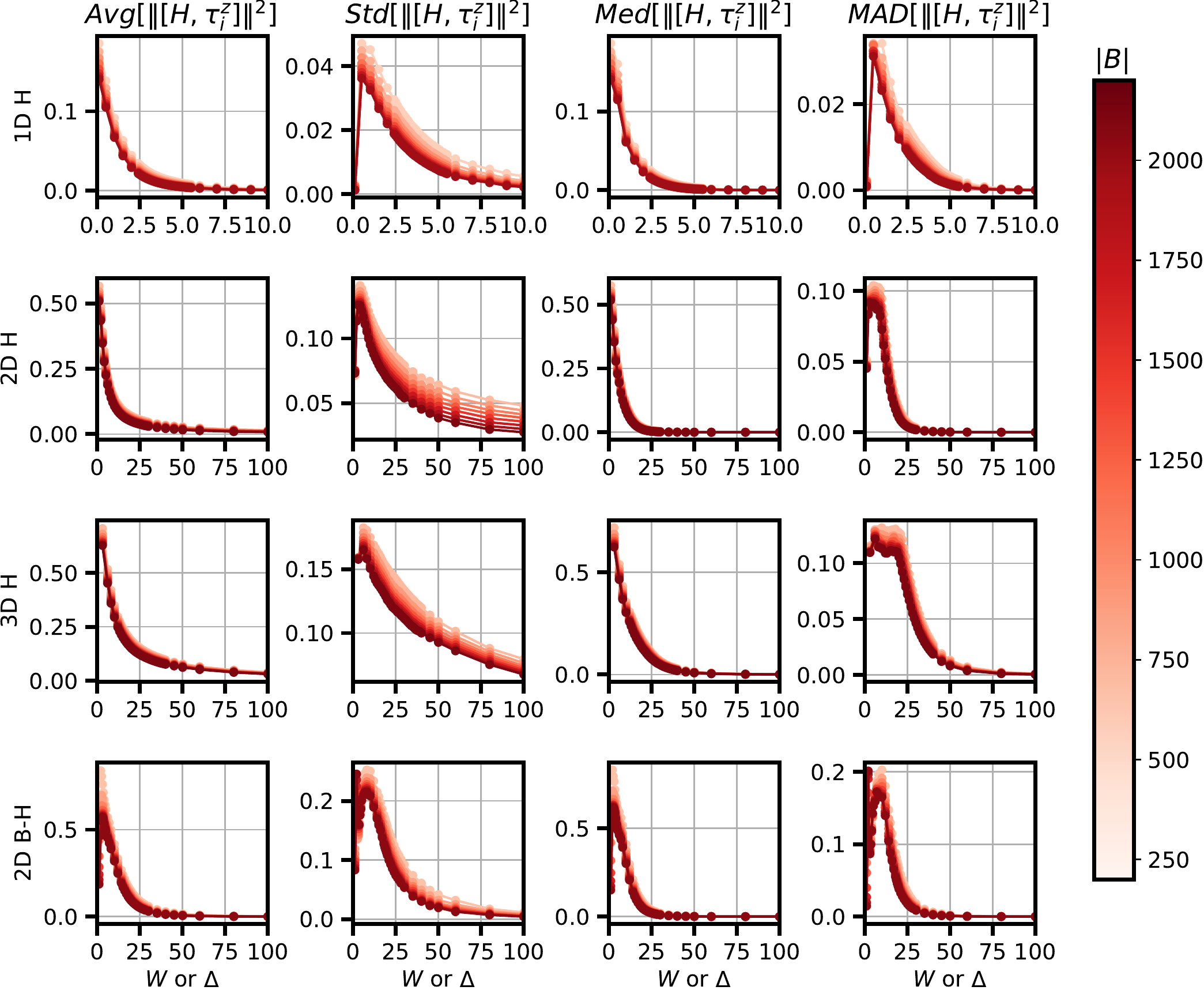}
\end{center}
\caption{Statistics of commutator norms for different models and basis sizes.} \label{fig:comnorm} 
\end{figure}

\begin{figure}[H]
\begin{center}
\includegraphics[height=0.4\textheight]{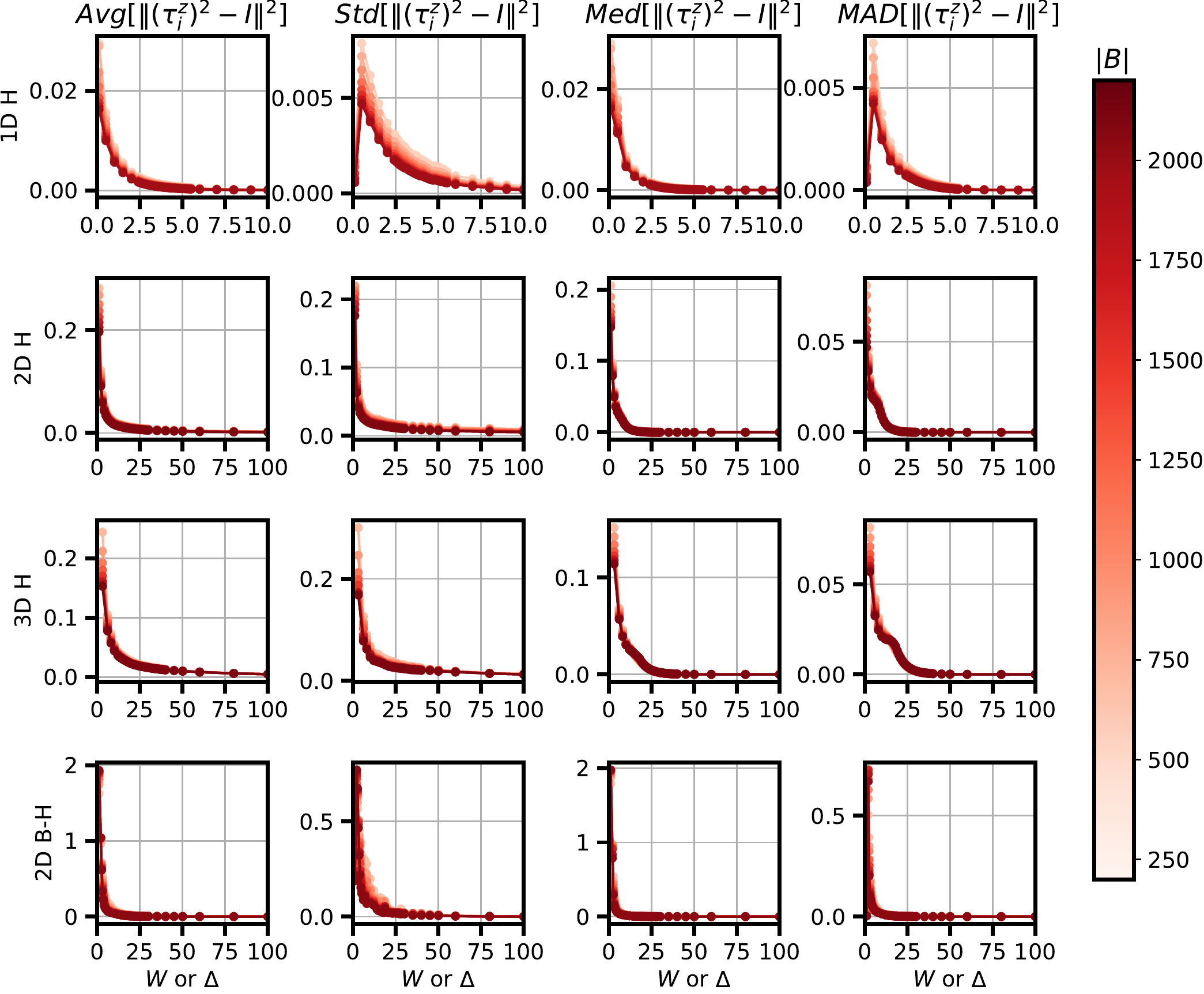}
\end{center}
\caption{Statistics of binarities for different models and basis sizes.} \label{fig:binarity}
\end{figure}

\begin{figure}[H]
\begin{center}
\includegraphics[height=0.4\textheight]{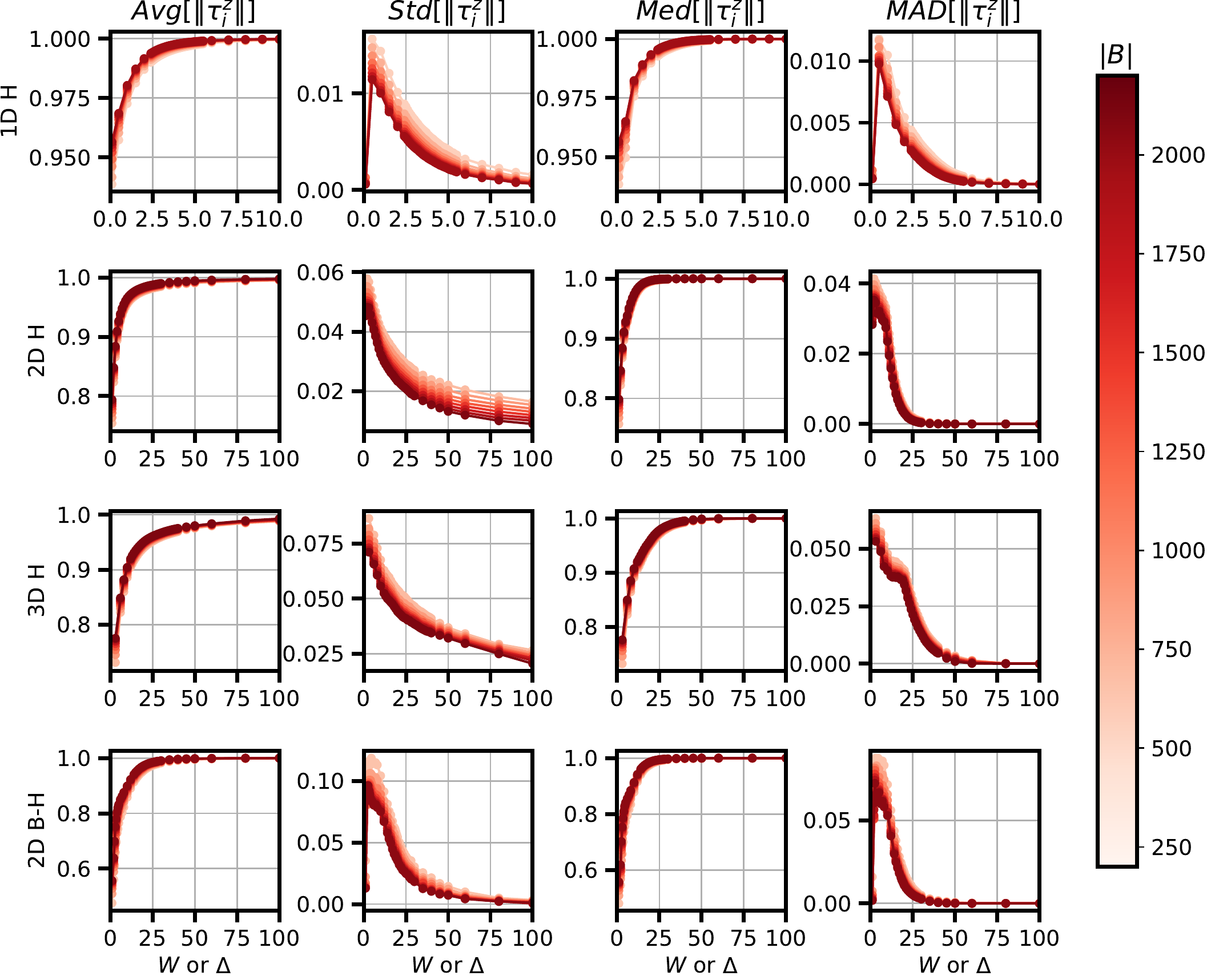}
\end{center}
\caption{Statistics of operator norms for different models and basis sizes.} \label{fig:opnorm}
\end{figure}

\begin{figure}[H]
\begin{center}
\includegraphics[height=0.4\textheight]{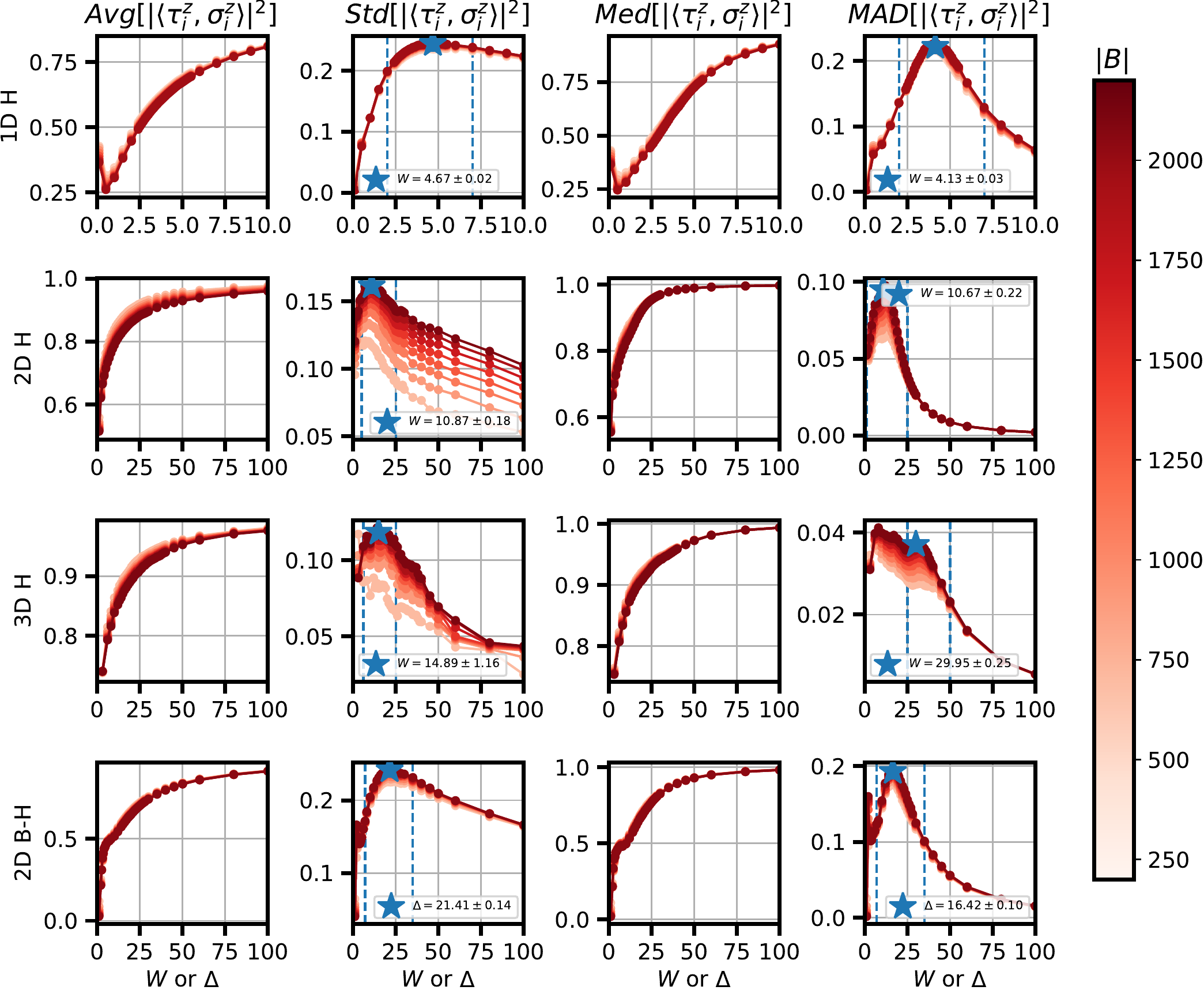}
\end{center}
\caption{Statistics of $|\langle \tau^z_i, \sigma^z_i \rangle|^2$ for different models and basis sizes.} \label{fig:zoverlap}
\end{figure}

\begin{figure}[H]
\begin{center}
\includegraphics[height=0.4\textheight]{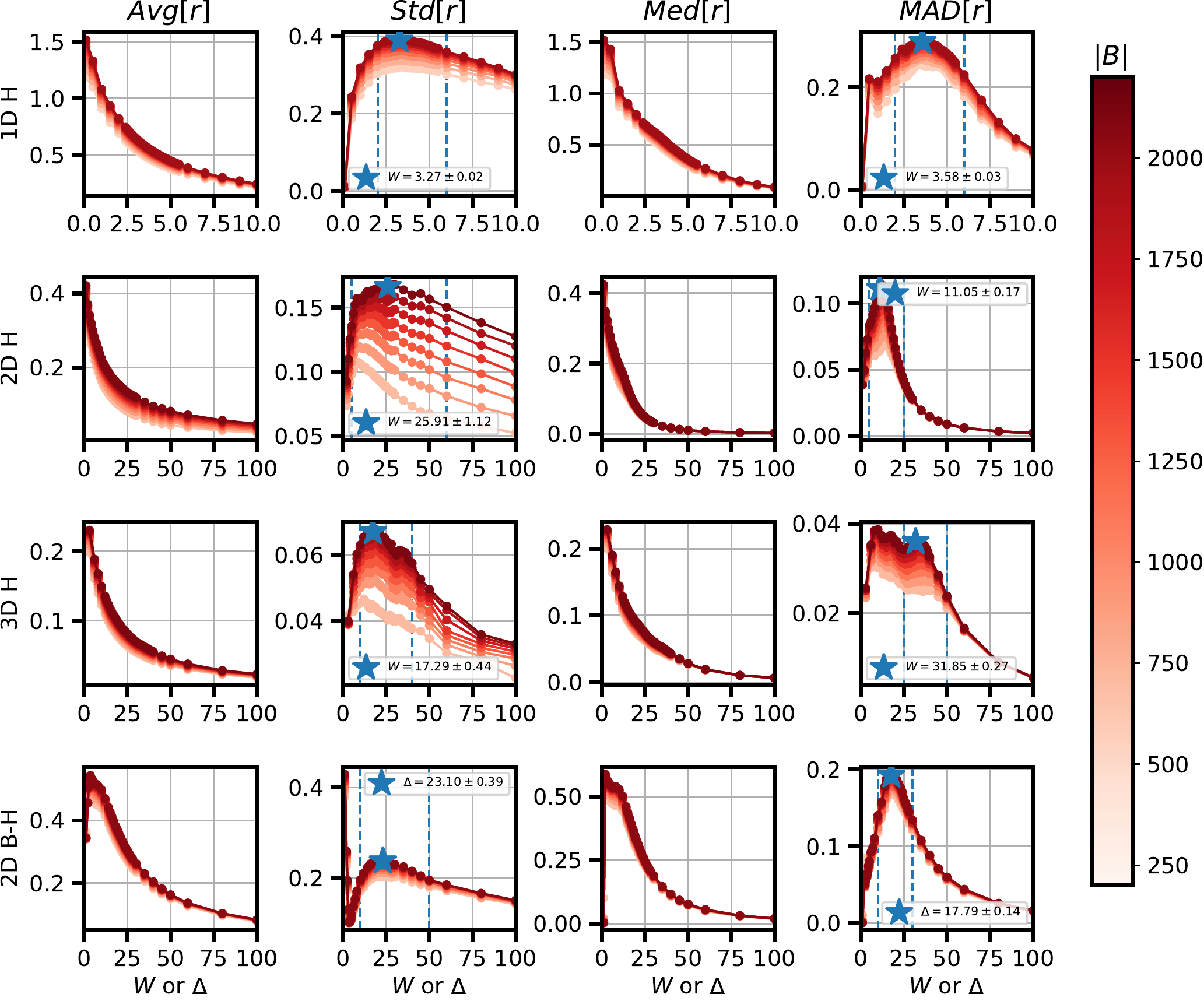}
\end{center}
\caption{Statistics of ranges for different models and basis sizes.} \label{fig:range}
\end{figure}

\begin{figure}[H]
\begin{center}
\includegraphics[height=0.4\textheight]{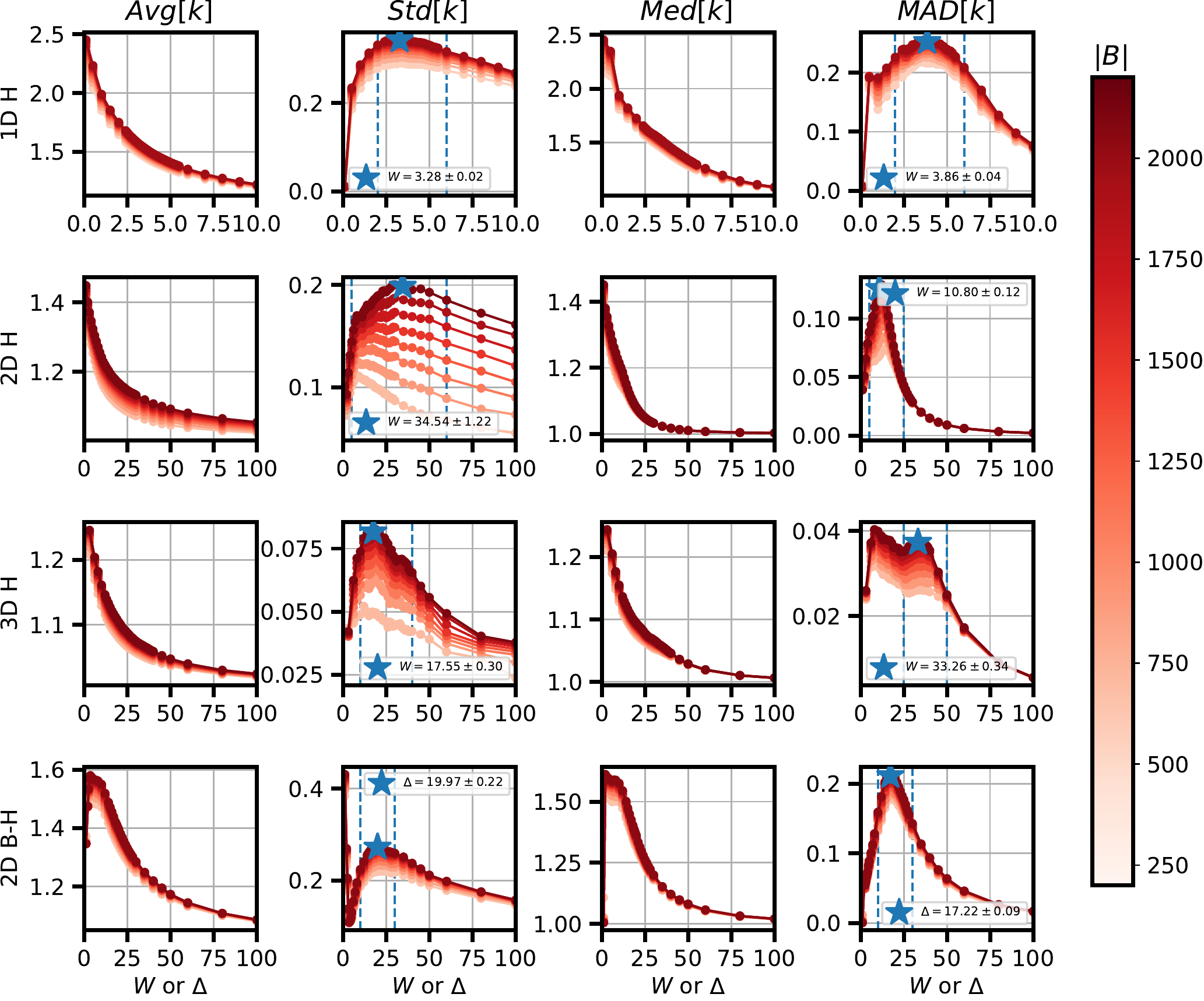}
\end{center}
\caption{Statistics of localities for different models and basis sizes.} \label{fig:locality}
\end{figure}

\begin{figure}[H]
\begin{center}
\includegraphics[height=0.4\textheight]{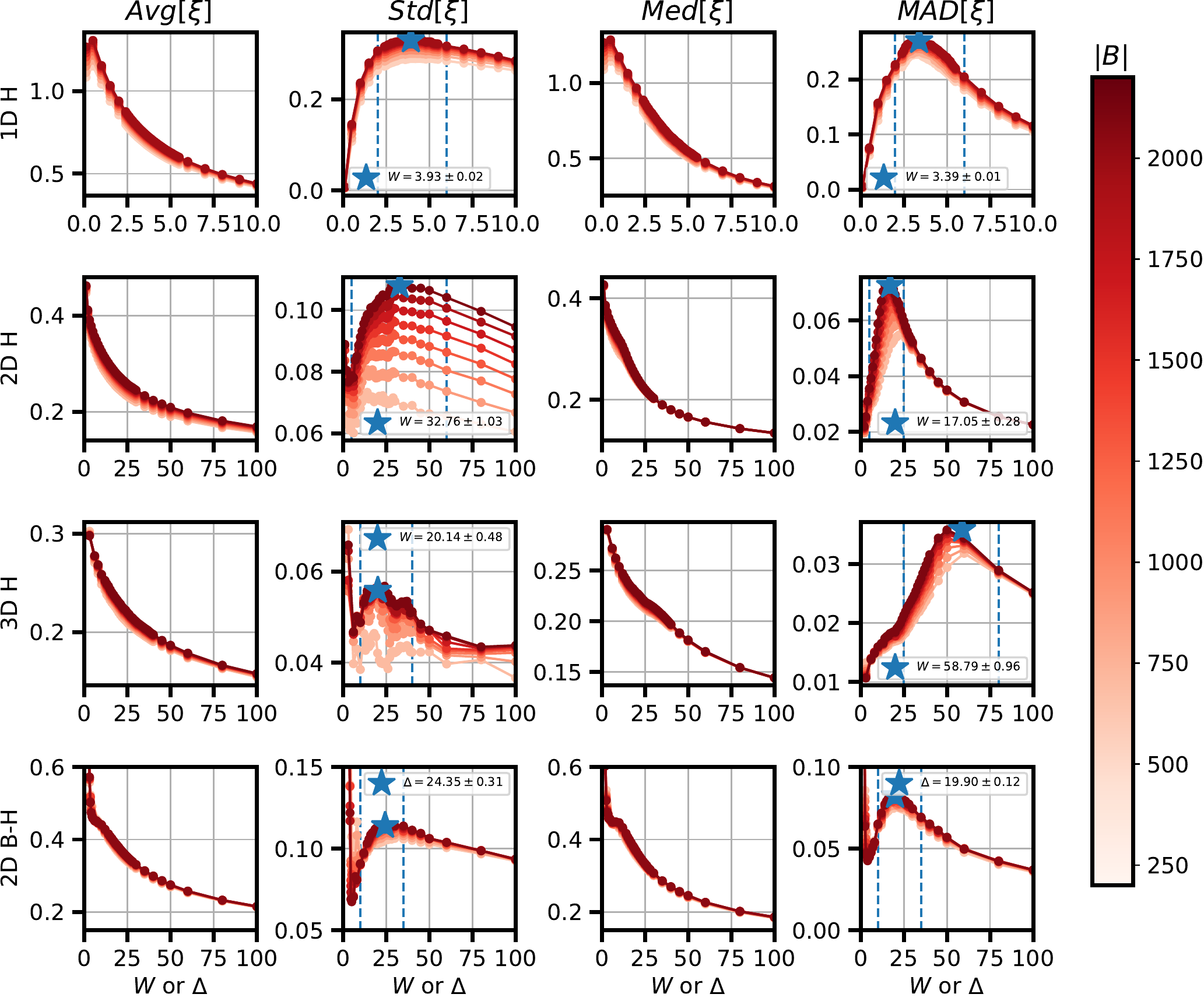}
\end{center}
\caption{Statistics of correlation lengths for different models and basis sizes.} \label{fig:corrlength}
\end{figure}

\begin{figure}[H]
\begin{center}
\includegraphics[height=0.4\textheight]{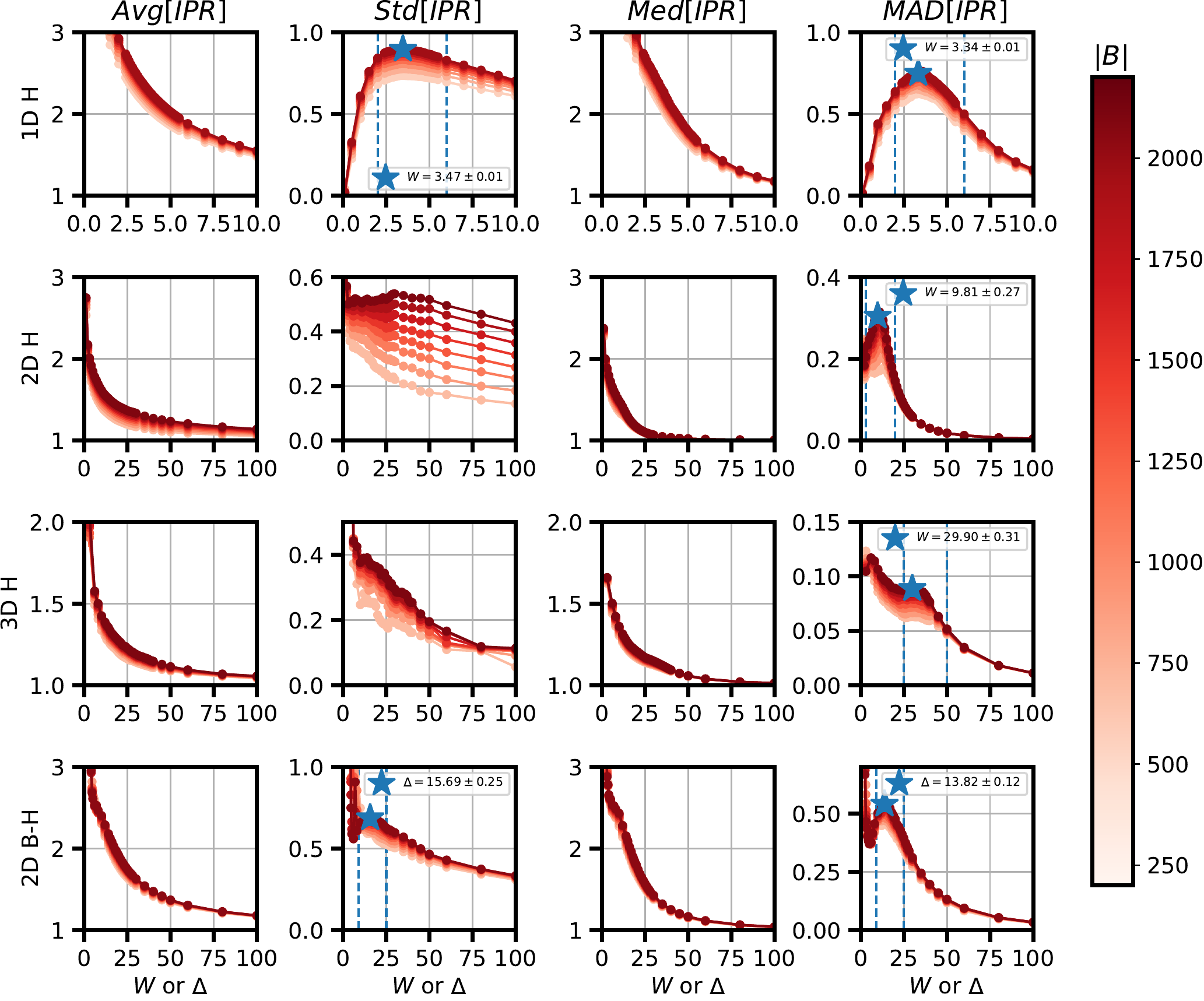}
\end{center}
\caption{Statistics of operator inverse participation ratios for different models and basis sizes.} \label{fig:opipr}
\end{figure}

\begin{figure}[H]
\begin{center}
\includegraphics[height=0.38\textheight]{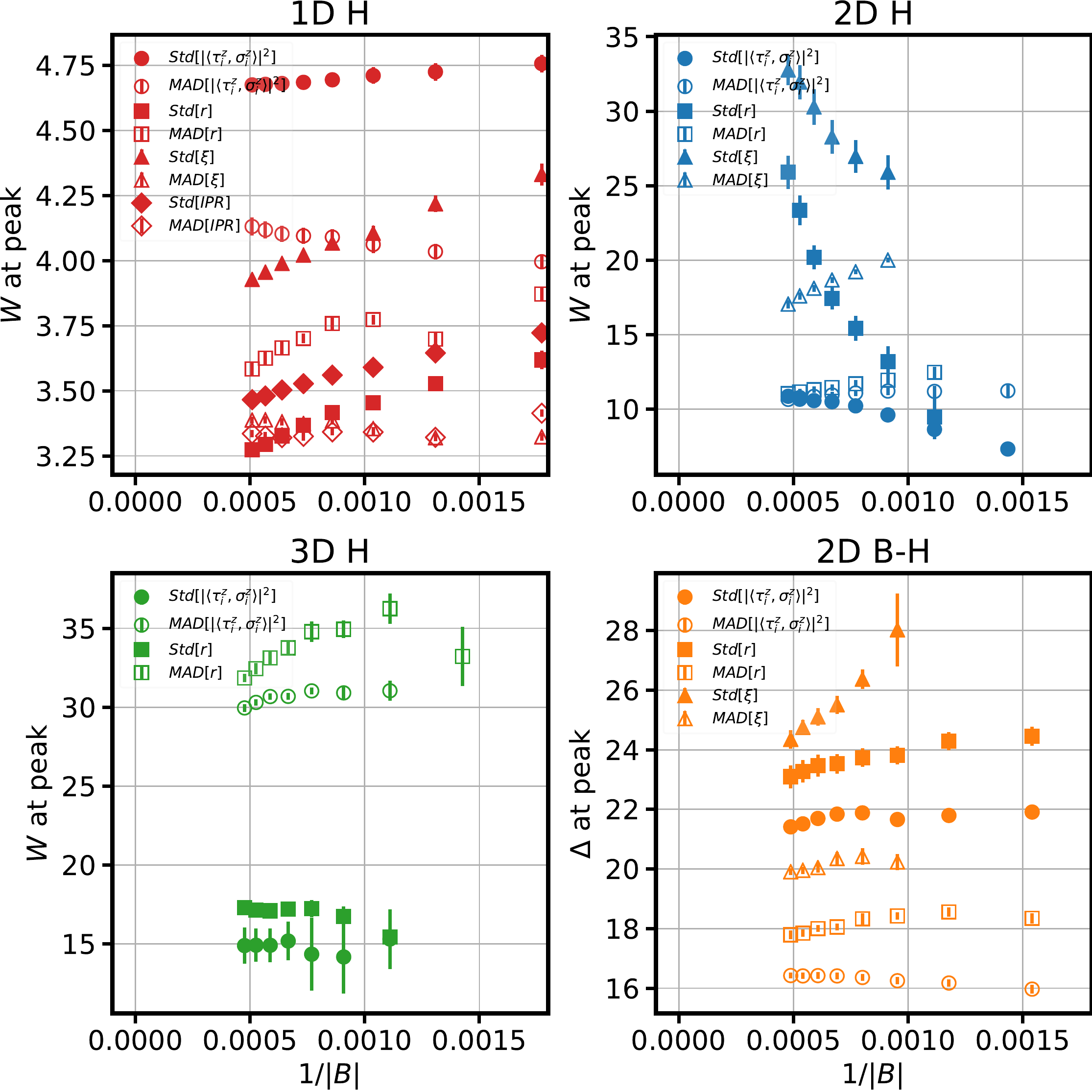}
\end{center}
\caption{The disorder strengths at which the standard deviations or MADs of various quantities are maximized versus $1/|B|$ for the four models studied.} \label{fig:criticalW}
\end{figure}

Figs.~\ref{fig:comnorm}--\ref{fig:opipr} show the averages, standard deviations, medians, and median absolute deviations (MAD, defined as the median distance from the median: $Med[|x-Med[x]|]$) of the previously discussed quantities versus disorder strength for the optimized $\tau^z_i$ operators that our algorithm produces. The medians and MADs are included for comparison since they are more robust to outliers than averages and standard deviations. Different colored lines represent calculations done in different basis sizes, i.e., at different steps in our expansion procedure. For visual guidance, we mark the approximate locations of local maxima in the plots with blue stars, which were obtained by fitting third-order polynomials to data in the windows marked by dashed blue lines. The locations of these maxima versus basis size are shown in Fig.~\ref{fig:criticalW}.

\subsection{Correlations between quantities}

\begin{figure}[H]
\begin{center}
\begin{tabular}{cc}
\includegraphics[width=0.4\textwidth]{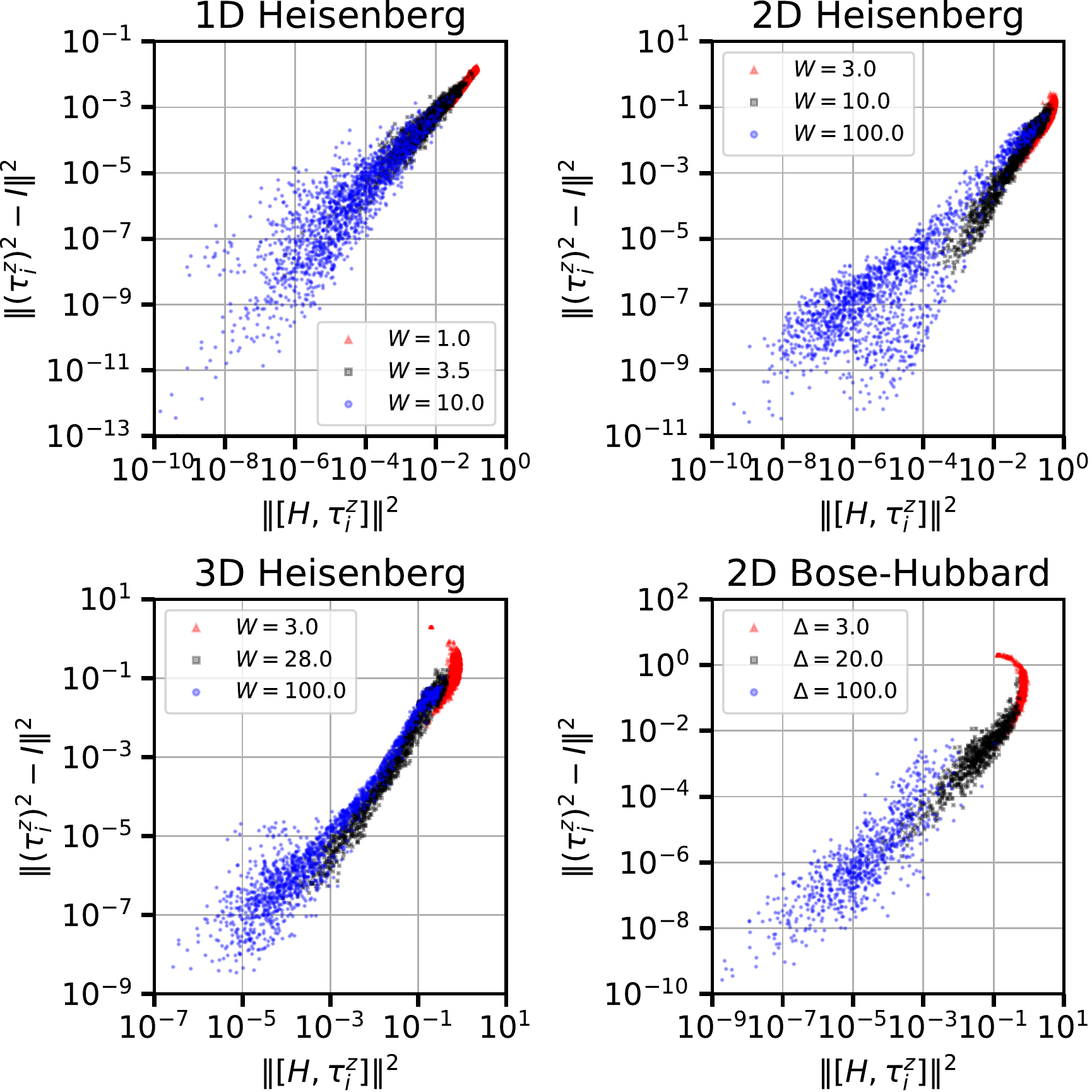} & \includegraphics[width=0.4\textwidth]{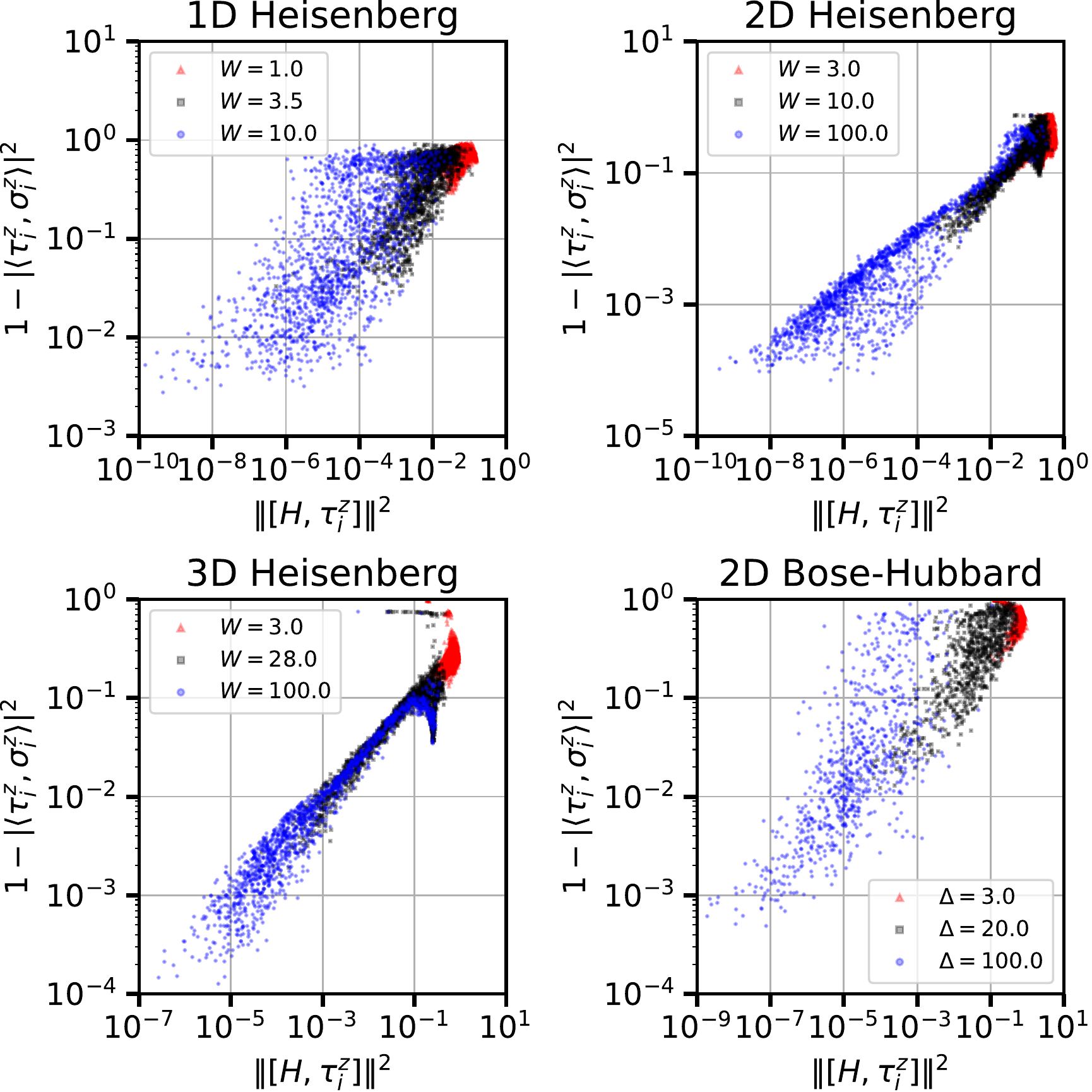}
\end{tabular}
\end{center}
\caption{Scatterplots of binarity versus commutator norm (left) and $1-|\langle \tau^z_i, \sigma^z_i \rangle|^2$ versus commutator norm (right) for $\tau^z_i$ obtained with our algorithm for the four models studied.} \label{fig:scatter1}
\end{figure}

\begin{figure}[H]
\begin{center}
\begin{tabular}{cc}
\includegraphics[width=0.4\textwidth]{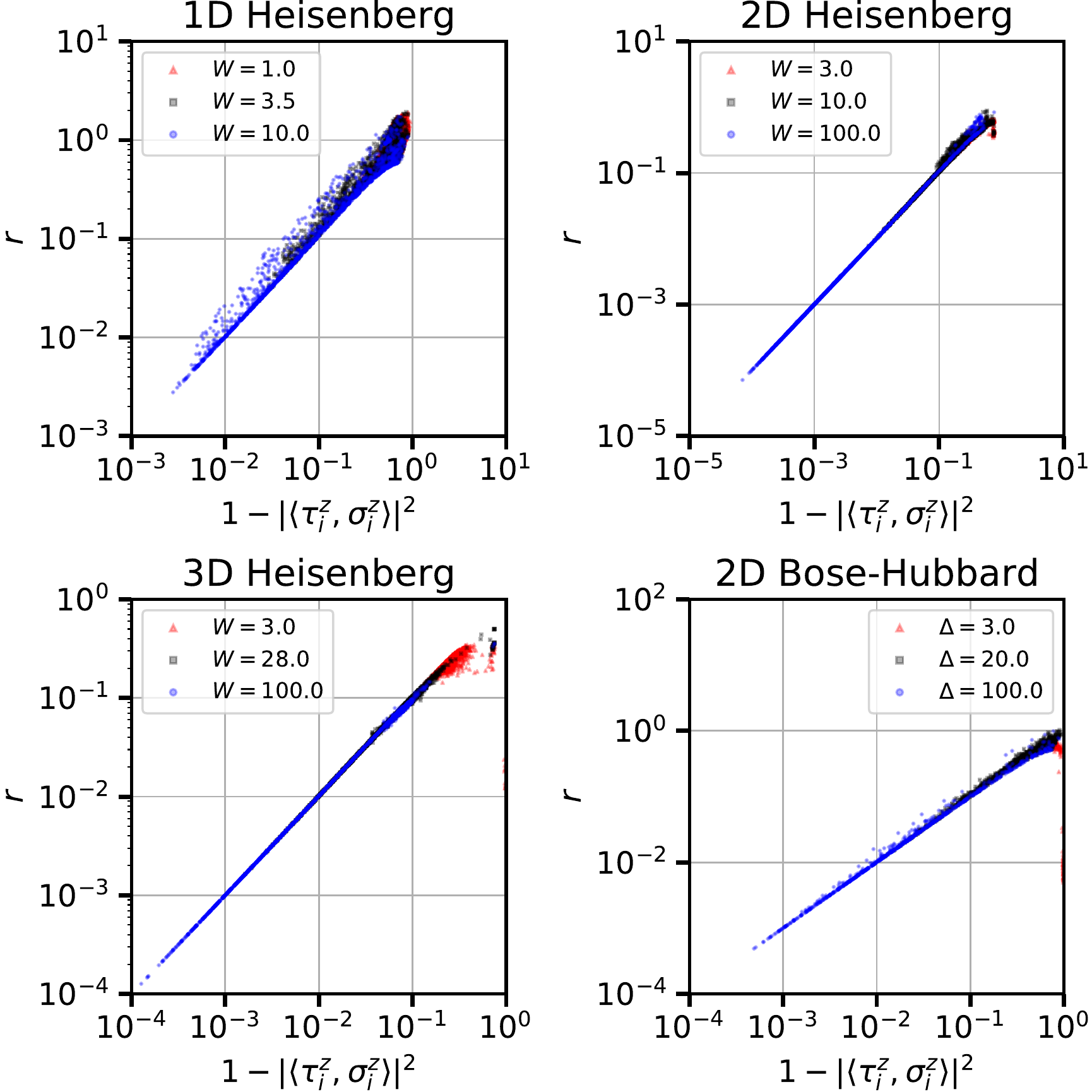} & \includegraphics[width=0.4\textwidth]{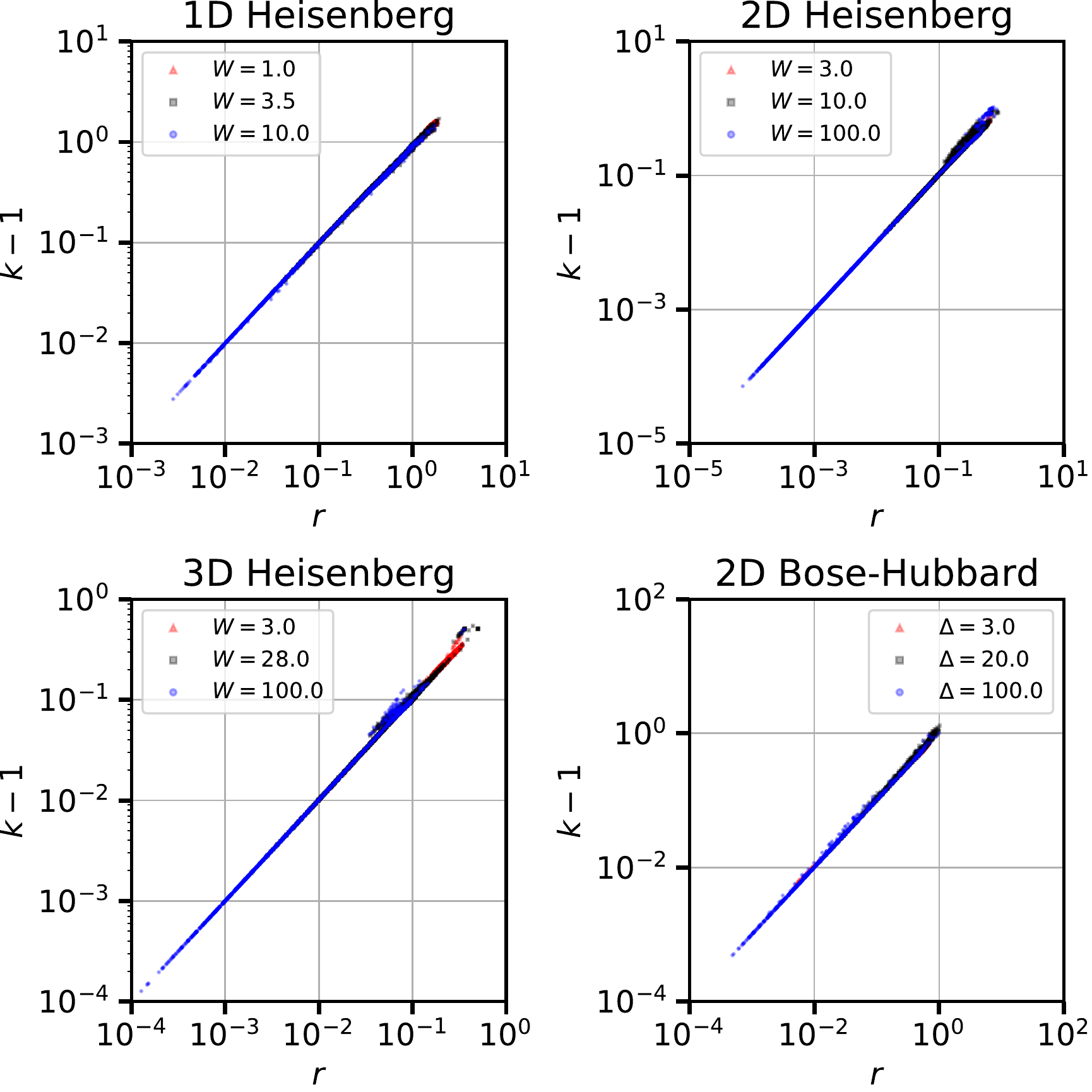}
\end{tabular}
\end{center}
\caption{Scatterplots of range $r$ versus $1-|\langle \tau^z_i, \sigma^z_i \rangle|^2$ (left) and $k-1$ versus $r$ (right) for $\tau^z_i$ obtained with our algorithm for the four models studied.} \label{fig:scatter2}
\end{figure}

Note that many of the quantities discussed above are strongly correlated with one another. The correlations between the commutator norm, binarity, $|\langle \tau^z_i, \sigma^z_i \rangle|^2$ overlaps, ranges, and localitites can be seen in the scatterplots shown in Figs.~\ref{fig:scatter1}~and~\ref{fig:scatter2}. The range $r$ and locality $k$ in particular are highly correlated so that $r \approx k-1$ to high accuracy.

\section{Analysis of correlation lengths}

\begin{figure}[H]
\begin{center}
\includegraphics[width=\textwidth]{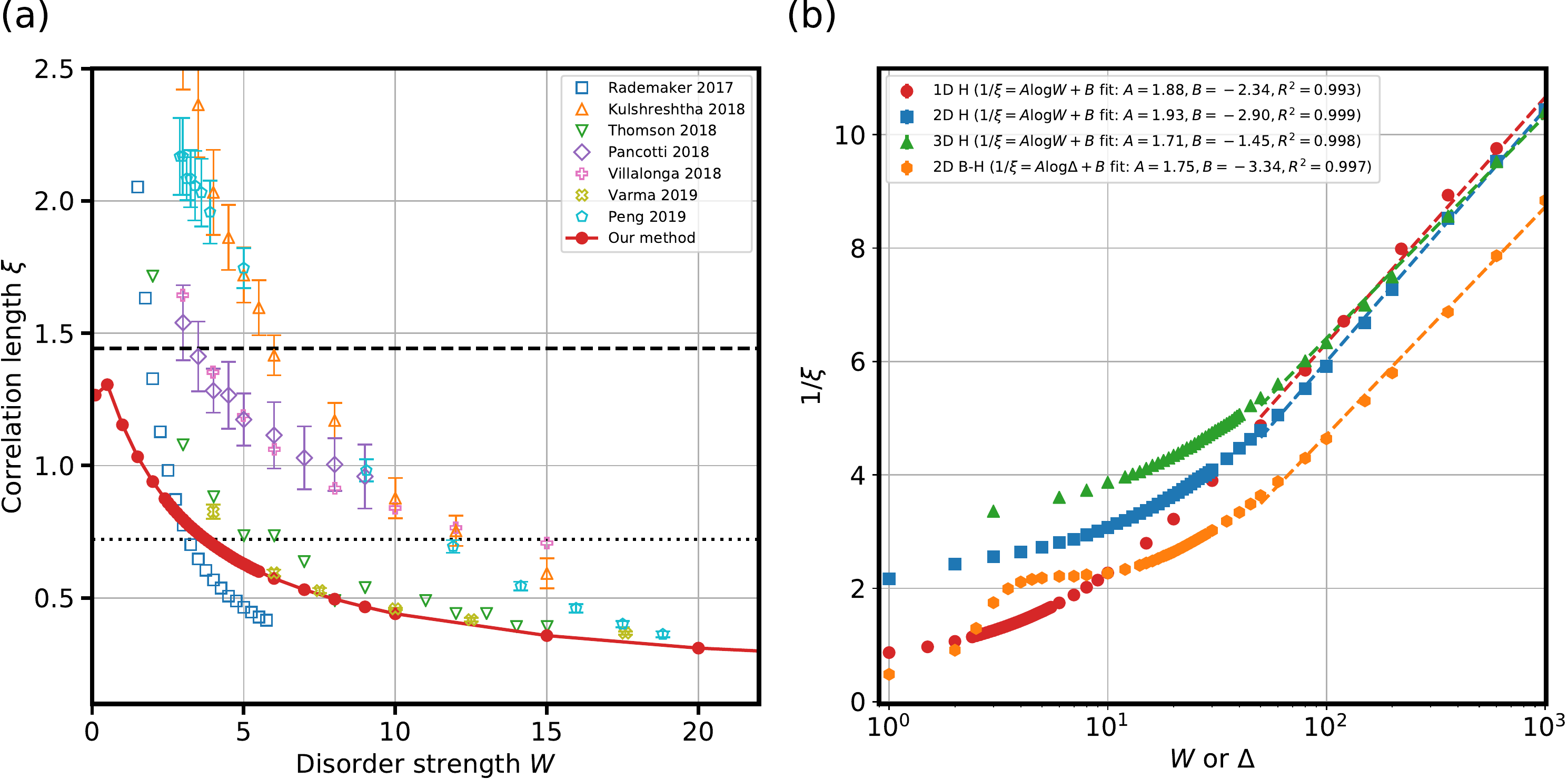}
\end{center}
\caption{(a) The average correlation lengths obtained with our method compared against values from past studies (described in text) for the 1D Heisenberg model. The dotted (dashed) horizontal line corresponds to a value of $1/\ln(4)$ ($1/\ln(2)$). (b) The inverse correlation lengths versus disorder strengths on a log-scale for the four models studied with fits to their large disorder strength ($\geq 50$) behavior.} \label{fig:corr_lengths}
\end{figure}

In past MBL studies, many correlation or localization lengths, defined in various ways, have been examined. Here we compare the correlation lengths of our $\tau^z_i$ operators obtained with our method against other $\ell$-bit lengthscales obtained in past studies of the 1D Heisenberg model \cite{Rademaker2017,Thomson2018,Kulshreshta2018,Pancotti2018,Peng2019} and the 2D hard-core Bose-Hubbard model \cite{Wahl2019}. We include this comparison for reference, but stress that these quantities are measured in different ways and might not be directly comparable. Fig.~\ref{fig:p_corr_length} shows the distributions of our correlation lengths for the different models studied. Fig.~\ref{fig:corr_lengths} (and Fig.~4 from the main paper) shows how our average correlation lengths compare with previous studies.

Rademaker 2017 \cite{Rademaker2017} obtained approximate $\ell$-bits obtained using a displacement transformation technique. Their data shown in Fig.~\ref{fig:corr_lengths} (taken from Fig.~10 of Ref.~\onlinecite{Rademaker2017}) was obtained by fitting an exponential decay to a quantity similar to the weight $w_{\vec{r}}$ of their approximate $\ell$-bits generated by a sixth-order displacement transformation on a length $L=20$ chain. Kulshreshta 2018 \cite{Kulshreshta2018} obtained $\ell$-bits using a hybrid ED-tensor network approach that involves matching eigenstates. Their average correlation lengths (taken from Fig.~3 of Ref.~\onlinecite{Kulshreshta2018}) were obtained by fitting to the weights of their $\ell$-bits for a length $L=14$ chain. Thomson 2018 \cite{Thomson2018} found approximate $\ell$-bits using a continuous unitary flow procedure, in which they restricted their Hamiltonian and $\ell$-bit ansatze to be composed of $1$ and $2$-local fermionic operators. Their average correlation lengths (taken from Fig.~1(c) of Ref.~\onlinecite{Thomson2018}) were obtained by fitting an exponential decay to coupling constants of their approximately diagonalized MBL Hamiltonian for a $L=100$ chain. Pancotti 2018 \cite{Pancotti2018} obtained approximate integrals of motion (that are often not binary $\ell$-bits) using the slow-operator method \cite{Kim2015}, exact diagonalization, and tensor networks. Their correlation lengths (taken from the inset of Fig.~2(b) of Ref.~\onlinecite{Pancotti2018}) were also obtained by fitting to the weights of their operators. Villalonga 2018 \cite{Villalonga2018} obtained approximate $\ell$-bits from one-particle orbitals using matrix product states. Their correlation lengths (taken from Fig.~4 of Ref.~\onlinecite{Villalonga2018}) were obtained by fitting to a quantity similar to the weight of their operators for a length $L=32$ chain. Varma 2019 \cite{Varma2019} obtained approximate $\ell$-bits using a Wegner-Wilson flow. The shown correlation lengths (taken from Fig.~1 of Ref.~\onlinecite{Varma2019}) are obtained from fitting an exponential to a transverse correlation function $|\langle n |\sigma^x_i \tau^x_j|n\rangle|$. Peng 2019 \cite{Peng2019} obtained $\ell$-bits using a ``quicksort-like'' algorithm involving ED for constructing binary $\ell$-bits that have high overlap with $\sigma^z_i$. Their correlation lengths (taken from Fig.~4(c) of Ref.~\onlinecite{Peng2019}) were obtained by fitting to the overlaps $\langle {\tau^z_i} , {\sigma^z_j} \rangle \sim \exp(-|i-j|/\xi)$ for a chain of length $L=12$. 

Interestingly, there is significant disagreement in the correlation lengths shown in Fig.~\ref{fig:corr_lengths}. Some studies predict that $\xi \approx 1/\ln(2)$ at the transition, though Ref.~\onlinecite{Varma2019} suggests that $\xi \approx 1/\ln(4)$ at the transition. Some of the data in the figure agree better with $1/\ln(2)$, though our data appear more consistent with $1/\ln(4)$. This discrepancy can potentially be attributed to whether the $\ell$-bits considered are ``edge'' or ``bulk'' spins, which has to do with whether the chain has periodic or open boundary conditions. In all of the studies mentioned above, Refs.~\onlinecite{Rademaker2017,Kulshreshta2018,Thomson2018,Pancotti2018,Villalonga2018,Varma2019,Peng2019}, the authors considered chains with open boundary conditions (this is implied, but not explicitly stated in Ref.~\onlinecite{Thomson2018}) while we considered the bulk of arbitrarily large chains. Moreover, except for Ref.~\onlinecite{Pancotti2018}, all of the mentioned studies found every $\ell$-bit (including edge $\ell$-bits) in the system for each disordered realization, while we only found a single $\ell$-bit at a time. This suggests that a significant fraction of previously considered $\ell$-bits might be ``edge'' spins, while our $\ell$-bits are ``bulk'' spins. Other potential explanations for the disagreements in the correlation length data include: the existence of multiple length-scales in MBL systems, finite-size effects, finite basis-size effects, differences in methodology, differences in correlation length definitions, algorithmic biases, and the non-uniqueness of $\ell$-bits.

For the 2D hard-core Bose-Hubbard model, we compared our average correlation lengths with the average correlation lengths obtained by Wahl 2019 \cite{Wahl2019} (see Fig.~4 in main paper). In their work, they used a shallow 2D tensor network to represent a short quantum circuit $U$ that could be used to construct binary $\ell$-bits of the form $\tau^z_i = U^\dagger \sigma^z_i U$. Their correlation lengths (taken from Fig.~6 of Ref.~\onlinecite{Wahl2019}) were obtained by fitting an exponential decay to a disordered-averaged eigenstate-averaged density-density correlation function. Our correlation lengths agree with Wahl 2019 at large disorder ($\Delta \geq 150$), though are larger at smaller disorder (Fig.~4 in the main paper). Also, as shown in Fig.~2(b), the average commutator norms of our approximate $\ell$-bits are orders of magnitude lower than their average commutator norms (scaled appropriately). This suggests that our method is able to construct more accurate $\ell$-bits with longer-range correlations than the shallow tensor network method used in Ref.~\onlinecite{Wahl2019}. 

Fig.~\ref{fig:corr_lengths}(b) shows the average correlation lengths that we measure over many decades of disorder strength. For all of the models studied, we find that the inverse correlation lengths scale logarithmically with disorder strength in the large disorder strength limit.

\section{Behavior of 2D hard-core Bose-Hubbard model at low disorder}

\begin{figure}[H]
\begin{center}
\includegraphics[height=0.5\textheight]{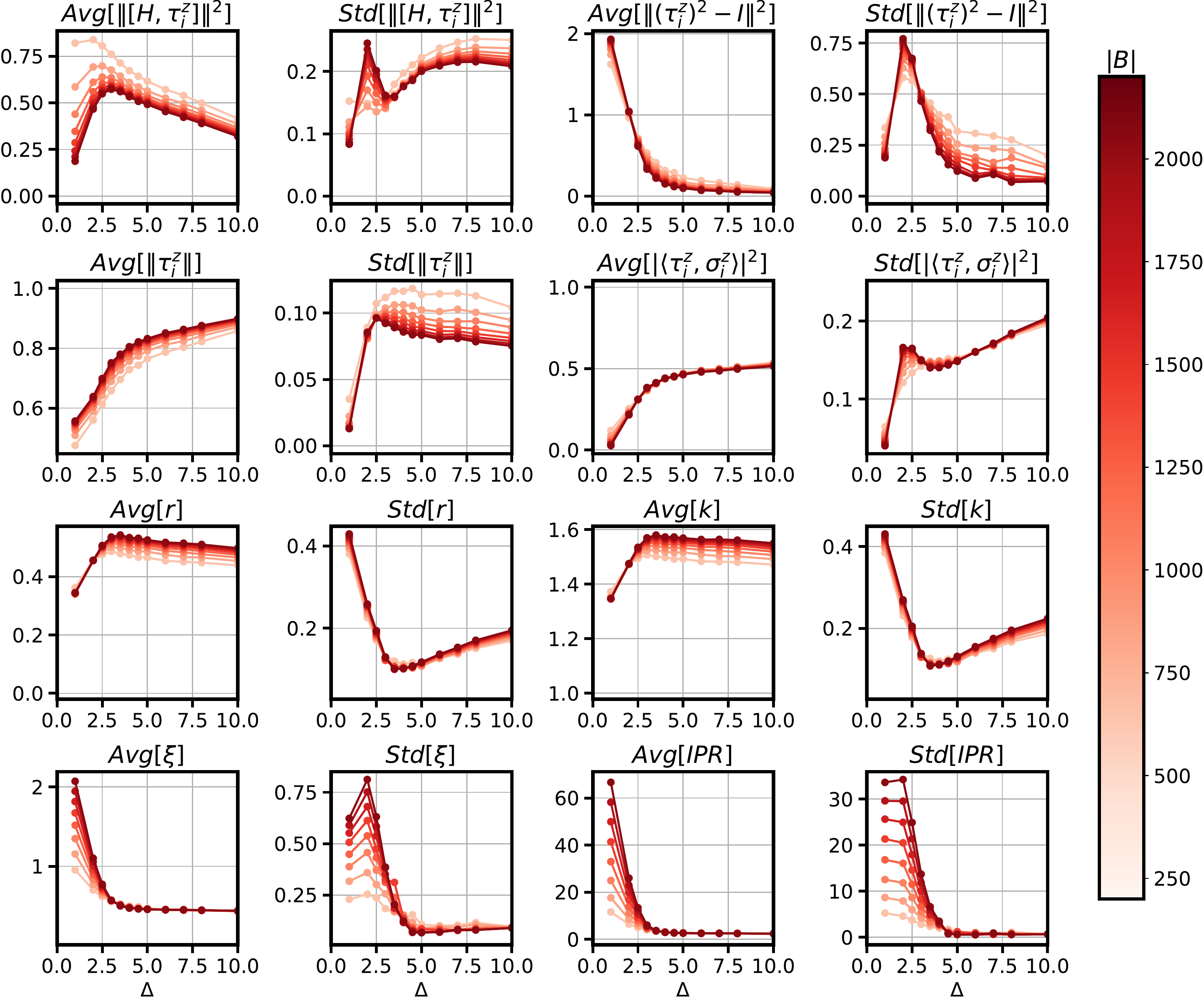}
\end{center}
\caption{A zoomed-in look at the averages and standard deviations of various quantities for the 2D hard-core Bose-Hubbard model at low disorder strength $\Delta$ where the algorithm fails to find binary operators.} \label{fig:zoomBH}
\end{figure}

Here we comment on the properties of the $\tau^z_i$ operators that we find at very low disorder strength $\lesssim 5$ for the 2D hard-core Bose-Hubbard model (see Fig.~\ref{fig:zoomBH}). At very low disorder, we find that many of the $\tau^z_i$ operators are essentially linear combinations of only single-site Pauli matrices so that $\tau^z_i \approx \sum_j c_j \sigma^z_j$ with $|c_j|^2$ peaked at site $i$ and spread over many sites (with occasional two-site Pauli strings contributing strongly). This causes $\tau^z_i$ to have a large $IPR$ and $\xi$ but small $|\langle \tau^z_i, \sigma^z_i \rangle|^2$, $r$, and $k$. Because of the small basis size, our algorithm is unable to reduce the binarity $\norm{(\tau^z_i)^2 - I}^2$ of the $\tau^z_i$ operator, as one can see in Fig~\ref{fig:zoomBH}. However, the algorithm is able to improve the commutator norm $\norm{[H, \tau^z_i]}^2$ by constructing a delocalized operator of the form $\tau^z_i \approx \sum_j c_j \sigma^z_j$. This particular delocalized operator commutes well with the Hamiltonian because it has high overlap with the total magnetization operator, $S_{tot}^z = \sum_{j=1}^N \sigma^z_j$, which is an exact integral of motion of the Hamiltonian. Note that for all four of the Hamiltonians considered $S_{tot}^z$ is an integral of motion and this tendency for producing $\tau^z_i$ that have high overlap with $S_{tot}^z$ at very low disorder exists. Yet, as we find empirically, this tendency is particularly strong in the Bose-Hubbard model. This might be due to the physics of the particular model or perhaps the nature of the disorder, which is generated from Gaussian instead of uniform sampling. Interestingly, the transition into this behavior is observed at low disorder, near $\Delta \approx 5$, close to the transition value of $\Delta_c^{exp} \approx 5.5(4)$ obtained by the experiment in Ref.~\onlinecite{Choi2016} (with non-hard-core bosons and interaction strength $U'=24.4$). 

We also note here that, for the Bose-Hubbard model, about half of the Pauli strings in the basis produced during our basis expansion heuristic had amplitudes $c_a$ in the $\tau^z_i = \sum_a c_a \mathcal{O}_a$ operators that were zero to machine precision. This was not the case for the other Hamiltonians and we believe might be due to the symmetries of the model. While this affects the effective basis sizes for the Bose-Hubbard model, it does not change any of our results or conclusions.

\section{Analysis of cumulative distribution functions}

\begin{figure}[H]
\begin{center}
\includegraphics[width=0.8\textwidth]{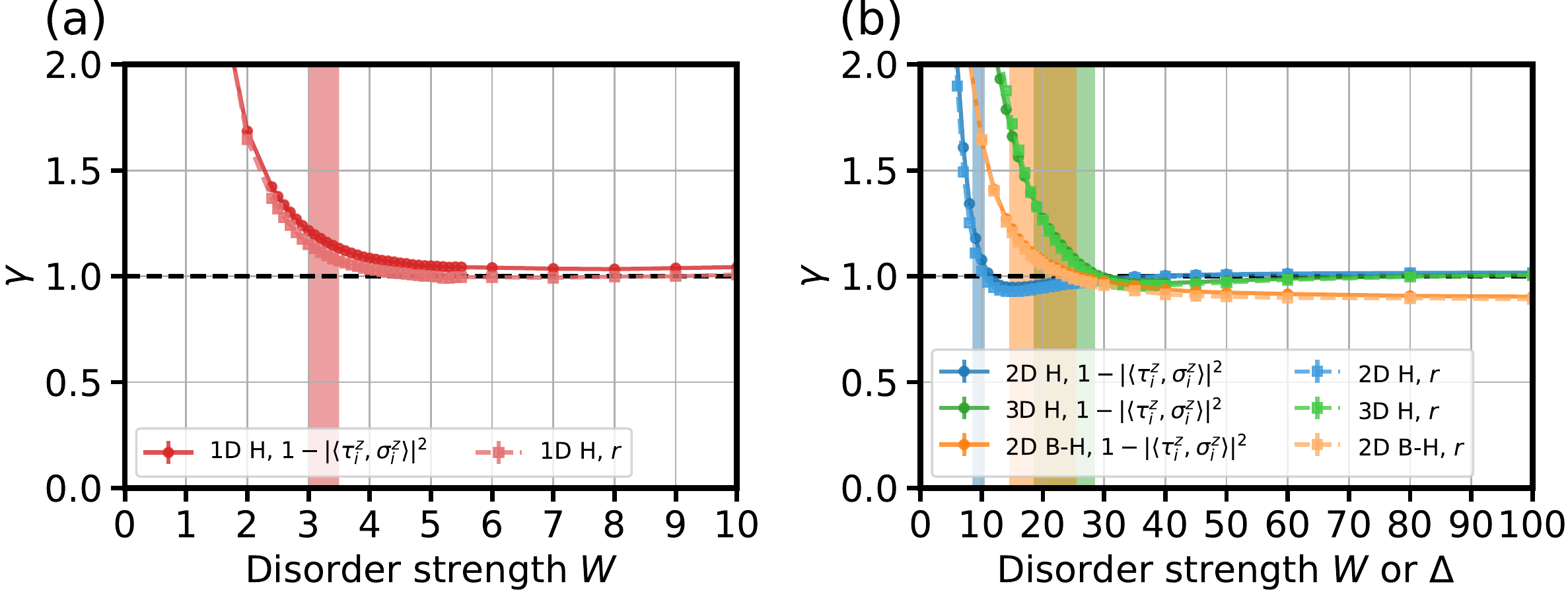}
\end{center}
\caption{The exponent $\gamma$ of the cumulative distribution function $CDF(x) \propto x^\gamma$ (obtained from fitting) versus disorder strength for $x=1 - |\langle \tau^z_i, \sigma^z_i \rangle|^2$ and $x = r$. (a) The exponents for the 1D Heisenberg model. (b) The exponents for the 2D and 3D Heisenberg models and the 2D Bose-Hubbard model. The approximate transition regions for the four models (see Fig.~3 in main text) are shaded for reference.} \label{fig:cdf_exponents}
\end{figure}

A key feature of the distributions of $|\langle \tau^z_i, \sigma^z_i \rangle|^2$ in the MBL phase is the presence of the sharp peak at $|\langle \tau^z_i, \sigma^z_i \rangle|^2 \approx 1$ (see Fig.~\ref{fig:p_zoverlap}), indicating a large probability of observing highly localized $\ell$-bits in the system. Here we analyze the cumulative distribution function (CDF) of $|\langle \tau^z_i, \sigma^z_i \rangle|^2$ for values of $|\langle \tau^z_i, \sigma^z_i \rangle|^2 \approx 1$. In particular, we fit the CDF of $x=1-|\langle \tau^z_i, \sigma^z_i \rangle|^2$ to a function of the form $CDF(x) = A x ^ \gamma$ for small $x$ and identified the exponent $\gamma$ for different disorder strengths (see Fig.~\ref{fig:cdf_exponents}). The idea is that, since the probability density function (PDF) is the derivative of the CDF, if $\gamma = 1$ as $x \rightarrow 0^+$ then $PDF(x\rightarrow 0^+) \propto \lim_{x \rightarrow 0^+} x^{\gamma - 1}$ is constant and there is a finite probability of observing a highly localized $\ell$-bit. For all of the models considered, at high disorder $\gamma \approx 1$. At disorder strengths near the transition regions (see Fig.~3 in the main paper), $\gamma$ begins to diverge, potentially suggesting a transition. Note that we observe the same behavior for $x=r$.

\begin{figure}[H]
\begin{center}
\includegraphics[height=0.35\textheight]{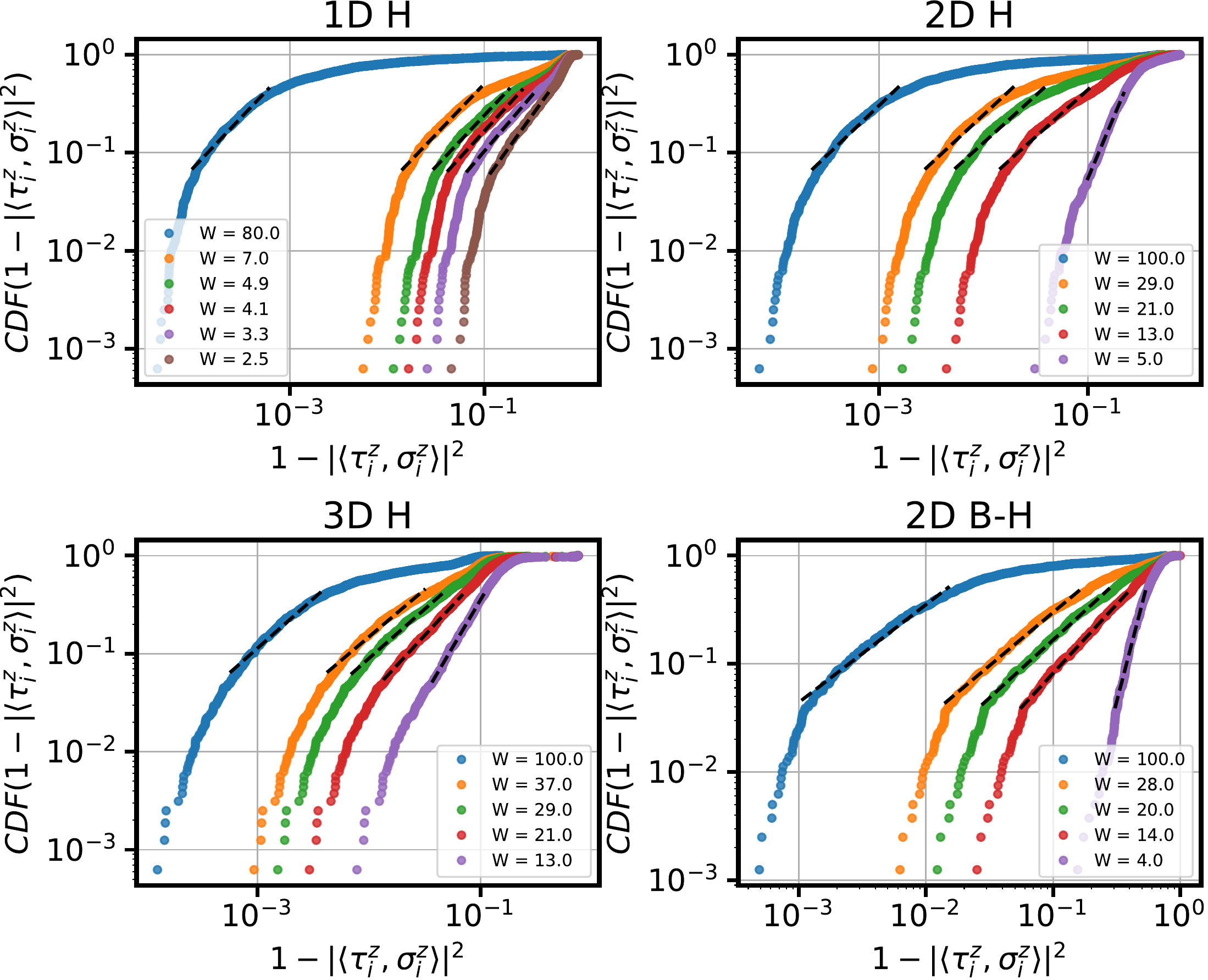}
\end{center}
\caption{Cumulative distribution functions (CDFs) of $1-|\langle \tau^z_i, \sigma^z_i \rangle|^2$ on a log-log scale superimposed with linear fits for different disorder strengths and different models.} \label{fig:cdfs}
\end{figure}

However, we would like to emphasize that the results in Fig.~\ref{fig:cdf_exponents}, while suggestive, are not entirely trustworthy. As can be seen by the fits used to obtain the values of $\gamma$, shown in Fig.~\ref{fig:cdfs}, the CDF is not obviously captured by a power-law form and has a noticeable change at very low $1-|\langle \tau^z_i, \sigma^z_i \rangle|^2$, which we do not fully understand. When we perform our fitting, we first sort the $N$ data points $x_1,\ldots, x_N$ (e.g., the $1-|\langle \tau^z_i, \sigma^z_i \rangle|^2$ values of the $\tau^z_i$ operators) in increasing order so that $x_1 \leq x_2 \leq \cdots \leq x_N$. Then, we define $CDF(x_n) = n/N$ as a discretized representation of the CDF. Finally, we perform linear fits on the data points $(\log_{10}(x_n), \log_{10}(CDF(x_n)))$ and determine the slope $\gamma$. We perform these fits using only points with $x_n$ in the range $c_0 \leq CDF(x_n) \leq c_1$. For the 1D, 2D, and 3D Heisenberg models, we set $c_0=0.05$ and $c_1=0.4$. For the 2D hard-core Bose-Hubbard model, we set $c_0=0.03$ and $c_1=0.35$. Setting $(c_0, c_1)$ to different values changes this fit, making our results somewhat arbitrary. Also, in Fig.~\ref{fig:cdf_exponents}, we omit $\gamma$ values obtained from bad linear fits with $R^2 < 0.95$, which occur for small values of $\Delta$ for the 2D Bose-Hubbard model. Note that the CDFs and fits for $x=r$ look similar. While these results are not completely reliable or systematic, we include them here for completeness. 

\section{Drift of $\ell$-bits}

\begin{figure}[H]
\begin{center}
\includegraphics[height=0.35\textheight]{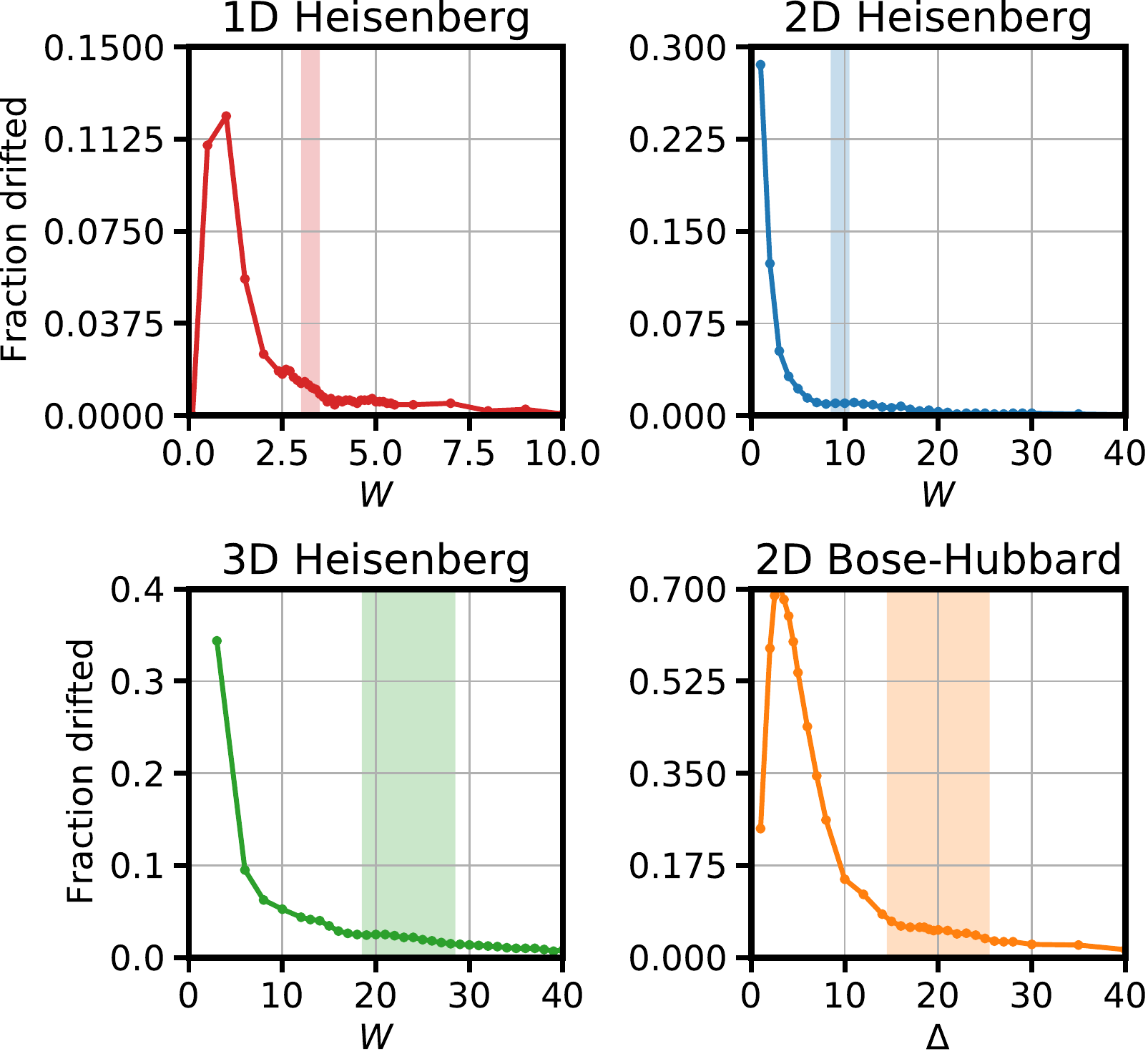}
\end{center}
\caption{The fraction of $\tau^z_i$ operators that drift away from their original site versus disorder strength for the four models studied. A $\tau^z_i$ operator has drifted away if its weight $w_{\vec{r}}$ is not maximal at the position where it was initialized. For reference, we shade the approximate transition regions (see Fig.~3 in the main paper).} \label{fig:lbit_drift}
\end{figure}

At low disorder, it is possible for our algorithm to find $\tau^z_i$ operators that are not located at the original site $i$ used to initialize the basis $B=\{ \sigma^z_i\}$. However, as shown in Fig.~\ref{fig:lbit_drift}, this algorithmic ``drift'' of $\ell$-bits away from their initial site is not significant for most disorder strengths considered. In particular, it only becomes significant in certain models and only for disorder strengths significantly below where we observe signatures of MBL transitions.

\section{Basis expansion heuristic behavior}

\begin{figure}[H]
\begin{center}
\includegraphics[height=0.4\textheight]{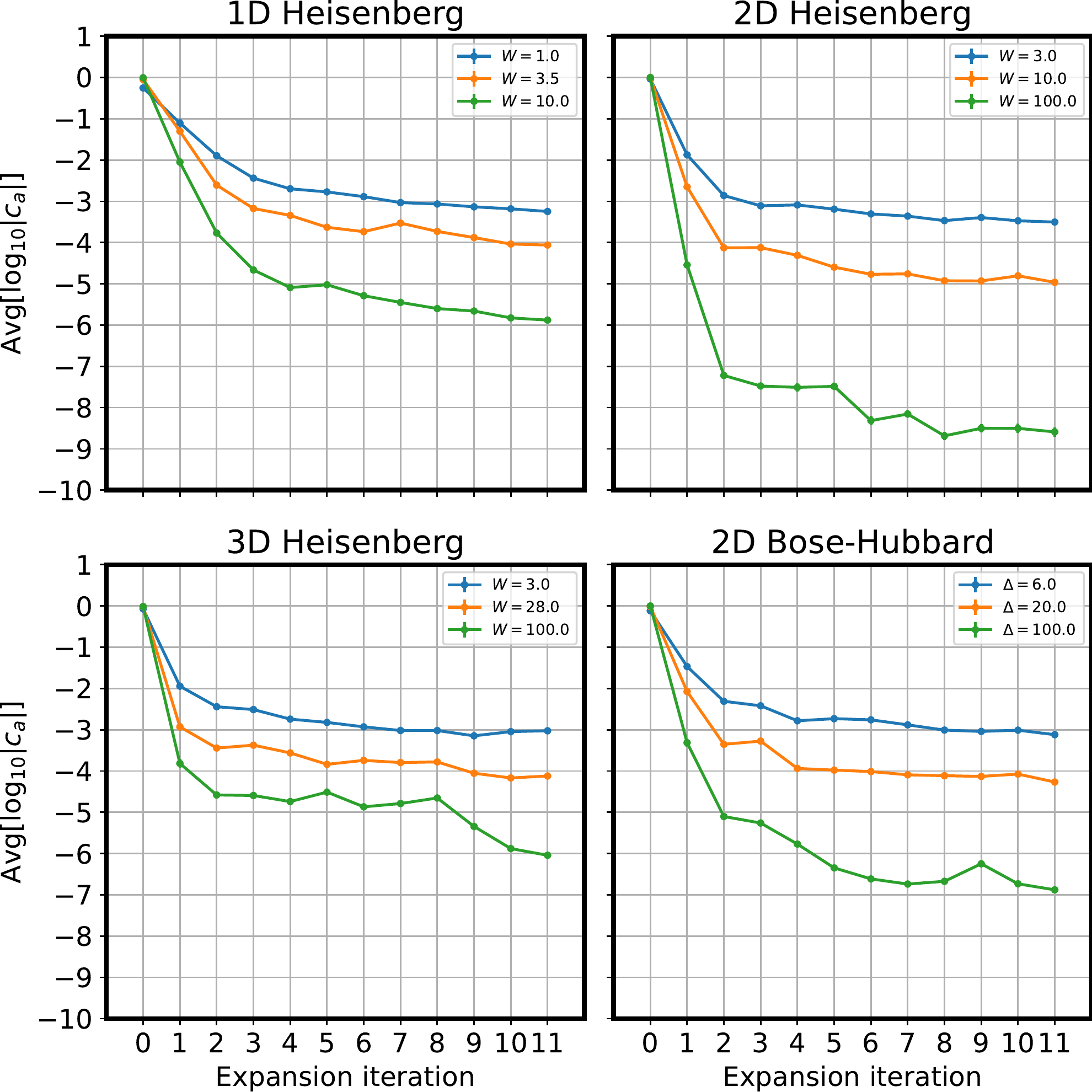}
\end{center}
\caption{Log-averages of the amplitudes $c_a$ of the Pauli strings $\mathcal{O}_a$ that were added at each expansion iteration of the algorithm. The amplitudes shown here were taken from the optimized $\tau^z_i = \sum_a c_a \mathcal{O}_a$ operator produced after the final (11th) expansion iteration. We show examples of single $\tau^z_i$ operators for each of the four models studied at low, high, and intermediate disorder strengths. The errorbars (standard errors of $\log_{10}|c_a|$) are smaller than the markers.} \label{fig:expansions}
\end{figure}

As described in the main text, the algorithm that we use to construct the approximate $\ell$-bit operators works by systematically expanding the basis of Pauli strings used to represent the $\ell$-bit. We know that our procedure is reasonable because the objective function (commutator norm plus binarity) systematically decreases after each expansion iteration. (One can see how the terms in the objective function decrease during the expansion iterations in Figs.~\ref{fig:scaling_comm_norm}~and~\ref{fig:scaling_binarity}.) That said, it is interesting to see whether our procedure is able to identify more important Pauli strings at earlier iterations. Indeed, we see this behavior in Fig.~\ref{fig:expansions}, which shows that the Pauli strings added in later iterations typically contribute less to the final optimized $\tau^z_i$ operators than the Pauli strings added in the earlier iterations.

\section{Estimation of transition regions}
\label{sec:transition_regions}

In this section we describe the method used to estimate the delocalization transition regions quoted in the main text for the various models studied.
As discussed in the main text, we observe signatures of bimodality in the distribution of the overlap of $\ell$-bits with single-site $\sigma^z_i$ operators, $|\langle \tau^z_i, \sigma^z_i\rangle|^2$ (see Fig. 3 of the main text).
This can be seen as the coexistence of highly localized and substantially delocalized $\ell$-bits: at disorder strengths higher than the transition values the overlap is close to 1, while below the transition the overlap is substantially smaller than 1; at the transition region, both behaviors coexist.
Note that other hints of bimodal behavior at the transition are also found in the literature~\cite{kjall2014many,Yu2016,Villalonga2018}.

\begin{figure}[H]
\begin{center}
\includegraphics[height=0.4\textheight]{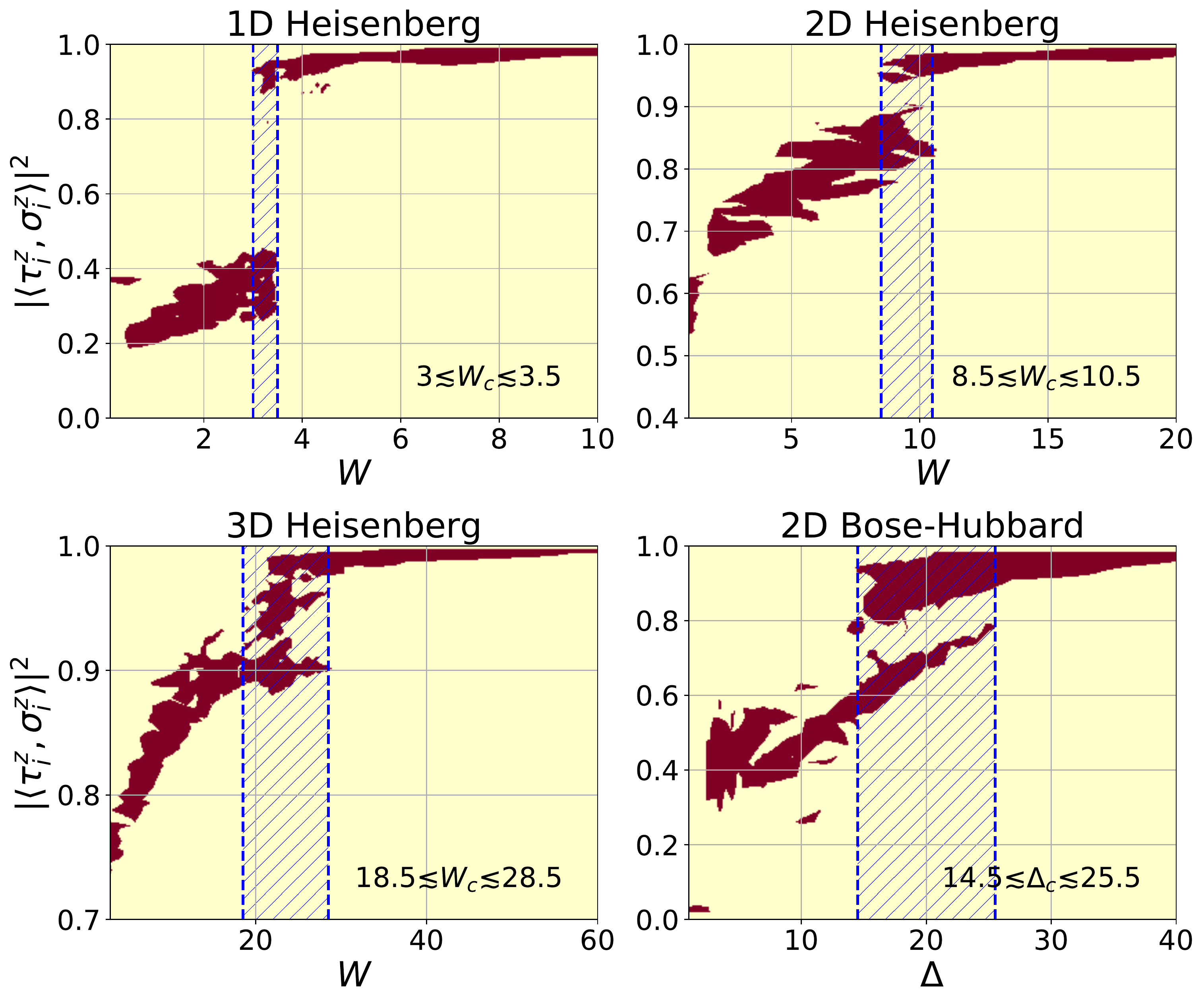}
\end{center}
\caption{Estimation of the transition regions of the four models studied.
A threshold $t_{\rm overlap} = 0.75$ is applied to the data presented in Fig.~3 of the main text, which results in the clear identification of a bimodal region around the transition.
The transition regions are shaded in the plot, and their approximate lower and upper bounds are written in the lower right corner of each panel.}
\label{fig:transition_regions}
\end{figure}

\begin{figure}[H]
\begin{center}
\begin{tabular}{cc}
\includegraphics[width=0.5\textwidth]{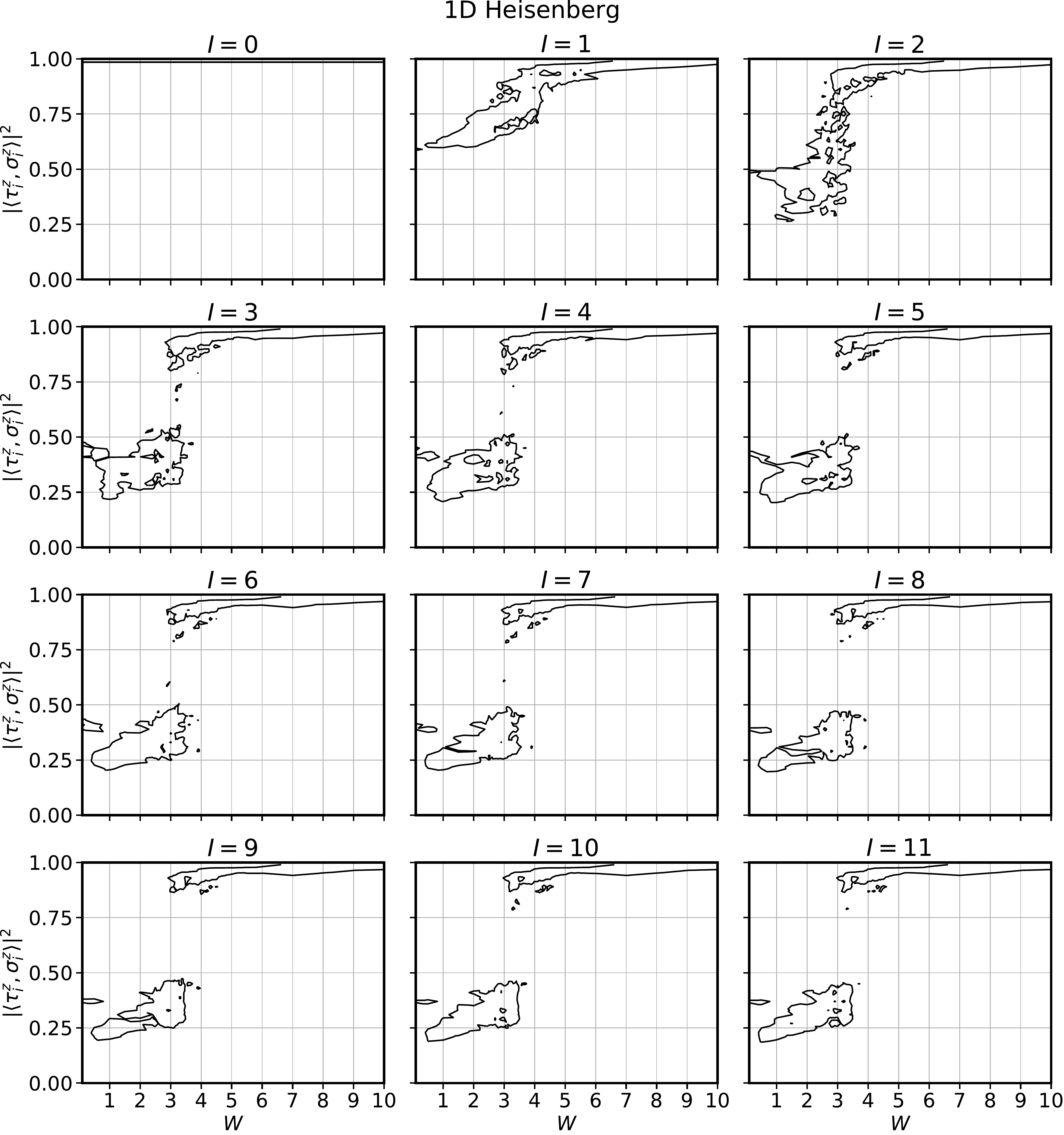} & \includegraphics[width=0.5\textwidth]{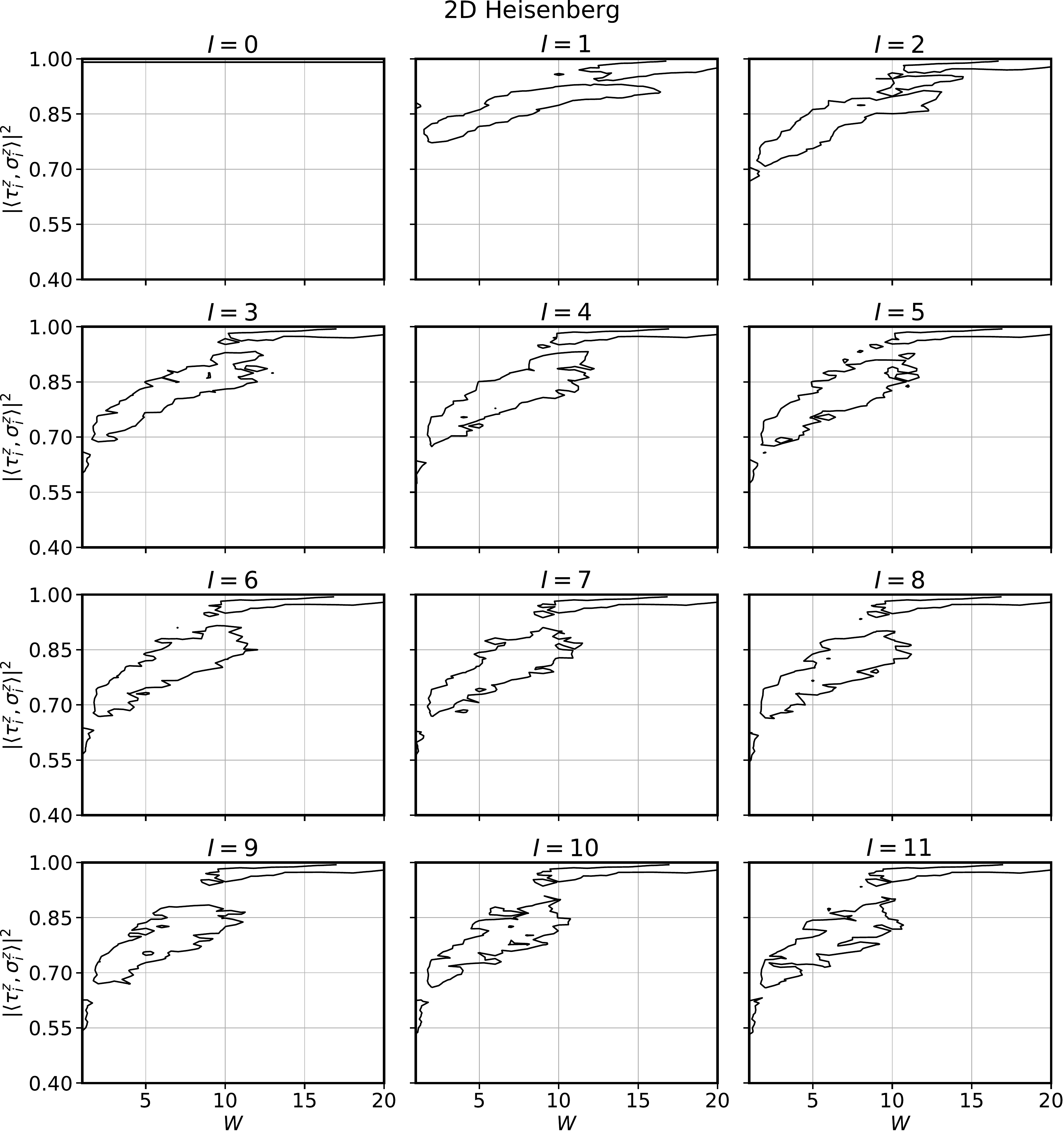} \\
\includegraphics[width=0.5\textwidth]{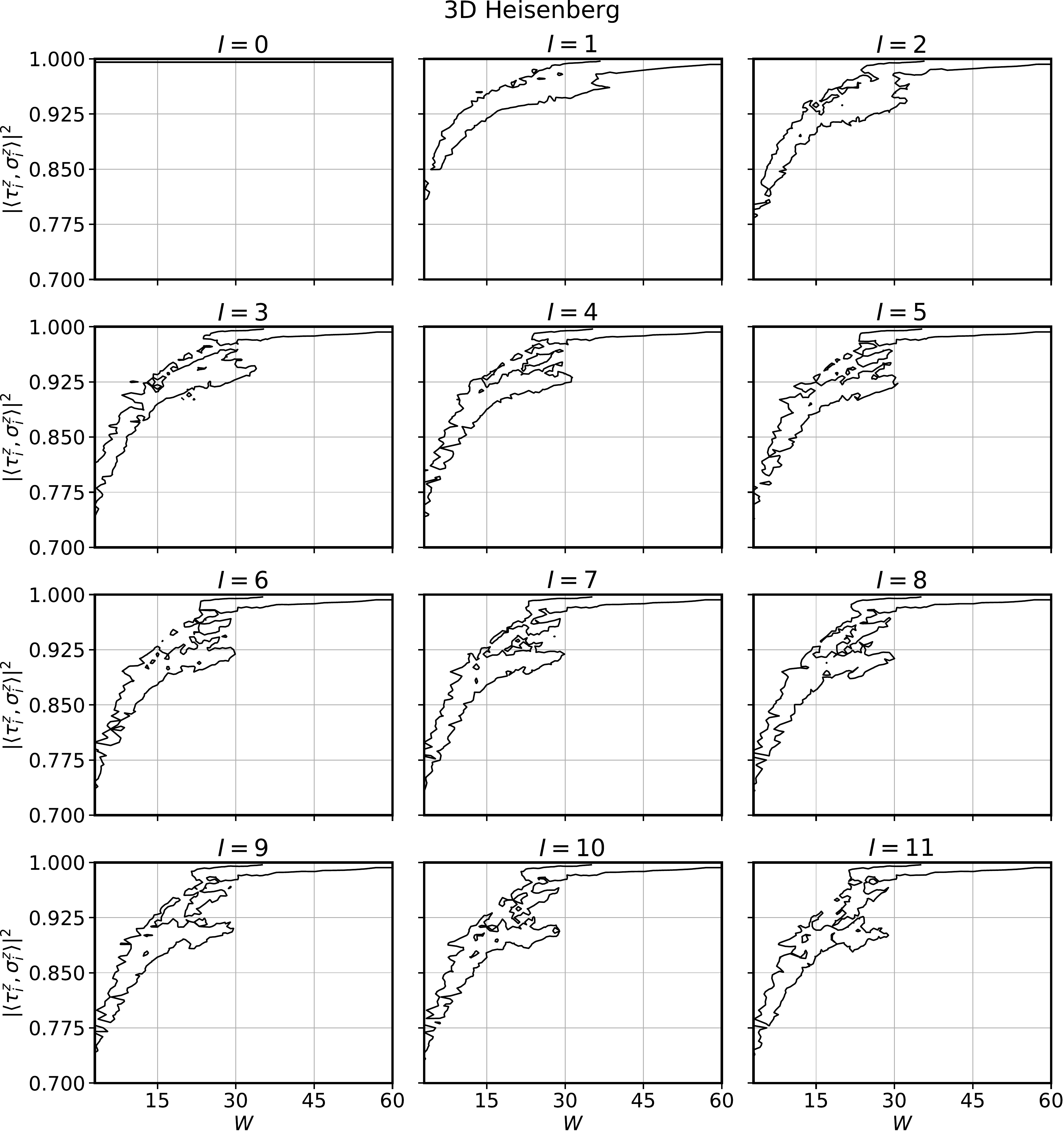} & \includegraphics[width=0.5\textwidth]{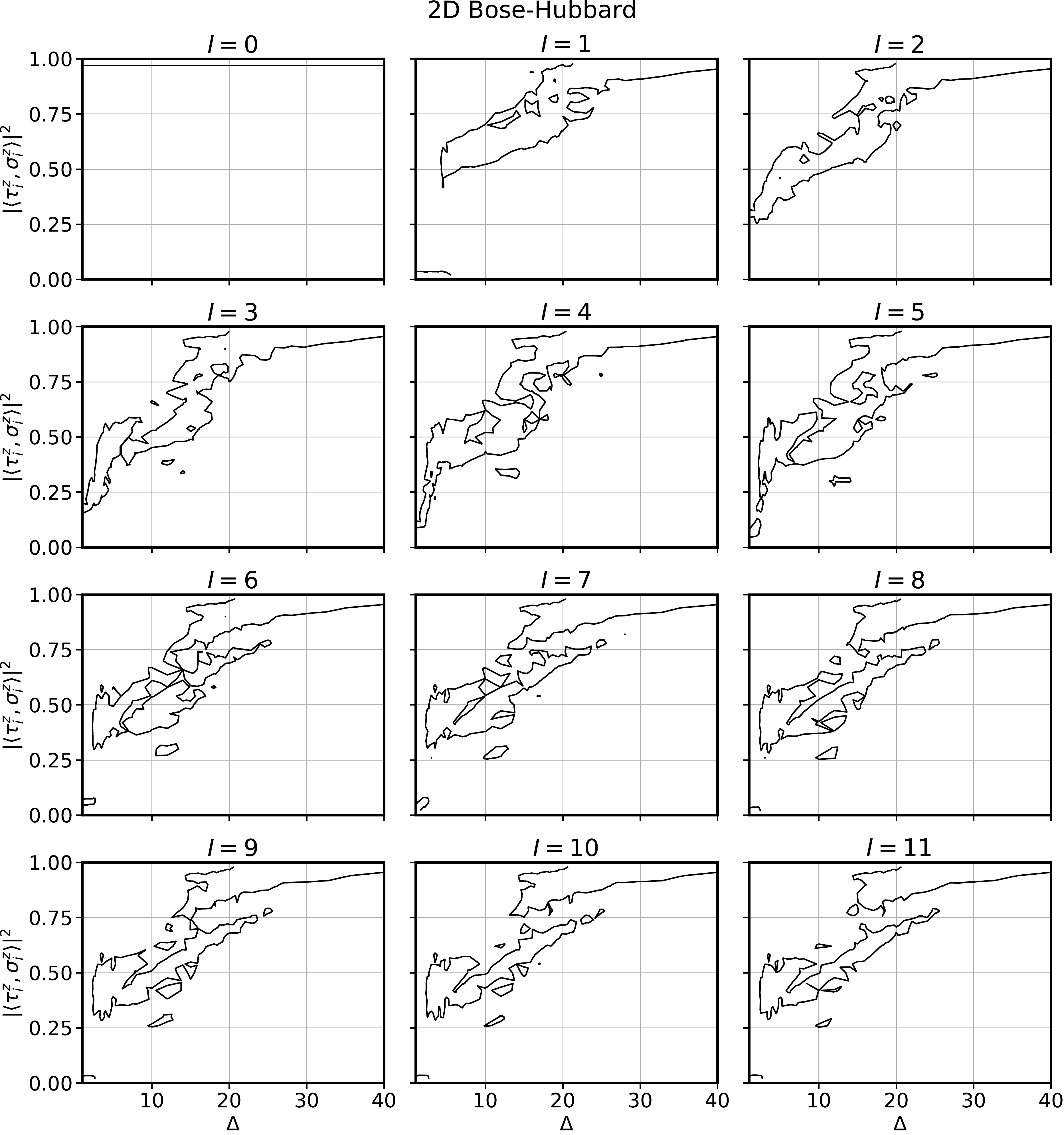}
\end{tabular}
\end{center}
\caption{Outlines of the regions with $f(W, |\langle \tau^z_i, \sigma^z_i \rangle|^2) > t_{\rm overlap} = 0.75$ for iterations $I=0,\ldots,11$ of the basis expansion heuristic.
Lines represent contour lines at $f(W, |\langle \tau^z_i, \sigma^z_i \rangle|^2) = 0.75$}
\label{fig:transition_regions_iterations}
\end{figure}

In order to estimate the transition regions in a systematic way, we perform an analysis on the data presented in Fig. 3 of the main text, which shows the histograms of the overlap $|\langle \tau^z_i, \sigma^z_i \rangle|^2$ at fixed disorder strength $W$ ($\Delta$ in the case of the 2D Bose-Hubbard), normalized so that their maximum is 1, and interpolated over a window of values of $W$ (resp. $\Delta$).
Let us denote this by $f(W, |\langle \tau^z_i, \sigma^z_i \rangle|^2)$.
A straightforward way to approximately identify the region of bimodality of $f$ is to establish a threshold $t_{\rm overlap}$ and plot the regions where $f > t_{\rm overlap}$; we anticipate unimodal behavior (at fixed $W$) above and below the transition, and bimodal behavior around the transition.
In Fig.\ref{fig:transition_regions} we show the result of applying a threshold $t_{\rm overlap} = 0.75$, and estimate transition regions by bounding the regions of bimodality; as for Fig.~3 of the main text, this figure is generated with data from the last iteration of the basis expansion, \emph{i.e.} $I = 11$.
Fig.~\ref{fig:transition_regions_iterations} shows how the thresholded regions change at each iteration of the basis expansion.

\section{Scaling of the commutator norm and the binarity with basis size}
\label{sec:scaling}

\begin{figure}[H]
\begin{center}
\includegraphics[height=0.5\textheight]{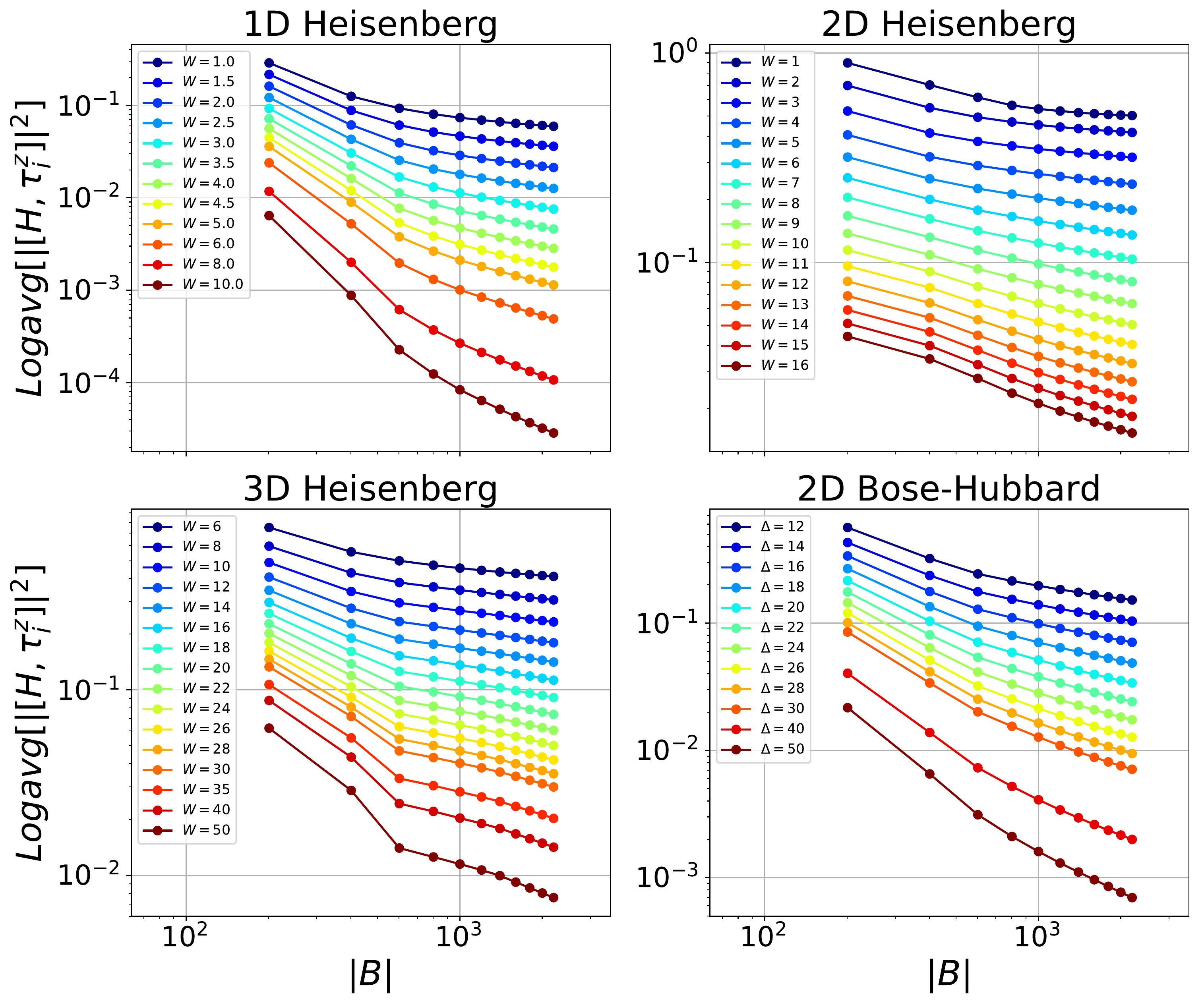}
\end{center}
\caption{Log-log plot of the log-averaged commutator norm as a function of basis size $|B|$ for a set of disorder strengths.}
\label{fig:scaling_comm_norm}
\end{figure}

In this section we study the scaling of the commutator norm and binarity with the basis size $|B|$.
In Fig.~\ref{fig:scaling_comm_norm} we show the values of the log-averaged commutator norm $\norm{[H, \tau^z_i]}^2$ as a function of $|B|$ for a subset of disorder strengths (which includes the transition region) for the four models studied.
Although the scaling presented in a log-log plot might be suggestive of a power law, we do not have enough decades of data to make any well-motivated claim about this scaling.

\begin{figure}[H]
\begin{center}
\includegraphics[height=0.5\textheight]{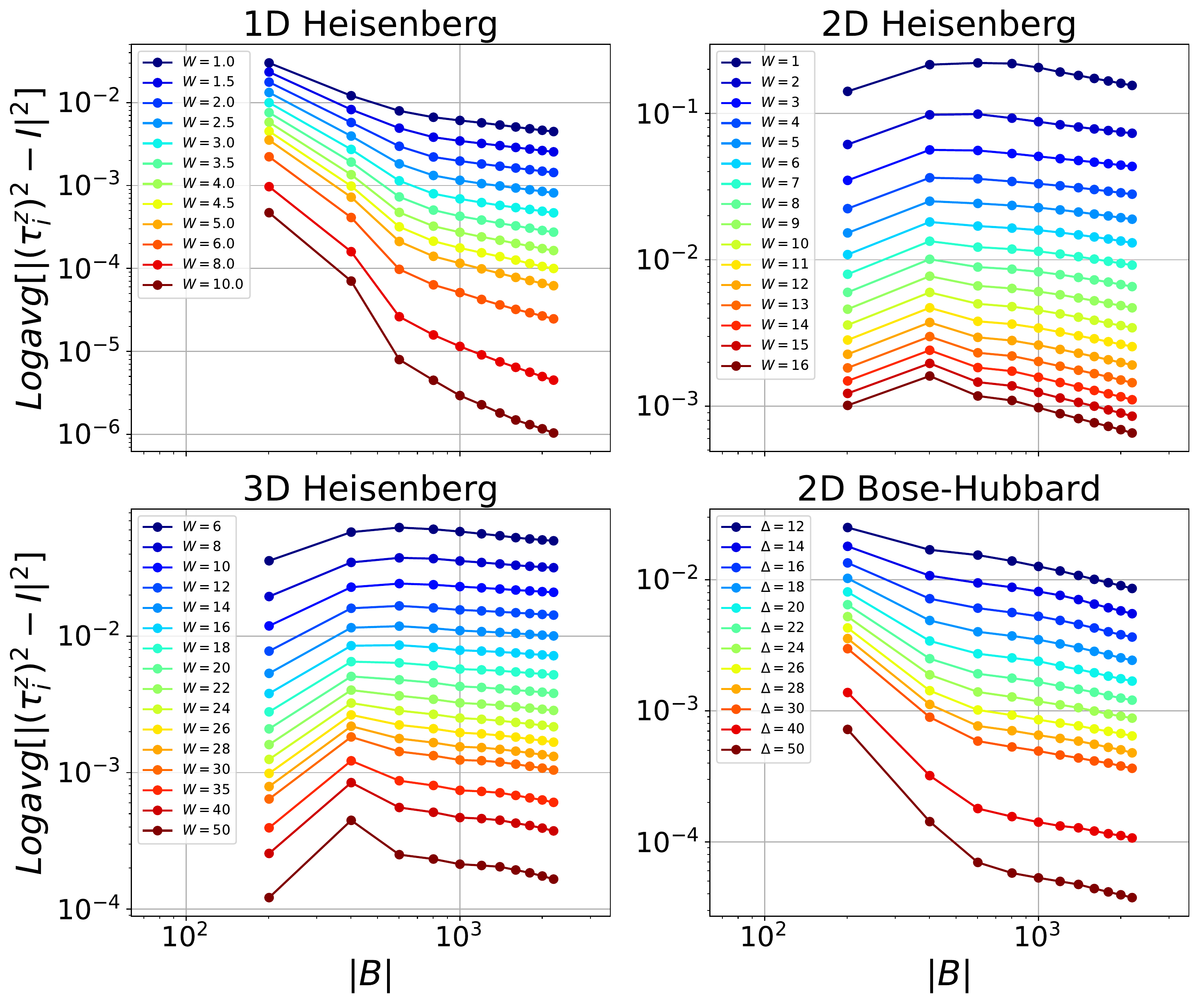}
\end{center}
\caption{Log-log plot of the log-averaged binarity as a function of basis size $|B|$ for a set of disorder strengths.}
\label{fig:scaling_binarity}
\end{figure}

Similar to the case of the commutator norm, we show in Fig.~\ref{fig:scaling_binarity} the values of the log-averaged binarity $\norm{(\tau^z_i)^2 - I}^2$ as a function of the basis size $|B|$ for a subset of disorder strengths for the four models studied.
The scaling of the binarity is similar to that one of the commutator norm.
Note that the binarity is exactly zero (\emph{i.e.}, optimal) before any basis expansion (not shown), \emph{i.e.}, when $I = 0$ we have $\tau^z_i = \sigma^z_i$.
For that reason, it is natural for the binarity to increase over the first few iterations and eventually decrease monotonically as it is optimized over larger basis sizes.
This behavior can be observed with the 2D and 3D Heisenberg models.

At very high disorder strength ($W = 1000$ or $\Delta = 1000$), the last iteration of the algorithm reaches a log-averaged commutator norm of about $10^{-22}$ for the 1D Heisenberg model, $10^{-10}$ for the 2D Heisenberg model, $10^{-7}$ for the 3D Heisenberg model, and $10^{-10}$ for the 2D Bose-Hubbard model.
Similarly, the binarity reaches log-averaged values of about $10^{-24}$, $10^{-10}$, $10^{-8}$, and $10^{-12}$, respectively.

\bibliography{refs}